%% file: choptuik.tex
\documentclass[a4paper,twoside,12pt]{article}
\usepackage{amsmath,amssymb}
\usepackage{verbatim}
\usepackage{bbm}
\usepackage{enumerate}
\usepackage{stmaryrd}
\usepackage{graphicx,color}
\usepackage{float}
\usepackage{caption}
\usepackage{pifont}
\usepackage{wasysym}

\chardef\myunderbar=`\_ 

\DeclareMathOperator{\RE}{Re}
\DeclareMathOperator{\IM}{Im}
\DeclareMathOperator{\image}{image}
\DeclareMathOperator{\diag}{diag}
\DeclareMathOperator{\Chebyshev}{Chebyshev}

\let\textlabel\label 
\newcounter{REMINDERCOUNTER}
\newcommand{\REMINDERSYMBOL}{\clubsuit}
\newcommand{\tagforcomp}{{}^{\REMINDERSYMBOL}}
\newcommand{\tagforcompref}[1]{{}^{\REMINDERSYMBOL\mathbf{\ref{#1}}}}
\newcommand{\tagforcomprefcheck}[1]{{}^{\REMINDERSYMBOL\mathbf{\ref{#1}}\hskip2pt\text{\ding{52}}}}
\newcommand{\tagforcomplabel}[1]{\begingroup\refstepcounter{REMINDERCOUNTER}\textlabel{#1}\tagforcompref{#1}\endgroup} 

\newcommand{\tagforB}{{}^\LHD}

\newcommand{\SELECTION}{\mathbf{S}}

\newcommand{\XiParity}{P}
\DeclareMathOperator{\TauDerivative}{TauDerivative}

\DeclareMathOperator{\OnePlusXiTimesXiDerivative}{OnePlusXiTimesXiDerivative}
\DeclareMathOperator{\TimesXi}{\Xi}
\DeclareMathOperator{\DivideByXiRegularized}{\Xi^{-1,\text{reg}}}
\DeclareMathOperator{\TF}{\mathbf{ToSeries}}

\newcommand{\DD}[1]{\hskip-3pt\not #1}
\DeclareMathOperator{\OOA}{J}
\DeclareMathOperator{\OOB}{K}
\DeclareMathOperator{\OO}{\mathbf{J}}

\DeclareMathOperator{\HH}{\mathbf{H}}

\DeclareMathOperator{\Band}{U}
\DeclareMathOperator{\BandEXEP}{U^{\text{sp}}_{-1,1}}
\DeclareMathOperator{\BandI}{V}

\newcommand{\Ric}{\mathrm{Ricci}}

\newcommand{\Corr}[1]{{#1}_{\mathrm{corr}}}
\newcommand{\Gauged}[1]{{#1}_{\mathrm{Gauged}}}
\newcommand{\Ref}[1]{{#1}_{\mathrm{ref}}}
\newcommand{\RefA}[1]{{#1}_{\mathrm{refA}}}
\newcommand{\RefB}[1]{{#1}_{\mathrm{refB}}}

\DeclareMathOperator{\LINDEC}{\mathbbm{L}}

\newcommand{\ks}{\mathfrak{s}}
\newcommand{\KS}{\boldsymbol{\ks}}

\newcommand{\Space}{\mathcal{V}}
\newcommand{\TauPer}{\Space_{\mathrm{TauTwoPiPeriodic}}}
\newcommand{\TauAntiPer}{\Space_{\mathrm{TauTwoPiAntiPeriodic}}}
\newcommand{\XiOdd}{\Space_{\mathrm{XiOdd}}}

\newcommand{\SPACE}{\boldsymbol{\mathcal{W}}}
\newcommand{\SSPACE}{\boldsymbol{\Space}}

\newcommand{\dd}{\mathrm{d}}
\newcommand{\p}{\partial}

\renewcommand{\epsilon}{USE varepsilon INSTEAD}

\newcommand{\C}{\mathbbm{C}}

\newcommand{\R}{\mathbbm{R}}
\newcommand{\Z}{\mathbbm{Z}}

\newcommand{\Zylinder}{{\mathfrak Z}}

{} 
{} 

\begin{document}

\def\thesubsection{\arabic{subsection}}

\newcommand{\thick}{\vskip 3mm \noindent \rule{137mm}{2pt} \vskip 3mm \noindent }
\newcommand{\thin}{\vskip 3mm \noindent \rule{137mm}{0.3pt} \vskip 3mm \noindent }
\newcommand{\step}{\vskip 3mm \noindent }

\noindent {\bf\Large Choptuik's critical spacetime exists}
\vskip 5mm
\noindent {\bf Michael Reiterer\footnotemark[1], Eugene Trubowitz\footnotemark[1]}
\vskip 1mm
\noindent Department of Mathematics, ETH Zurich, Switzerland
\footnotetext[1]{Part of this work was carried out while the authors were visiting NYU Abu Dhabi, in fall 2011.}
\vskip 4mm
\noindent {\bf Abstract:} About twenty years ago, Choptuik studied numerically the gravitational collapse (Einstein field equations) of a massless scalar field in spherical symmetry, and found strong evidence for a universal, self-similar solution at the threshold of black hole formation. We prove rigorously the existence of a real analytic solution, that we interpret as the solution observed by Choptuik. Our construction covers an open neighborhood of the past light cone of the singularity. The proof is computer assisted. Starting from an explicit approximate solution, we show that nearby there is a true solution. The source code and a high precision data file (about 80 significant decimal digits, with rigorous error bounds) are included. We do not study perturbations.
\renewcommand{\contentsname}{\large Contents}
\setcounter{tocdepth}{2}
\tableofcontents

\subsection{Introduction}\label{introsection}
\begin{figure}[H]
  \centering
\includegraphics[trim=170.5pt 176pt 40pt 45pt,clip]{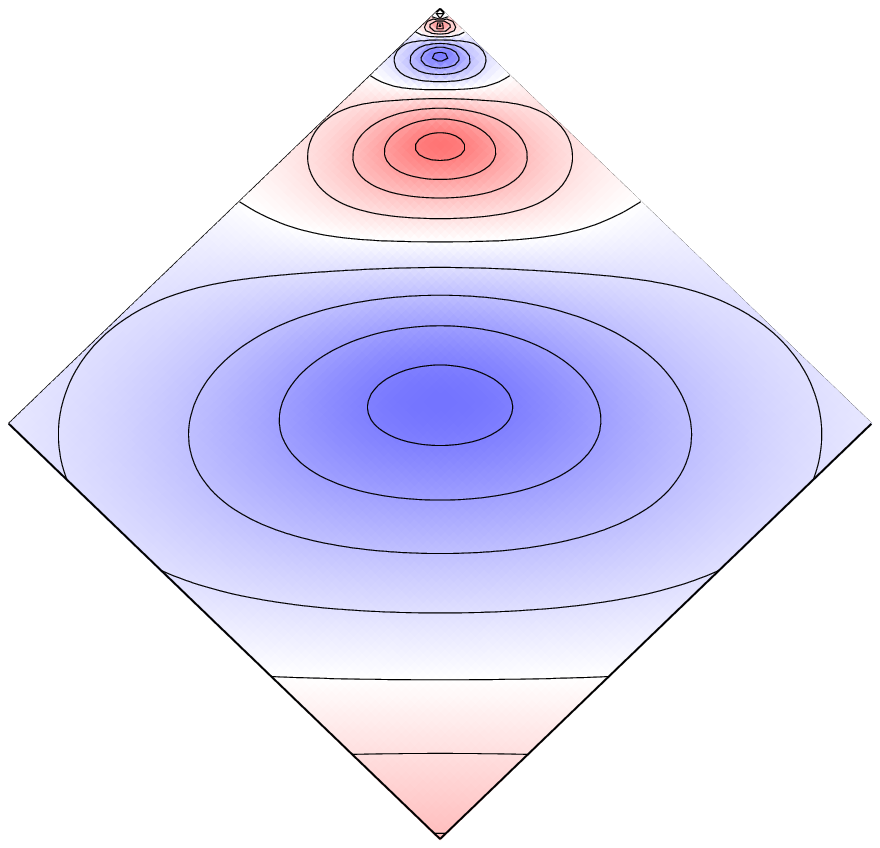}
\caption{Scalar field $\phi$ for Choptuik's spacetime.}\label{scfscf}
\end{figure}
The Einstein field equations for a four-dimensional metric $g$ with signature $(-,+,+,+)$, coupled to a massless scalar field $\phi$, are $\Ric_g = 2\,\dd \phi \otimes \dd \phi$ and $\Box_g \phi = 0$. If one imposes spherical symmetry on $g$ and $\phi$, the problem reduces to a two-dimensional problem. 
The self-similar solution $(g,\phi)$ that Choptuik \cite{C} observed numerically, and whose existence we prove in this paper, has the nontrivial scalar field $\phi$ in Figure \ref{scfscf}.

Figure \ref{scfscf} was generated from an approximation to the true solution, indistinguishable from the true solution at image resolution.
Regions with $\phi<0$ in blue, and with $\phi>0$ in red, alternate. The contour lines are at $\phi=0,\pm \tfrac{1}{2},\pm 1,\pm \tfrac{3}{2},\pm 2$. The self-similarity is captured by a diffeomorphism
\begin{equation}\label{dfdh33h4jh4k}
\Theta:\; \text{Choptuik spacetime}\; \to\; \text{Choptuik spacetime}
\end{equation}
that maps every blue (red) region to the next red (blue) region above it, $\phi \circ \Theta = - \phi$, and that rescales all lengths by a constant $e^{-K}$, equivalently $\Theta^{\ast}g = e^{-2K} g$. Here $K \approx 1.72$ and we prove
\begin{equation}\label{dnkdfkjhfkdhfkdhdfdk1}
\begin{aligned}
|K - &      17227262011139106749857559727881918622210\\
& \phantom{x}3458805781088634788403570540271455648855 \cdot 10^{-80}| \leq 10^{-80}
\end{aligned}
\end{equation}
consistent with the numerical result $3.445452402(3)$ for $2K$ in \cite{MG2}. The very good review \cite{GM} contains references to other numerical results.

\begin{figure}[hb]
  \centering
  \vskip 3mm
\input{penrose.pstex_t}
\caption{The $(u_-,u_+)$ plane. The shaded region is the domain of Figure \ref{scfscf}. At the future singular point, the scalar curvature blows up. Note that the dashed line is actually closer to $u_-=0$ than shown here.}\label{penrosediag}
\vskip 3mm
\end{figure}
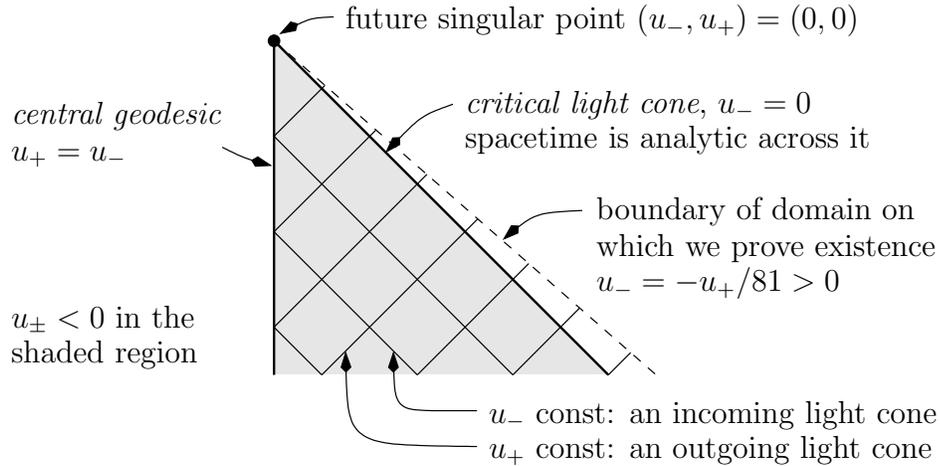

We have to specify the coordinates used to generate Figure \ref{scfscf}. Every point in Figure \ref{scfscf} is a two-sphere. Lines at $\pm 45$ degrees are level sets of an ingoing or outgoing spherically symmetric null coordinate $u_-$ or $u_+$, respectively.

The key property that distinguishes the coordinates $u_{\pm}$ is that \emph{$\Theta$ is exactly a linear rescaling of Figures \ref{scfscf} and \ref{penrosediag}} about the singular point: There is a constant $\mu>0$ such that $u_{\pm} \circ \Theta = e^{-2\pi \mu} u_{\pm}$.
This property, and Figure \ref{penrosediag}, determine $u_{\pm}$  up to a multiplicative constant, see Section \ref{sec:gicst}.
The constant $\mu \approx 0.168$ is itself a geometric invariant of the Choptuik spacetime. We prove
\begin{equation}\label{dnkdfkjhfkdhfkdhdfdk2}
\begin{aligned}
|\mu - & 1683070789634499695101349790428574207210\\
       & 0199080892966476395293134873313662587505 \cdot 10^{-80}| \leq 10^{-80}
\end{aligned}
\end{equation}

\begin{figure}[H]
  \centering
\input{SIGMANEW.pstex_t}
\caption{Principle of causality.}\label{figSIGMA}
\end{figure}
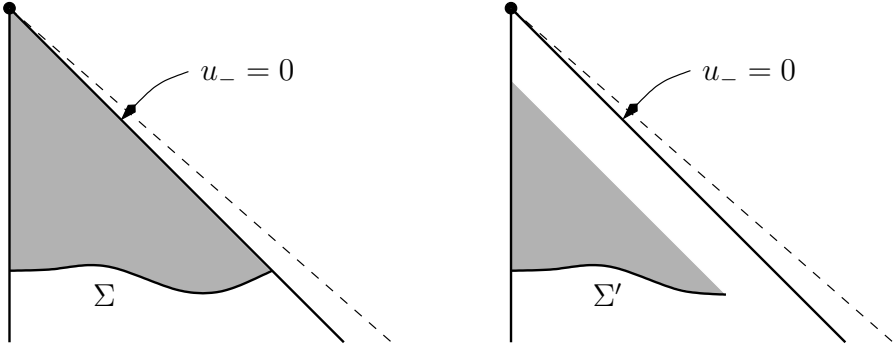

The critical light cone $u_-=0$ plays a crucial role. For example, by causality, any initial data set for the Einstein-scalar-field system that coincides with Choptuik's spacetime on a spacelike hypersurface $\Sigma$ as in Figure \ref{figSIGMA} also contains the shaded region above it, and is singular. No such conclusion can be drawn for $\Sigma'$. For this reason, \emph{it is essential to construct (an open neighborhood of) the past light cone of the singularity}.

We prove that  $(g,\phi)$ is real analytic on the open domain bounded by the dashed line $u_- = -u_+/81>0$ (the value $81$ has no special significance). \emph{This includes real analyticity across $u_-=0$, a basic feature of Choptuik's spacetime.} We now state our result in a logically complete way.
\vskip 2mm
\noindent
{\bf Main result.} \emph{Let $M = \{(u_-,u_+)\in \R^2\;|\;u_+ < 0,\; u_+ < u_- < -u_+/81\}$. There are constants $K,\mu > 0$ that satisfy \eqref{dnkdfkjhfkdhfkdhdfdk1} and \eqref{dnkdfkjhfkdhfkdhdfdk2}, and real analytic functions $\phi,\zeta,Q: M \to \R$, that satisfy
\begin{align*}
\phi\circ \Theta & = -\phi\\
\zeta\circ \Theta & = \phantom{-}\zeta+ K\\
Q\circ \Theta & = \phantom{-}Q & \text{with}\;\;\Theta: M \to M,\;\; (u_-,u_+)\mapsto e^{-2\pi \mu}(u_-,u_+)
\end{align*}
such that on the four-dimensional manifold $M\times S^2$, one has $\mu + Q\xi^2>0$ with $\xi=\frac{u_+-u_-}{u_++u_-}$, and the spherically symmetric Lorentzian metric
\begin{equation*}
g = \exp(-2\zeta) \left(
- \frac{2\left(\dd u_- \otimes \dd u_+ + \dd u_+ \otimes \dd u_-\right)}{\mu^2(u_++u_-)^2}
 + \frac{\xi^2}{(\mu + Q\xi^2)^2}\,g_{\text{standard $S^2$}}
\right)
\end{equation*}
where $g_{\text{standard $S^2$}}$ is the metric on the unit sphere $S^2 \subset \R^3$, satisfies:
\begin{enumerate}[$\bullet$]
\item $\Theta^{\ast}g = e^{-2K}g$, the spacetime is `discretely self-similar'.\\
Here, $\Theta$ is extended trivially to a map $M\times S^2\to M\times S^2$.
\item $\Ric_g = 2\,\dd \phi \otimes \dd \phi$ and $\Box_g \phi = 0$.
\item The scalar curvature $R_g = 2g^{-1}(\dd \phi,\dd \phi)$ is not identically zero.\\
It satisfies $R_g \circ \Theta = e^{2K} R_g$ and is therefore unbounded.
\item The boundary $u_+=u_-$ is a removable standard polar coordinate singularity, because the functions $f=\phi,\zeta,Q$ extend real analytically by reflection, $$f(u_-,u_+) = f(u_+,u_-)$$
\item The solution is close to the high-precision data in the file \texttt{RefAplusB.dat}. For a quantitative statement, see equations
\eqref{ref5ref5ref5ref5}, 
\eqref{corrinb},
\eqref{kdkkkskkskksksks}. Also see \eqref{eq23eq23just}.
\end{enumerate}
}
\vskip 2mm

We do not study perturbations about the Choptuik spacetime, that are relevant for many of the (conjectured) properties of the solution. Our paper can be the  starting point for a rigorous investigation of this kind. This would be interesting, because current numerical results seem to be inconclusive, see \cite{MG1} and \cite{CHLP}, and Sections 3.7 and 5.3 of \cite{GM}.

It has been shown for the Einstein-scalar field system in spherical symmetry that the solution disperses for sufficiently small initial data, and forms trapped spheres for sufficiently large initial data, see \cite{Chr1}, \cite{Chr2}, \cite{Chr3}. 

Since we construct the solution through a contraction mapping, it would be possible to state not only an existence, but also a local uniqueness result of a technical kind. This would allow us to \emph{define} the \emph{Choptuik spacetime} as that unique solution. We have taken the freedom to refer to the  \emph{Choptuik spacetime}, even though we have not formally stated local uniqueness.
\begin{figure}[ht]
  \centering
\includegraphics[trim=170.5pt 176pt 40pt 45pt,clip]{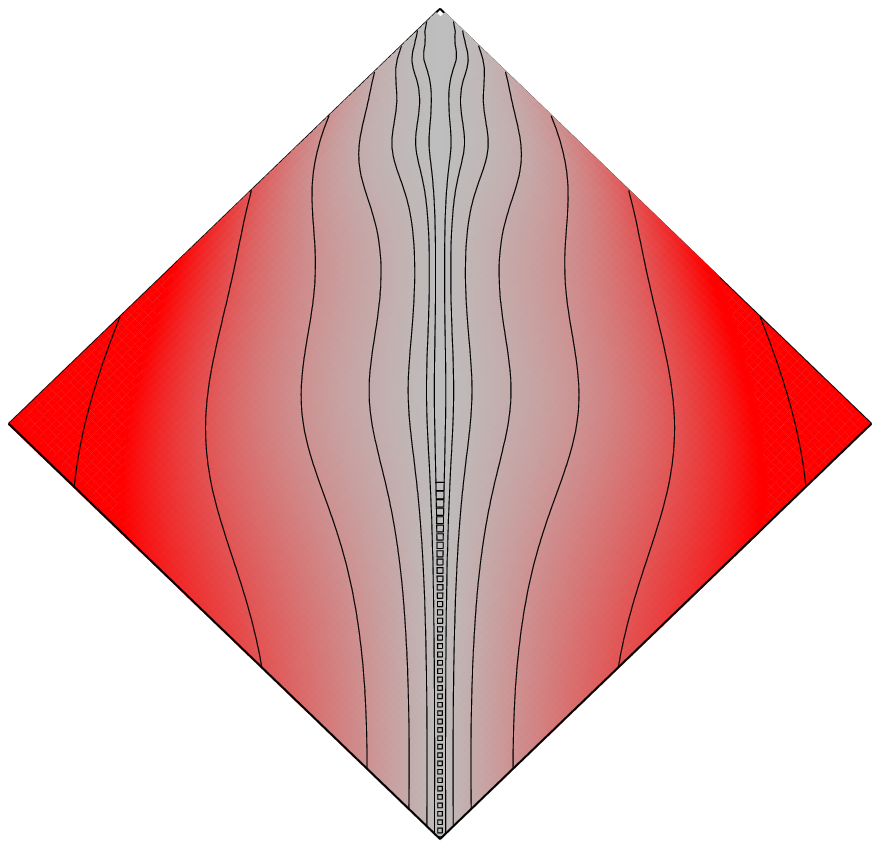}
\caption{Lines of constant area radius $e^{-\zeta} |\xi|/(\mu + Q\xi^2)$.}\label{scfgjhghjgscf}
\end{figure}

The rest of Section \ref{introsection} are detailed (but informal) overviews:\\
\rule{40pt}{0pt} Section \ref{setupanalysis}: setup and analysis\\
\rule{40pt}{0pt} Section \ref{basicstrat}: basic strategy\\
\rule{40pt}{0pt} Section \ref{allthingscomp}: role of the computer\\
\rule{40pt}{0pt} Section \ref{sec:gicst}:  the invariant $\mu$ and the coordinates $(\tau,\xi)$
\subsubsection{Overview -- setup and analysis}\label{setupanalysis}
We reduce the spherically symmetric four-dimensional problem to a two-dimensional problem, stated in Section \ref{2d4d}. Contrary to the introduction, we \emph{do not} use $u_{\pm}$ as coordinates, and we \emph{do not} use $\phi,\zeta,Q$ as unknowns:
\begin{description}
\item{\emph{Coordinates:}} We use $(\tau,\xi)$ given by the change of coordinates
$$
u_{\sigma} = -(1+\sigma \xi) \exp(-\mu\tau)\qquad \sigma = \pm$$
Note that $(\tau,\xi)\circ \Theta = (\tau+2\pi,\xi)$, and one can use Fourier series.
\item{\emph{Unknowns:}} A number $\mu>0$ (see below) and $\omega = (\omega_1,\omega_2,\omega_3,\omega_4)$ given by the change of variables  \eqref{ekjehee}. Advantage: The field equations become first order and quadratically nonlinear, a property that we consistently exploit. Each $\omega_i$ is $2\pi$ periodic or antiperiodic in $\tau$, all are $4\pi$ periodic.
\item{\emph{The role of $\mu$:}} For the analysis, $\mu>0$ is not a constant, but an unknown on the same footing as the $\omega_i$. This `gain' of one degree of freedom is, in many respects,
a compensation for the `loss' of one degree of freedom due to translation invariance of the problem in $\tau$.
This becomes most explicit in \eqref{dfdhfkhfdfkskkskkks}.
\end{description}
Note that the ubiquitous, characteristic differential operators \eqref{khdkjhdkhdkdjjdjdjjdjdjjdj} are \emph{quasilinear} but not semilinear, because they depend on the unknown $\mu$. It seems that this `minimal' quasilinearity cannot be removed from the problem -- not without introducing a complication somewhere else. Fortunately, in the space of analytic functions that we work in, the quasilinearity plays a minor role. This becomes explicit in the boundedness of the operator in \eqref{fdkhfdkhfjfjjfjjfxxx}.

\begin{figure}[H]
  \centering
\input{complexdomain.pstex_t}
\caption{Domain of analyticity. Here $\pm 1$ are \emph{not} the foci.}\label{hfkjhdkkddkdd}
\end{figure}
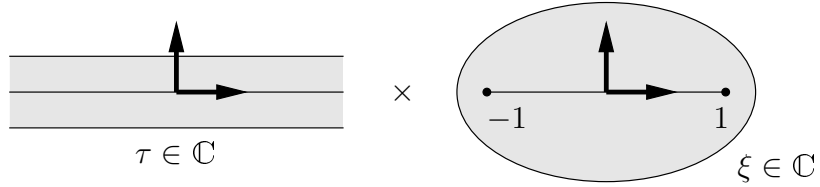

We represent the $\omega_i$ as combined Fourier-Chebyshev series:
\begin{alignat*}{5}
\text{\emph{$4\pi$-Fourier series} in $\tau$} & &&\;\;\to\;\; \text{index $m\in \Z$}\\
\text{\emph{Chebyshev series} in $\xi$}& \text{, \;i.e.~$2\pi$-Fourier series in $\arccos \xi$} && \;\;\to\;\;  \text{index $n\in \Z$}
\end{alignat*}
We work in a space in which the coefficients decay as $\mathcal{O}((\kappa_1)^{-|m|}(\kappa_2)^{-|n|})$ in the high frequency limit $|m|+|n|\to \infty$, with $\kappa_1,\kappa_2>1$. Such functions are analytic on the open subset of $\C\times \C$ in Figure \ref{hfkjhdkkddkdd}, the infinite strip $|\IM(\tau)| < 2 \log \kappa_1$ times the open ellipse with radii $\tfrac{1}{2}(\kappa_2\pm 1/\kappa_2)$.

We use $\kappa_1 = 65/64$ and $\kappa_2=5/4$. Thus, the domain that we obtain is \emph{smaller} than the one shown in Figure \ref{hfkjhdkkddkdd}. When we restrict to real $\xi$, we get  $|\xi| < 41/40$. This explains the earlier inequality $u_-<-u_+/81$.

The obstruction to using `big' $\kappa_1$ and $\kappa_2$ is that one has to avoid singularities (e.g.~poles) that the complex analytic extension of the Choptuik spacetime \emph{may} have. 
This motivates our use of Chebyshev series instead of power series in $\xi$, because a disk with radius bigger than $1$, big enough to contain the interval $[-1,1]$, \emph{may} already contain singularities.

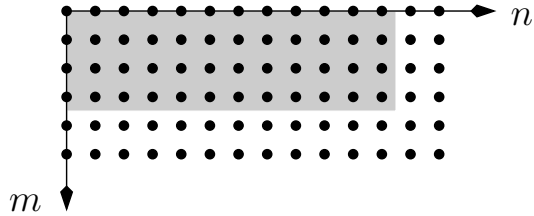
\begin{figure}[H]
  \centering
\input{lowhigh.pstex_t}
\caption{Images of $\KS$ (shaded) and of $1-\KS$ (non-shaded). 
}\label{kskskskswiiiwiwi}
\end{figure}

We use a low-high frequency decomposition of the identity, $\KS + (1-\KS) = 1$. Here $\KS$ and $1-\KS$ are projection operators to low and high frequencies, respectively. In particular, \emph{$\KS$ has a finite dimensional image}. These projections divide the construction into the computer and analytic estimates parts:
\begin{align*}
\text{things related to $\KS$}\;\;\; & \to\;\;\; \text{computer}\\
\text{things related to $1-\KS$}\;\;\; & \to\;\;\; \text{analytic estimates}
\end{align*}
The shaded region in  Figure \ref{kskskskswiiiwiwi} indicates the image of $\KS$.  Every {\small $\bullet$} in  Figure \ref{kskskskswiiiwiwi} stands for a finite number of real degrees of freedom. Only non-negative indices $m,n$ are shown, because they determine all, by reality constraints. We actually use a much bigger region: $250\times 750$.

To work in Fourier-Chebyshev \emph{frequency space} is very useful, because:
\begin{itemize}
\item It reveals the \emph{elliptic} nature of the problem. In fact, the principal part of the equations is \emph{morally} multiplication by
$$im/2 + \mu(n+1) \qquad \text{($i$ is the imaginary unit)}$$
when $m,n\geq 0$, with an inverse that is $\mathcal{O}((m+n)^{-1})$ as $m+n\to \infty$. We say \emph{morally}, because the actual operators  in \eqref{dkhddjjjjjff}, while diagonal in $m$, are only upper triangular in the index $n$.
\item It disposes of the geometric boundaries (central geodesic and especially the critical light cone) in a \emph{seamless} way, and one does not specify initial data anywhere. That's good, because the Choptuik spacetime is supposed to be `universal' after all. By contrast, for alternative approaches not based in frequency space, the geometric boundaries may be unpleasant to deal with.
\end{itemize}

The common slogan in general relativity that `the constraints are satisfied if they are satisfied initially' \emph{does not apply here}, because there is no initial data in the first place. Not surprisingly, though, the usual differential identities yield a second elliptic system, this one linear and homogeneous, that is used to show that the constraints are satisfied. All objects related to the constraint equations are marked by $\sharp$.

 Our analysis uses $\ell^1$ norms. This has several advantages:
\begin{itemize}
\item Natural estimate for series convolution.
\item Transparent operator norm. For non-weighted $\ell^1$, it is the sup of the $\ell^1$ norms of the `columns'. For example, in finite, say two dimensions,
$$\left\|\begin{pmatrix}
a & b \\
c & d
\end{pmatrix}\right\|_{\ell^1 \leftarrow \ell^1} = \sup \left\{\left\| \begin{pmatrix} a \\ c \end{pmatrix} \right\|_{\ell^1}, \left \| \begin{pmatrix} b \\ d \end{pmatrix} \right\|_{\ell^1}\right\}\qquad a,b,c,d\in \R$$
If the $\ell^1$ is weighted, then the sup has to be \emph{inversely} weighted.
\end{itemize}
And particularly for coding (think finite dimensions here):
\begin{itemize}
\item The $\ell^1$ norm can be evaluated over the rationals, for rational components and weights.
For this reason, we even use $\|z\|_{\C} = |\RE z| + |\IM z|$.
\item If an operator is constructed column by column, its $\ell^1$ operator norm can be evaluated \emph{without ever storing the whole operator at once}.
\end{itemize}
\subsubsection{Overview -- basic strategy}\label{basicstrat}
From an abstract point of view, we construct a nontrivial solution $x$ (think metric and scalar field) to a nonlinear system $C(x)=0$, where
$$C(x) = Ax + B(x,x)$$
The unknown $x$ lies in an infinite dimensional real vector space (of Fourier-Chebyshev coefficients), $A$ is a linear operator, and $B$ is a symmetric bilinear operator. That the Einstein equations, with or without scalar field, can be written as a system with only quadratic nonlinearities is a basic fact, but often unappreciated.

Even though the vector spaces are infinite dimensional,
imagine that there are
\emph{as many equations as unknowns}, i.e.~that $x$ and $C(x)$ are in vector spaces of `equal dimensions'. We thereby ignore the constraint equations in this overview. The computer is used to construct an approximate nontrivial solution $\Ref{x}$, i.e. one for which $C(\Ref{x})$ is `small'. With $x=\Ref{x}+\Corr{x}$, the system $C(x)=0$ becomes the following system
for the correction $\Corr{x}$:
\begin{equation}\label{iiwiiwiissos}
\big[A+2B(\Ref{x},\,\cdot\,)\big]\Corr{x} =  - C(\Ref{x}) -B(\Corr{x},\Corr{x})
\end{equation}
with the notation $B(a,\,\cdot\,)b = B(a,b)$. The term $-B(\Corr{x},\Corr{x})$ can be attenuated arbitrarily, by using a sufficiently good $\Ref{x}$, because it is the only term that scales \emph{quadratically} with the distance of $\Ref{x}$ from the hypothetical solution $x$. Somewhat more concretely, one will seek $\Corr{x}$ in a ball with small radius $\mathcal{R}>0$, centered at the origin, on which $-B(\Corr{x},\Corr{x})$ will have Lipschitz constant $\sim \mathcal{R}$, and the better $\Ref{x}$, the smaller one can choose $\mathcal{R}$, the smaller the Lipschitz constant.

Thus, a good $\Ref{x}$ yields an \emph{essentially linear problem} for $\Corr{x}$, and the key task is to show that the linear operator $A+2B(\Ref{x},\,\cdot\,)$ has an inverse, and control it.
We first reduce the task to that of inverting a certain \emph{finite square} matrix $X$, and a robust method to invert a matrix is given by the Neumann series.
In fact, if one can find an \emph{approximate inverse} $Y$,
 i.e.~such that the operator norm $\|1-XY\|<1$, then the Neumann series
\begin{equation} \label{HHFKFKFHKHSKKKSJHKSS}
\textstyle Y \frac{1}{1-(1-XY)} = Y \sum_{k=0}^{\infty} (1-XY)^k
\end{equation}
is an inverse of $X$. (It is a right-inverse to begin with, which for finite square matrices is also a left-inverse.) This method gives one a lot of flexibility in the construction of $Y$. See Section \ref{allthingscomp} for a detailed discussion.

\emph{The notation used in this overview is local.}
To connect this abstract discussion with the actual construction, we note that the role of $C(x)=0$ is played by $\SELECTION \Omega^+(\mu,\omega)=0$, cf.~\eqref{fkdkkdkkkeieiruus}. With tiny modifications, equation \eqref{iiwiiwiissos} becomes the key equation \eqref{equivdkhjdhdkkkdk}.
For equation \eqref{HHFKFKFHKHSKKKSJHKSS}, see \eqref{hffjdsjkskskahhhh287383}.
\subsubsection{Overview -- role of the computer}\label{allthingscomp}
Computer-assisted analytical proofs have a tradition in dynamical systems, beginning with Lanford's \cite{L} in 1982. This paper may be the first in general relativity.

We use the computer for the primitive purpose of \emph{number crunching} -- neither for symbolic manipulations, nor for proof verification.  Our code is portable and yields reproducible results. It \emph{uses integer arithmetic only}. The C source code is in the directory \texttt{sourcecode} (arXiv ancillary files).
\vskip 4mm

The computer is used for the following tasks:
\vskip 2mm
\noindent
\rule{20pt}{0pt} (1a)\, To \emph{construct} a reference (an approximation to Choptuik's solution).\\
\rule{20pt}{0pt} (1b)\, To \emph{estimate} how good the reference is.\\
\rule{20pt}{0pt} (2a)\, To \emph{construct} an approximate inverse (of a finite square matrix).\\
\rule{20pt}{0pt} (2b)\, To \emph{estimate} how good the approximate inverse is.\\
\rule{20pt}{0pt} \hskip6.5pt(3)\, To \emph{estimate} the norm of an operator in Section \ref{dflfklfjkfhkkkkkkkkkkkkdhkhdk}.
\vskip 2mm

From a logical point of view, (1a) and (2a) are not strictly required: \emph{it would suffice to provide the outputs of (1a) and (2a) in files.} For (1a) we do exactly that, see the file \texttt{RefAplusB.dat}. For (2a), such a file would be quite big, and therefore the code is used to construct the approximate inverse \emph{column by column, never all of it at once}.

We have included a tool that can generate better and better references, if used appropriately, and given enough computing resources.
See \texttt{ImproveRef} in the file \texttt{choptuik.c}. We have used it to construct \texttt{RefAplusB.dat}. This tool is \emph{not part of the existence proof}.
\vskip 3mm

From a practical point of view, assuming $\sim 2010$ equipment:
\vskip 2mm
\noindent
\rule{20pt}{0pt} (1a), (1b) take a few dozen CPU hours, and may be memory intense.\\
\rule{20pt}{0pt} (2a), (2b), (3) take a few thousand CPU hours.
\vskip 2mm
\noindent
The time consuming tasks (2a), (2b), (3) are \emph{embarrassingly parallel}, they split into many individual and completely independent tasks. Every individual task can also be run on a personal computer.

The bottleneck is the convolution of 2-dim arrays of integers, which we have implemented using integer multiplication from the GMP library, which uses FFT.
We use the GCC compiler, and OpenMP. We ran the code on the Brutus HPC cluster at ETH Zurich.

\vskip 6mm
We discuss (2a), (2b) in some detail. Given a real matrix $X$, the task is to construct an approximate inverse $Y$, see Section \ref{basicstrat}. The matrices are $N\times N$, where $N=654375$ with our choice of parameters.
This number arises roughly as $250\cdot 750 \cdot 2 \cdot (0.25+0.5+0.5+0.5)  = 656250$ with:
\vskip 2mm
\noindent
\rule{20pt}{0pt} $250$ number of Fourier indices, $0 \leq m < 250$, see $\KS$ in Section \ref{setupanalysis}\\
\rule{20pt}{0pt} $750$ number of Chebyshev indices, $0 \leq n < 750$, see $\KS$ in Section \ref{setupanalysis}\\
\rule{20pt}{0pt} $2$ for real and imaginary parts\\
\rule{20pt}{0pt} $0.25$ for every field restricted by a $\Z_2\times \Z_2$ symmetry\\
\rule{20pt}{0pt} $0.5$ for every field restricted by a $\Z_2$ symmetry
\vskip 2mm
\noindent
Reality constraints for the zeroth Fourier coefficient yield a value $N$ smaller than $656250$. This counting applies to the rows of $X$. For the columns of $X$, one of these degrees of freedom is dropped (a gauge condition that removes translational symmetry in $\tau$), and one is added for $\mu$. See \eqref{dfdhfkhfdfkskkskkks}. Of course, the role of columns and rows is reversed for $Y$. 

Each entry of the matrix $X$ is a \emph{dyadic rational}, i.e.~an integer divided by a power of 2.
The matrix $X$ is not stored entry by entry, but arises naturally as a sum of compositions of a small set of operators $O$. Each $O$ is implemented through an efficient algorithm, of which only convolution is non-trivial, and each $O$ is equivalent to a matrix with dyadic rational entries.

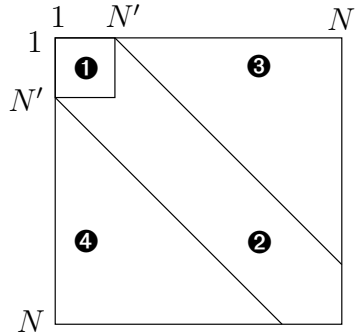
\begin{figure}[H]
  \centering
\input{matrixX.pstex_t}
\caption{Matrix $X$, with
$N'\approx N/100$.}\label{dfkfhkhkfhjjjjffh}
\end{figure}

There is no useful linear order for the rows and columns of $X$, because there
are \emph{two} frequency indices $m$ and $n$. This makes it hard to visualize $X$.
We will simplify our discussion below by pretending, \emph{very informally}, that there is just one frequency index, and that the rows and columns are arranged in order of `increasing frequency'. With this simplification, the matrix $X$ has the structure in Figure \ref{dfkfhkhkfhjjjjffh}.
The square submatrix \ding{202} is dense and some of its entries are large.
The band \ding{203} is close to the identity, but \emph{not very close} in the operator norm.
The entries in \ding{204} are extremely small in the operator norm.
The entries in \ding{205} are identically zero.
To summarize, $X$ is \emph{close to a band matrix in the operator norm}. However, the width of the band is in the thousands.

In (2a), the construction of $Y$, only \ding{202} and \ding{203} are taken into account. We first apply Gaussian elimination to \ding{202}, with partial pivoting and rounding to dyadic rationals at each step.
The output of Gaussian elimination is combined with \ding{203} to a first naive approximate inverse $Y_{\text{naive}}$.
Finally, a truncated Neumann series is used to generate from $Y_{\text{naive}}$ a better approximate inverse $Y$. (For efficiency, our implementation differs slightly from this description.)

This approach to (2a) exploits the approximate sparseness of $X$, and it made the calculation feasible with the resources available to us. In principle, with enough resources, one could apply Gaussian elimination to all of $X$ at once.

In (2b) we calculate the operator norms $\|Y\|$ and $\|1-XY\|$. This is the only place where \ding{204} has to be used.
\emph{Only once we have checked $\|1-XY\|<1$ do we know that the construction in (2a) was actually successful.} We can then use the Neumann series inequality $\|X^{-1}\| \leq \|Y\|/(1-\|1-XY\|)$.

Tasks (2a) and (2b) are embarrassingly parallel, because both the construction of $Y$ and the evaluation of $\|Y\|$ and $\|1-XY\|$ can be done column by column, see the discussion of the $\ell^1$ operator norm in Section \ref{setupanalysis}.

Tasks (2a) and (2b) have to be repeated for the $\sharp$ system, that is used to show that the constraints vanish. Here, $N=280125 \approx 250\cdot 750\cdot 2 \cdot (0.25+0.5)$.

\subsubsection{Overview -- the invariant $\mu$ and the coordinates $(\tau,\xi)$} \label{sec:gicst}
The purpose of this section is to motivate the geometric ansatz that we use to construct the Choptuik spacetime. It provides important intuition. However, it is \emph{not part of the existence theorem}. First, we need three facts. 

Let $G$ be the set of \emph{germs}
of $C^{\infty}$ functions $\R \to \R$ at the origin $x = 0$, with $f(0) = 0$ and $f'(0) > 0$.
Germs are convenient for making local statements about functions, without having to worry about domains; for a definition see \texttt{Wikipedia}. 
The set $G$ is a group under composition $(\circ)$. Two elements $f,g\in G$ are called \emph{conjugate} if and only if there is a $\kappa \in G$ such that $\kappa \circ f = g \circ \kappa$. For all $f,g,\kappa\in G$:
\vskip 2mm
\noindent {\bf Fact 1.} \emph{If $f,g$ are conjugate, then $f'(0) = g'(0)$.}\\
\noindent {\bf Fact 2.} \emph{If $f'(0)\neq 1$, then $f$ is conjugate to the linear function $x \mapsto f'(0)x$.}\\
\noindent {\bf Fact 3.} \emph{If $f$ is linear with $f'(0)\neq 1$, and $f \circ \kappa = \kappa \circ f$, then $\kappa$ is linear.}
\vskip 2mm
\noindent
In words: The derivative at the origin is a conjugation invariant (Fact 1), and if it is not equal to one, then it is the only invariant (Fact 2).
\vskip 2mm
The rest of this Section \ref{sec:gicst} is not rigorous. We blur the distinction between a function and its germ, and we use implicitly some very general qualitative properties of the Choptuik spacetime.
\vskip 2mm
The Choptuik spacetime has a conjugation invariant as in Fact 1. To see this, first introduce a set
of \emph{incoming null coordinates}. Namely, let $\mathrm{NullCoord}_-$ be the set of all $u: \text{(Choptuik spacetime)} \to \R$ for which:
$$
\left\{\begin{aligned}
&\text{$u$ is a spherically symmetric incoming null coordinate, with $\dd u \neq 0$}\\
&\text{$u=0$ is the critical light cone of the Choptuik spacetime}\\
&\text{$u<0$ in the interior, $u>0$ in the exterior}
\end{aligned}\right.
$$
The group $G$ acts on $\mathrm{NullCoord}_-$, in fact
$$ G \times \mathrm{NullCoord}_- \to \mathrm{NullCoord}_-\qquad  (f,u) \mapsto f \circ u$$
is well defined as a map, and is a group action. This action has an important property:  for any two $u$, $u'$ in $\mathrm{NullCoord}_-$, there exists a unique $f \in G$ with $f\circ u = u'$. Henceforth, this property is called \emph{regularity}.

Let $\Theta$ be the self-similarity diffeomorphism \eqref{dfdh33h4jh4k} of the Choptuik spacetime.
If $u$ is in $\mathrm{NullCoord}_-$, so is $u \circ \Theta^{-1}$. By regularity, there is a map
$$
\mathrm{NullCoord}_- \to G,\; u \mapsto f_u \qquad \text{defined by}\qquad u \circ \Theta^{-1} = f_u \circ u$$
We have $f_{\kappa \circ u} = \kappa \circ f_u \circ \kappa^{-1}$ for all $\kappa \in G$. By regularity:
\vskip 2mm
\noindent {\bf By Fact 1.} $f_u'(0)$ is an invariant (the same for every $u \in \mathrm{NullCoord}_-$).
\vskip 2mm
\noindent The constant $\mu$ used in this paper is defined by $e^{2\pi \mu} = f'_u(0)$. We anticipate that for the Choptuik spacetime, $\mu > 0$ and $f'_u(0)>1$.  Therefore:
\vskip 2mm
\noindent {\bf By Fact 2.} There exist $u \in \mathrm{NullCoord}_-$ such that $f_u$ is linear.\\
\noindent {\bf By Fact 3.} If $f_u$ and $f_{u'}$ are linear, then $f_u=f_{u'}$ and $u=cu'$ for a $c>0$.
\vskip 2mm
Fix a $u_- \in \mathrm{NullCoord}_-$ for which $f_{u_-}$ is linear, $u_- \circ \Theta^{-1} = e^{2\pi \mu}u_-$. Let $u_+$ be the spherically symmetric outgoing null coordinate that satisfies $u_+=u_-$ along the central geodesic. Then $u_+ \circ \Theta^{-1} = e^{2\pi \mu}u_+$, because it is true on the central geodesic. The 2D coordinate pair $(u_-,u_+)$ is unique up to joint multiplication by a positive constant.

The 2D coordinate pair $(\tau,\xi)$ used in this paper can now be defined by
\begin{equation} \label{coordinatesxitau}
u_{\sigma} = -(1+\sigma \xi) \exp(-\mu\tau)\qquad \sigma = \pm
\end{equation}
By construction, $\xi=0$ is the central geodesic, $\xi=1$ is the critical light cone, and $(\tau,\xi) \circ \Theta = (\tau+2\pi,\xi)$. The pair $(\tau,\xi)$ is unique up to adding a constant to $\tau$.
We compare the two coordinate systems in a table:
\begin{center}
\begin{tabular}{|c|c|c|}
\hline
& Self-similarity diffeomorphism $\Theta$ & Null directions\\
&     (with $\phi \circ \Theta = -\phi$) & (2D causal structure) \\
\hline
\hline
\rule{0pt}{13pt}$(u_-,u_+)$ & $(u_-,u_+)\circ \Theta = e^{-2\pi\mu}(u_-,u_+)$ & $\ker \dd u_{\sigma}$\\
\hline
\rule{0pt}{13pt}$(\tau,\xi)$ & $(\tau,\xi) \circ \Theta = (\tau+2\pi,\xi)$ & $\ker [\sigma\,\dd \xi - \mu(1+\sigma \xi) \dd \tau]$\\
\hline
\end{tabular}
\end{center}
In the first case, $\Theta$ depends on $\mu$, but the 2D causal structure doesn't.\\
In the second case, $\Theta$ doesn't depend on $\mu$, but the 2D causal structure does.

Either way, the constant $\mu>0$ appears \emph{somewhere}. Since it is a geometric invariant of the Choptuik spacetime, $\mu$ cannot be chosen freely before the construction. Rather, it is an unknown and it has to be constructed.

\subsection{Einstein equations coupled to scalar field}

The equations for a metric $g$, coupled to a massless scalar field $\phi$, are
\begin{align*}
\Ric_g & = k\,\dd \phi \otimes \dd \phi && \text{Einstein field equations}\\
\Box_g \phi & = 0 && \text{wave equation}
\end{align*}
where $\Ric_g$ is the Ricci curvature of $g$, the one form $\dd \phi$ is the differential of $\phi$, and $\Box_g$ is the Laplace-Beltrami operator for $g$. All values for the constant $k>0$ are equivalent, by rescaling $\phi$. This rescaling is necessary to compare papers that use different values for $k$. This paper uses  $k=2$.

The four-dimensional problem reduces, in the case of spherical symmetry, to a two-dimensional problem:
\begin{center}
\input{reduction.pstex_t}
\end{center}
From a pedagogical perspective, the direction from 4D to 2D (Section \ref{4d2d}) is easier to understand, because it starts from the basic equations. From a logical perspective, only the direction from 2D to 4D (Section \ref{2d4d}) is required for the existence theorem that we prove.

We use a frame formalism, because it yields first order, quadratically nonlinear equations. The 2D formulation has more equations than unknowns. The equations are dependent, see the differential identities  in Section \ref{dkhfkhdfkhfieuhfkdf}.
\subsubsection{From 4D to 2D}\label{4d2d}
Let $(\tau,x^1,x^2,x^3)$ be Cartesian coordinates on $\R^4$. Spherical symmetry means invariance under the standard action of the matrix group $\mathrm{O}(3)$ on the coordinates $(x^1,x^2,x^3)$, with $\tau$ fixed. It is convenient to start with Cartesian coordinates, because (as opposed to polar coordinates) they are regular along the line $(x^1,x^2,x^3)=0$, the set of fixed points of the action. Set $$\xi = \sqrt{(x^1)^2+(x^2)^2+(x^3)^2}$$

Let $\mu > 0$ be a \emph{constant}, and let $Q$ and $\zeta$ be real valued \emph{functions} that depend only on $\tau$ and $\xi$. Introduce a frame $(\mathbf{e}_0,\mathbf{e}_1,\mathbf{e}_2,\mathbf{e}_3)$ and a spherically symmetric metric $g$ with signature $(-,+,+,+)$:
\begin{equation}\label{ekhfkjfhkfhfkhf}
\begin{aligned}
\mathbf{e}_0 & = \frac{\p}{\p \tau} + \mu \sum_{k=1}^3 x^k \frac{\p}{\p x^k}\\
\mathbf{e}_i & = \mu \frac{\p}{\p x^i} + Q \sum_{k=1}^3 x^k\Big(x^k \frac{\p}{\p x^i} - x^i \frac{\p}{\p x^k}\Big) \qquad \text{with $i=1,2,3$}\\
g^{-1} & = \exp(2\zeta)\;\Big(-\mathbf{e}_0 \otimes \mathbf{e}_0 + \mathbf{e}_1 \otimes \mathbf{e}_1 + \mathbf{e}_2 \otimes \mathbf{e}_2 + \mathbf{e}_3 \otimes \mathbf{e}_3\Big)
\end{aligned}
\end{equation}
A spherically symmetric spacetime can always be brought into this form, by appropriate choice of the two free functions $Q$ and $\zeta$. We omit a more precise statement or a proof, because it is logically unnecessary for this paper.

We assume $\mu + \xi^2 Q > 0$, which implies that the frame is non-degenerate:
$$\mathbf{e}_0 \wedge \mathbf{e}_1 \wedge \mathbf{e}_2 \wedge \mathbf{e}_3 = \mu (\mu + \xi^2 Q)^2\, \frac{\p}{\p \tau}\wedge \frac{\p}{\p x^1}\wedge \frac{\p}{\p x^2}\wedge \frac{\p}{\p x^3}$$

\begin{figure}[H]
  \centering
\input{causal.pstex_t}
\caption{Forward light cones. Exact directions depend on $\mu$.\newline
However, $\xi=1$ is a null hypersurface for all $\mu$, because it is a level set of $u_-$.}\label{hfkjhjfkhkfhfkhfjfd}
\end{figure}
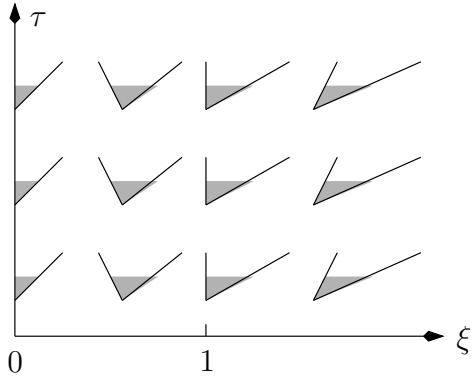

By construction, the functions $u_{\sigma} = -(1+\sigma \xi) \exp(-\mu \tau)$ with $\sigma = \pm$ both solve the Eikonal equation $g^{-1}(\dd u_{\sigma},\dd u_{\sigma}) = 0$. This determines the causal structure of the restriction of the metric to the 2-dimensional plane $(\tau,\xi) \mapsto (\tau,\xi,0,0)$. \emph{This causal structure depends only on the constant $\mu$}, not on $Q$ or $\zeta$.
See Figure \ref{hfkjhjfkhkfhfkhfjfd}.
Introduce the two radial null vector fields
$$D^{\sigma} = \sigma \mathbf{e}_0 + \sum_{k=1}^3 \frac{x^k \mathbf{e}_k}{\xi}$$
In particular $D^{\sigma}(\tau) = \sigma$ and $D^{\sigma}(\xi) = \mu(1+\sigma \xi)$. Let $\phi$ be another function that depends only on $\tau$ and $\xi$, it plays the role of the scalar field coupled to the Einstein equations. The functions $\zeta,Q,\phi$ are not themselves useful for us. \emph{Instead, the basic unknown functions of $\tau$ and $\xi$ that are used throughout the paper (together with the unknown constant $\mu$) are}
\begin{equation}\label{ekjehee}
\begin{aligned}
\omega_1 & = 2\xi Q + D^-\big(\zeta + \log(\mu + \xi^2Q)\big)
+ D^+\big(\zeta + \log(\mu + \xi^2Q)\big)\\
\omega_2^{\sigma} & = D^{\sigma}\big(\zeta + \log(\mu + \xi^2Q)\big) - \sigma \mu\\
\omega_3^{\sigma} & = D^{\sigma}(\zeta) - \sigma \mu\\
\omega_4^{\sigma} & = D^{\sigma}(\phi)
\end{aligned}
\end{equation}
with $\sigma = \pm$. A useful property is the invariance of these seven functions under a global conformal transformation of the metric, $\zeta \to \zeta + \text{const}$.

The equations \eqref{ekjehee} can be rewritten as
\begin{subequations}
\begin{align}
\label{dhjhfkfhfe1} Q & = (\omega_1 - \omega_2^--\omega_2^+)/(2\xi)\\
\label{dhjhfkfhfe2} D^{\sigma}\big(\log(\mu+Q\xi^2)\big) & = \omega_2^{\sigma} - \omega_3^{\sigma}\\
\label{dhjhfkfhfe3} D^{\sigma} (\zeta) & = \omega_3^{\sigma} + \sigma \mu \\
\label{dhjhfkfhfe4} D^{\sigma}( \phi) & = \omega_4^{\sigma}
\end{align}
\end{subequations}
This can be used to calculate $\Ric_g(\mathbf{e}_m,\mathbf{e}_n) - 2\mathbf{e}_m(\phi)\mathbf{e}_n(\phi)$ with $m,n=0,1,2,3$ and $\exp(-2\zeta) \Box_g \phi$ in terms of just $\omega_1$, $\omega_2^{\sigma}$, $\omega_3^{\sigma}$, $\omega_4^{\sigma}$. This lengthy but straightforward calculation is not included. The result is
\begin{subequations} \label{dfkhdjhsdshkfd}
\begin{multline}
\Ric_g(\mathbf{e}_i,\mathbf{e}_j) -  2\mathbf{e}_i(\phi)\mathbf{e}_j(\phi)  = \frac{\delta_{ij}}{4\xi}\big( \Omega_1^- + \Omega_1^+ + \Omega_2^- + \Omega_2^+ - \Omega^-_{2\sharp} - \Omega^+_{2\sharp}\big) \\
+ \frac{x^ix^j}{4\xi^3}\big(\Omega^-_1 + \Omega^+_1 - \Omega^-_2 - \Omega^+_2 + \Omega^-_{2\sharp} + \Omega^+_{2\sharp} + 2\Omega^-_3 + 2\Omega^+_3\big)
\end{multline}
for all $i,j=1,2,3$ and
\begin{equation}
\begin{aligned}
\Ric_g(\mathbf{e}_0,\mathbf{e}_0) - 2 \mathbf{e}_0(\phi)\mathbf{e}_0(\phi) & = \frac{1}{2\xi}\big( -\Omega_2^- - \Omega_2^+ + \Omega^-_{2\sharp} + \Omega^+_{2\sharp} - \Omega^-_3 - \Omega^+_3\big)\\
\Ric_g(\mathbf{e}_0,\mathbf{e}_i) - 2\mathbf{e}_0(\phi)\mathbf{e}_i(\phi) & = \frac{x^i}{2\xi^2}\big(-\Omega^-_1 + \Omega^+_1 - \Omega^-_3 + \Omega^+_3\big)\\
\exp(-2\zeta) \Box_g\phi & = \frac{1}{2\xi}(\Omega^-_4 + \Omega^+_4)
\end{aligned}
\end{equation}
\end{subequations}
Here, by definition:
\begin{subequations}\label{eezuuirzr}
\begin{align}
& \begin{aligned}
\Omega_1^{\sigma}(\mu,\omega) & = \xi (D^{\sigma}+\sigma \mu)(\omega_1) + \mu(2\omega_1) \\
& \hskip 4mm + \tfrac{1}{4}\xi (\omega_1 \omega_1 - 2\omega_2^{-\sigma} \omega_1 - 6 \omega_2^{\sigma} \omega_1 + 4 \omega_3^{\sigma}\omega_1  + \omega_2^{-\sigma} \omega_2^{-\sigma}\\
&\hskip 16mm  - 2\omega_2^{\sigma}\omega_2^{-\sigma}  - 4 \omega_3^{\sigma} \omega_2^{-\sigma}   + \omega_2^{\sigma} \omega_2^{\sigma} + 4 \omega_3^{\sigma} \omega_2^{\sigma}  - 4 \omega_4^{\sigma} \omega_4^{\sigma})
\end{aligned}\displaybreak[0]\\
& \begin{aligned}
\Omega_2^{\sigma}(\mu,\omega) & = \xi (D^{\sigma}+\sigma \mu)(\omega_2^{-\sigma}) + \mu (\omega_1 + \omega_2^{-\sigma} +   \omega_2^{\sigma})\\
& \hskip 4mm + \tfrac{1}{4}\xi(\omega_1 \omega_1 - 2\omega_2^- \omega_1 - 2\omega_2^+ \omega_1  + \omega_2^- \omega_2^- - 6 \omega_2^+ \omega_2^- + \omega_2^+ \omega_2^+)
\end{aligned}\displaybreak[0]\\
& \begin{aligned}
\Omega^{\sigma}_{2\sharp}(\mu,\omega) & = \xi (D^{\sigma}+\sigma \mu)(\omega_2^{\sigma}) + \mu (2\omega_2^{\sigma} - 2 \omega_3^{\sigma}) \\
& \hskip 4mm + \tfrac{1}{4}\xi(-4 \omega^{\sigma}_2 \omega^{\sigma}_2 + 8 \omega^{\sigma}_3 \omega^{\sigma}_2  - 4 \omega^{\sigma}_4 \omega^{\sigma}_4)
\end{aligned}\displaybreak[0]\\
& \begin{aligned}
\Omega_3^{\sigma}(\mu,\omega)& = \xi (D^{\sigma}+\sigma \mu)(\omega_3^{-\sigma}) + \mu ( - \omega_1) \\
& \hskip 4mm + \tfrac{1}{4}\xi(-\omega_1 \omega_1 + 2\omega_2^- \omega_1 +2 \omega_2^+ \omega_1  - \omega_2^- \omega_2^- + 2\omega_2^+ \omega_2^- \\
& \hskip 63mm  - \omega_2^+ \omega_2^+ - 4\omega_4^+ \omega_4^- )
\end{aligned}\displaybreak[0]\\
& \begin{aligned}
\Omega_4^{\sigma}(\mu,\omega) & = \xi (D^{\sigma}+\sigma \mu) (\omega_4^{-\sigma}) + \mu (\omega_4^{-\sigma} + \omega_4^{\sigma}) \\
& \hskip 4mm + \tfrac{1}{4}\xi(-4 \omega_4^+\omega_2^- - 4 \omega_4^- \omega_2^+)
\end{aligned}
\end{align}
\end{subequations}
\step
{\bf Claim.} \emph{If $\Ric_g = 2\,\dd \phi\otimes \dd \phi$ and $\Box_g \phi = 0$, then the ten functions $\Omega^{\sigma}_i$ with $i \in \{1,2,2\sharp,3,4\}$ and $\sigma = \pm$ all vanish.}
\vskip 4mm
\noindent
In fact, $[D^-,D^+] = \mu(D^-+D^+)$
and equations \eqref{ekjehee} imply that $\Omega_1^{\sigma} - \Omega_2^{\sigma} - \Omega_{2\sharp}^{\sigma}$ and $-\Omega_2^+ + \Omega_2^-$ and $-\Omega_3^+ + \Omega_3^-$ and $-\Omega_4^+ + \Omega_4^-$ all vanish. By $\Ric_g = 2\, \dd \phi\otimes \dd \phi$ and $\Box_g \phi = 0$, all $\Omega$'s vanish.
\vskip 2mm
\noindent
{\bf Remark.} The principal parts of the right hand sides of \eqref{eezuuirzr} have the directions indicated in Figure \ref{hfkjhjfkhkfhfkhfjfd}, and
their coefficients vanish at $\xi=0$. This \emph{degeneration} at $\xi=0$ is standard for polar coordinates.
\subsubsection{From 2D to 4D}\label{2d4d}
 A periodic 2D problem (called {\bf 2Dprob}) is defined below, every solution of which yields a 4D spacetime as in Section \ref{4d2d}. The periodicity of {\bf 2Dprob} corresponds to the fundamental self-similarity diffeomorphism $\Theta$ of the 4D spacetime. \emph{This section is logically independent from Section \ref{4d2d}.}
\vskip 2mm
\noindent {\bf Prerequisites (2Dpre).} \emph{Let $\xi_{\ast} > 1$ be constant. Let
\begin{equation} \label{eueueziu33333}
(\tau,\xi) \;\;\in \;\;\Zylinder = (\R/4\pi\Z)\times (-\xi_{\ast},\xi_{\ast})
\end{equation}
The set $\Zylinder$ is a cylinder. The unknowns are a constant $\mu > 0$ and four real analytic functions $\omega_1,\omega_2,\omega_3,\omega_4: \Zylinder \to \R$ that satisfy (a) through (d):
\begin{enumerate}[(a)]
\item Under $(\tau,\xi) \mapsto (\tau+2\pi,\xi)$, they transform as $$(\omega_1,\omega_2,\omega_3,\omega_4) \mapsto (\omega_1,\omega_2,\omega_3,-\omega_4)$$ That is, three functions are invariant, one changes its sign.
\item If $\omega_4$ is invariant under $(\tau,\xi) \mapsto (\tau+T,\xi)$, then $T \in 4\pi \Z$.\\
In particular, $\omega_4$ does not vanish identically.
\item The function $\omega_1$ satisfies $P\omega_1 = -\omega_1$. See \eqref{defpdefpdefp}.
\item The inequality $2\mu + \xi(\omega_1 - \omega_2 + P\omega_2) > 0$ holds on $\Zylinder$. See \eqref{defpdefpdefp}.
\end{enumerate}
Here and below, $P$ is defined by
\begin{equation}\label{defpdefpdefp}
(Pf)(\tau,\xi) = f(\tau,-\xi)
\end{equation}
for every function $f$.  Set
 \begin{equation}\label{khdkjhdkhdkdjjdjdjjdjdjjdj}
 \textstyle D^{\sigma} = \sigma \frac{\p}{\p \tau} + \mu(1+\sigma \xi) \frac{\p}{\p \xi}
 \qquad\qquad \sigma = \pm
 \end{equation}
 They are well defined as vector fields on $\Zylinder$. Set $\omega_i^- = \omega_i$ and $\omega_i^+ = -P\omega_i$ for $i=2,3,4$. Adopt equations \eqref{eezuuirzr} as definitions. The $\Omega$'s are now real valued functions \emph{on} $\Zylinder$.}
 \vskip 5mm
\noindent {\bf Periodic 2D problem (2Dprob).} \emph{Find $\mu$ and $(\omega_1,\omega_2,\omega_3,\omega_4)$ that satisfy {\bf 2Dpre} and that satisfy
$\Omega_1^+=\Omega_2^+=\Omega^+_{2\sharp} =\Omega^+_3 = \Omega^+_4=0$ identically on $\Zylinder$.}
\vskip 5mm
\noindent {\bf Claim.} \emph{Every solution $\mu$, $(\omega_1,\omega_2,\omega_3,\omega_4)$ to {\bf 2Dprob} yields a unique real analytic solution $(g,\phi)$ to $\Ric_g = 2\,\dd \phi\otimes \dd \phi$ and $\Box_g \phi = 0$, as in Section \ref{4d2d}, on the open set $\R \times B_{\xi_{\ast}}(0) \subset \R \times \R^3$, that satisfies:
\begin{itemize}
\item The constants $\mu$ in Sections \ref{4d2d} and \ref{2d4d} are identified, and $\omega_1$, $\omega_2^{\pm}$, $\omega_3^{\pm}$, $\omega_4^{\pm}$ in Sections \ref{4d2d} and \ref{2d4d} are identified as functions of $\tau,\xi$ for $\xi\geq 0$.
\item $\zeta(0,0)=0$ and $\phi(2\pi,0) = - \phi(0,0)$.
\end{itemize}
Furthermore, this unique solution satisfies for all $(\tau,x) \in \R \times B_{\xi_{\ast}}(0)$:
\begin{subequations}
\begin{align}
\label{zetazetammm}
Q\circ \Theta & = \phantom{+}Q\\
\label{zetazeta1}
\zeta\circ \Theta & = \phantom{+}\zeta + K \qquad \text{with}\qquad K = 2\pi \mu - \textstyle \int_0^{2\pi} \dd \tau\, \omega_3(\tau,0)\\
\label{zetazeta2}
\rule{0pt}{12pt}\phi\circ \Theta  & = -\phi
\end{align}
\end{subequations}
where $\Theta: (\tau,x) \mapsto (\tau+2\pi,x)$ is a global conformal isometry of $g$.
}
\vskip 2mm
\noindent To check this, suppose $\mu$, $(\omega_1,\omega_2,\omega_3,\omega_4)$ solve {\bf 2Dprob}. Note that $PD^{\sigma}f = -D^{-\sigma}Pf$ for every function $f$. Recall that $P\omega_1 =-\omega_1$ and $P\omega_i^{\sigma} = -\omega_i^{-\sigma}$ with $i=2,3,4$, by (c). Therefore, equations \eqref{eezuuirzr} imply $P\Omega_i^{\sigma} = - \Omega_i^{-\sigma}$. It now follows from {\bf 2Dprob} that $\Omega_1^-=\Omega_2^-=\Omega^-_{2\sharp} =\Omega^-_3 = \Omega^-_4=0$ on $\Zylinder$. That is, all $\Omega$'s vanish on $\Zylinder$. With this piece of information, we can introduce two real analytic functions $Q,\phi$ on $\Zylinder$, and one real analytic function $\zeta$ on $\R\times (-\xi_{\ast},\xi_{\ast})$, the universal cover of $\Zylinder$, as follows:
\begin{itemize}
\item Let $Q: \Zylinder \to \R$ be given by \eqref{dhjhfkfhfe1}. It satisfies \eqref{zetazetammm} and $PQ=Q$.
\item Let $\zeta: \R\times (-\xi_{\ast},\xi_{\ast}) \to \R$ be the unique solution to \eqref{dhjhfkfhfe3} and \mbox{$\zeta(0,0)=0$}. Such a solution exists, because the one form $\alpha$ on $\Zylinder$ given by $\alpha(D^{\sigma}) = \omega_3^{\sigma} + \sigma \mu$ is closed:
\begin{align*}
\xi (\dd \alpha)(D^-,D^+) & = \xi D^-(\alpha(D^+)) - \xi D^+(\alpha(D^-)) - \xi \alpha([D^-,D^+]) \\&  =\xi  (D^--\mu)(\omega_3^+) - \xi (D^++\mu)(\omega_3^-) = \Omega_3^--\Omega_3^+ = 0
\end{align*}
Thus, $\dd \alpha$ vanishes when $\xi\neq 0$, and by continuity at $\xi=0$. Then $f = \zeta \circ \Theta - \zeta$ satisfies $\dd f = 0$, and is equal to a constant $K$. The equation $(\frac{\p}{\p \tau} \zeta)(\tau,0) = \mu - \omega_3(\tau,0)$ implies \eqref{zetazeta1}. The difference $f = P\zeta - \zeta$ satisfies $\dd f = 0$ and $f(\tau,0) = 0$, and therefore $P\zeta = \zeta$ identically.
\item Let $\phi : \R\times (-\xi_{\ast},\xi_{\ast}) \to \R$ be the unique solution to \eqref{dhjhfkfhfe4} and $\phi(2\pi,0) = - \phi(0,0)$. Such a solution exists, because the one form $\alpha(D^{\sigma}) = \omega_4^{\sigma}$ is closed by  $\Omega_4^--\Omega_4^+ = 0$. The sum $f = \phi\circ \Theta + \phi$ satisfies $\dd f = 0$ and $f(0,0)=0$, which implies \eqref{zetazeta2}. It follows that $\phi(\tau+4\pi,\xi) = \phi(\tau,\xi)$, i.e.\,$\phi$ is a function on the cylinder, $\phi: \Zylinder \to \R$. Finally, $P\phi = \phi$.
\end{itemize}
Denote by the same symbols $Q,\zeta$ and $\phi$ the corresponding functions on $\R \times B_{\xi_{\ast}}(0) \subset \R \times \R^3$. Define a frame $(\mathbf{e}_0,\mathbf{e}_1,\mathbf{e}_2,\mathbf{e}_3)$ and a metric $g$ by \eqref{ekhfkjfhkfhfkhf}. The condition $\mu + \xi^2Q>0$ follows from (d). The frame, the metric and $\phi$ are real analytic (here $PQ=Q$ and $P\zeta=\zeta$ and $P\phi=\phi$ are used). While \eqref{dhjhfkfhfe1}, \eqref{dhjhfkfhfe3}, \eqref{dhjhfkfhfe4} hold by definition of $Q,\zeta,\phi$, equation \eqref{dhjhfkfhfe2} is now a consequence of equation \eqref{dhjhfkfhfe1}, of the fact that the definitions of $D^{\pm}$ in Sections \ref{4d2d} and \ref{2d4d} are consistent, and of $\Omega_1^{\sigma} - \Omega_2^{\sigma} - \Omega_{2\sharp}^{\sigma} = 0$. Therefore, equations \eqref{ekjehee} hold. The vanishing of all the $\Omega$'s and equations \eqref{dfkhdjhsdshkfd} imply that $\Ric_g - 2 \,\dd \phi\otimes \dd \phi=0$ and $\Box_g \phi=0$ when $\xi\neq 0$, and when $\xi=0$ by continuity of the left hand sides. The uniqueness part of the claim is automatic, because every step of the construction is forced.
\subsubsection{The $\sharp$ system}\label{dkhfkhdfkhfieuhfkdf}
Suppose $\mu$, $(\omega_1,\omega_2,\omega_3,\omega_4)$ satisfy {\bf 2Dpre}, but not necessarily  {\bf 2Dprob}. A straightforward calculation with $[D^-,D^+] = \mu(D^-+D^+)$ and $-D^-=PD^+P$ shows that the $\Omega$'s satisfy a homogeneous differential identity.
Namely, with $\Omega_i = \Omega_i^-$ for all $i \in \{1,2,2\sharp,3,4\}$ (equation may continue on the next page):
\begin{align*}
& 0 = \xi \begin{pmatrix}
\tfrac{1}{2}(1-P)(D^++\mu)\Omega_1 \\
(D^++\mu)\Omega_{2\sharp} - P(D^++\mu)\Omega_2
\end{pmatrix}
+ \mu \begin{pmatrix}
\tfrac{1}{2}(1+P)\Omega_1\\
\Omega_1+\Omega_2+ P\Omega_2-2 P\Omega_3
\end{pmatrix} \displaybreak[0]\\
& + \frac{\xi}{2} \left\{
\begin{pmatrix}2P\omega_2 - P\omega_3 + \omega_1 - 2\omega_2 + \omega_3\\
P\omega_2 + \omega_1 - \omega_2\end{pmatrix}\frac{\Omega_1+P\Omega_1}{2}  \right. \\ & \hskip13mm \left.
 +\begin{pmatrix}P\omega_2 - P\omega_3 + \omega_2 - \omega_3\\
 P\omega_2 + \omega_1 - \omega_2\end{pmatrix}\frac{\Omega_1-P\Omega_1}{2} \right. \displaybreak[0] \\ & \hskip13mm\left.
 +\begin{pmatrix}-P\omega_3 - 3\omega_1 + \omega_3\\
 -P\omega_2 - \omega_1 - 7\omega_2 + 4\omega_3\end{pmatrix}\frac{\Omega_2+P\Omega_2}{2} \right. \displaybreak[0] \\ & \hskip13mm\left.
 +\begin{pmatrix}-P\omega_2 - P\omega_3 - \omega_2 - \omega_3\\
 -P\omega_2 - \omega_1 + \omega_2 - 4\omega_3\end{pmatrix}\frac{\Omega_2-P\Omega_2}{2} \right. \displaybreak[0] \\ & \hskip13mm\left.
+ \begin{pmatrix}P\omega_3 - \omega_1 - \omega_3\\
3P\omega_2 - \omega_1 + \omega_2\end{pmatrix} \frac{\Omega_{2\sharp}+P\Omega_{2\sharp}}{2} \right. \displaybreak[0] \\ &\hskip13mm \left.
 +\begin{pmatrix}P\omega_2 + P\omega_3 + \omega_2 + \omega_3\\
 3P\omega_2 - \omega_1 + \omega_2\end{pmatrix} \frac{\Omega_{2\sharp}-P\Omega_{2\sharp}}{2} \right. \displaybreak[0] \\ & \hskip13mm\left.
 +\begin{pmatrix}2\omega_1\\
 4\omega_2\end{pmatrix}\frac{\Omega_3+P\Omega_3}{2}
 +\begin{pmatrix}-2P\omega_2 - 2\omega_2\\
 -4\omega_2\end{pmatrix}\frac{\Omega_3-P\Omega_3}{2} \right. \displaybreak[0] \\ & \hskip13mm\left.
 +\begin{pmatrix}2P\omega_4 - 2\omega_4\\
 -4\omega_4\end{pmatrix}\frac{\Omega_4+P\Omega_4}{2} 
 +\begin{pmatrix}2P\omega_4 + 2\omega_4\\
 4\omega_4\end{pmatrix}\frac{\Omega_4-P\Omega_4}{2}\right\}
\end{align*}
{\bf Claim.} \emph{Suppose $\mu$ and $(\omega_1,\omega_2,\omega_3,\omega_4)$ satisfy {\bf 2Dpre}. If $$(1+P)\Omega_1^+=\Omega_2^+=\Omega_3^+=\Omega_4^+=0\qquad \text{on $\Zylinder$}$$ then $P\Omega_1=-\Omega_1$ and (this will be referred to as the $\sharp$ system):}
\begin{equation}\label{kdjskkkkksjsjhd}
\begin{aligned}
& 0 = \xi \begin{pmatrix}
\tfrac{1}{2}(1-P)(D^++\mu)\Omega_1\\
(D^++\mu)\Omega_{2\sharp}
\end{pmatrix}
+ \mu \begin{pmatrix}
0\\
\Omega_1
\end{pmatrix}\\
& + \frac{\xi}{2} \left\{
 \begin{pmatrix}P\omega_2 - P\omega_3 + \omega_2 - \omega_3\\
 P\omega_2 + \omega_1 - \omega_2\end{pmatrix} \Omega_1
+ \begin{pmatrix}P\omega_3 - \omega_1 - \omega_3\\
3P\omega_2 - \omega_1 + \omega_2\end{pmatrix} \frac{\Omega_{2\sharp}+P\Omega_{2\sharp}}{2}
\right. \\ &\hskip50mm \left.
 +\begin{pmatrix}P\omega_2 + P\omega_3 + \omega_2 + \omega_3\\
 3P\omega_2 - \omega_1 + \omega_2\end{pmatrix} \frac{\Omega_{2\sharp}-P\Omega_{2\sharp}}{2} \right\}
\end{aligned}
\end{equation}
\subsection{Fourier-Chebyshev series}
We reformulate {\bf 2Dprob} in Section \ref{2d4d} as a new problem {\bf 2DprobSeries} for Fourier-Chebyshev series, see page \pageref{dd384g8r784rh49}, and introduce handy notation for all the objects in {\bf 2DprobSeries}, to prepare for the analysis in Section \ref{sec:ANALYSISanalysis}.
\step
This section is essentially a collection of definitions. Reading through them linearly makes sense from a logical point of view, but is not advisable, because the motivation for each definition lies in the definitions that come after it.
\thin
{\bf Series.} For every $v=(v_{mn})$ in the real vector space
$$
\Space  = \bigg\{(v_{mn})_{m,n\in \Z}\subset \C\,\bigg|\, \begin{aligned}
& v_{mn} = \overline{v_{-m,-n}}\text{ and } v_{mn} = v_{m,-n} \text{ and there}\\
& \text{are $\alpha,\beta > 0$ with $\textstyle\sup_{m,n \in \Z}|e^{\alpha |m| + \beta |n|}v_{mn}| < \infty$}
\end{aligned}\bigg\}
$$
we introduce the Fourier in $\tau$, Chebyshev in $\xi$ series
\begin{equation}\label{eq:operators}
\TF(v) = \Big(\textstyle\sum_{m,n \in \Z} v_{mn}\exp(\tfrac{1}{2}im\tau+in\theta)\Big)_{\theta = \arccos \xi}
\end{equation}
It is a function of $(\tau,\xi)$, or a function of $(\tau,\theta)$ that is even in $\theta$.
\thin
{\bf Properties of \eqref{eq:operators}.} Let $v \in \Space$.
\begin{itemize}
\item There are $\alpha,\beta > 0$ as above, depending on $v$, such that  \eqref{eq:operators} converges and is analytic jointly in $(\tau,\xi)$ on the open, product subset of $\C\times \C$:
$$|\IM(\tau)| < 2\alpha \qquad \text{and}\qquad \frac{(\RE \xi)^2}{(\cosh\beta)^2} + \frac{(\IM\xi)^2}{(\sinh \beta)^2} < 1$$
The ellipse contains the real interval $\xi \in [-1,1]$. Cf.~Figure \ref{hfkjhdkkddkdd}. 
\item The series \eqref{eq:operators} is real when $\tau$ and $\xi$ are real.
\item  In terms of the polynomials  $\Chebyshev_n(\xi) = \cos (n \arccos \xi)$,
\begin{align*}
& \TF(v) \\
& = \textstyle\sum_{m\in \Z} v_{m0}\exp(\tfrac{1}{2}im\tau) +
 2 \textstyle\sum_{m \in \Z,n >0} v_{mn} \exp(\tfrac{1}{2}im\tau)\, \Chebyshev_n(\xi)
\end{align*}
\end{itemize}%
\thin
{\bf Convention:} In this section, $\TF$ is thought of as a map from $\Space$ to maps $(\R/4\pi \Z) \times [-1,1] \to \R$. Advantage: all functions produced by $\TF$ have the same domain. By analytic continuation, the identities that we state will automatically hold on appropriate domains in $\C \times \C$.
\thin
{\bf Fields and multifields.} We refer to elements of $\Space$ as \emph{fields}, and to tuples of fields as \emph{multifields}. For fields, we define the real vector spaces:
\begin{align*}
\TauPer & = \big\{v\in \Space \,\big|\, v_{mn} = 0 \text{ for all odd $m$}\big\}\\
\TauAntiPer  & = \big\{v\in \Space \,\big|\, v_{mn} = 0 \text{ for all even $m$}\big\}\\
\Gauged{\Space} & = \big\{v\in \Space \,\big| \RE(v_{10}) = 0\big\} \\ 
\XiOdd  & = \big\{v\in \Space \,\big|\, v_{mn} = 0 \text{ for all even $n$}\big\}
\end{align*}
For multifields:
\begin{align*}
\SPACE & = \Space_{\mathrm{TP}}\times\Space_{\mathrm{TP}}\times\Space_{\mathrm{TP}}\times\Space_{\mathrm{TP}}\times\Space_{\mathrm{TAP}} && \subset \Space^5\\
\SSPACE & = 
\big(\Space_{\mathrm{TP}} \cap \XiOdd \big)\times 
\Space_{\mathrm{TP}} \times
\Space_{\mathrm{TP}}  \times
\Space_{\mathrm{TAP}}  && \subset \Space^4\\
\Gauged{\SSPACE} & = 
\big(\Space_{\mathrm{TP}}  \cap \XiOdd \big)\times 
\Space_{\mathrm{TP}}  \times
\Space_{\mathrm{TP}}  \times
\big(\Space_{\mathrm{TAP}}  \cap \Gauged{\Space}\big)
 && \subset \Space^4\\
\SSPACE^{\sharp} & = 
\big(\Space_{\mathrm{TP}}  \cap \XiOdd \big)\times 
\Space_{\mathrm{TP}}  && \subset \Space^2
\end{align*}
with the local notation $\Space_{\mathrm{TP}} = \TauPer$,\,
$\Space_{\mathrm{TAP}} = \TauAntiPer$.
\thin
Every field $v \in \Space$ is determined by its $m,n\geq 0$ elements. Precisely, the map
\begin{align*}
\Space & \to  \bigg\{(v_{mn})_{m,n \geq 0}\subset \C\,\bigg|\, \begin{aligned}
& v_{0n} \in \R \text{ and there are $\alpha,\beta > 0$}\\
& \text{with $\textstyle\sup_{m,n \geq 0}|e^{\alpha m + \beta n}v_{mn}| < \infty$}
\end{aligned}\bigg\} \\
v & \mapsto v|_{m,n \geq 0}
\end{align*}
is bijective, an isomorphism. \emph{This useful fact will often be used implicitly.}
\thin
{\bf Auxiliary linear operator $\Band$.}
Let $f = (f_{mn})_{m,n\geq 0} \subset \C$ be a sequence with $f_{0n}\in \R$, and for which $|f_{mn}|$ has a polynomial bound in $m$ and $n$. For $k \geq 0$ define
$\Band[k,f]: \Space  \to \Space $ by
$$\forall m,n \geq 0:\qquad \big(\Band[k,f]v\big)_{mn} = f_{mn} v_{m,n+k}$$
It is a well defined $\Space \to \Space$ map. It acts \emph{diagonally} on the index $m$, and therefore
leaves the subspaces
$\TauPer $ and
$\TauAntiPer $ invariant.
\step As a special case, introduce the operator $\BandEXEP: \Space \to \Space$ by
$$\forall m,n\geq 0:\qquad (\BandEXEP v)_{mn} = \begin{cases}
0 & \text{if $n=0$}\\
v_{m,n-1} & \text{if $n > 0$}
\end{cases}$$
\thin
{\bf Auxiliary linear operator $\BandI$.}
Let $f$ be as above, with $|f_{mn}|\leq 1$ for all $m,n$.
For $k>0$, define the linear map
$\BandI[k,f]: \Space \to \Space$ by
\begin{align*}
\forall m,n\geq 0:\quad (\BandI[k,f]v)_{mn}
& = \textstyle\sum_{L=0}^{\infty} (-1)^L (\Band[k,f]^Lv)_{mn}\\
& = \textstyle\sum_{L=0}^{\infty} (\Band[kL,f_L]v)_{mn}
\end{align*}
with $(f_L)_{mn} = (-1)^{L} \textstyle\prod_{\ell=0}^{L-1}f_{m,n+k\ell}$. Note that
$$\BandI[k,f]\big(1 + \Band[k,f]\big)
= \big(1 + \Band[k,f]\big)\BandI[k,f] = 1$$
In particular, $1+\Band[k,f]$ is invertible if $k>0$ and $|f_{mn}|\leq 1$ for all $m,n$.
\thin
{\footnotesize To see that $\BandI[k,f]: \Space \to \Space$ is well defined, note that $|(f_L)_{mn}|\leq 1$ for all $L,m,n$. Therefore, if $\alpha,\beta > 0$ are such that $\sup_{m,n\geq 0} |e^{\alpha m + \beta n}v_{mn}| < \infty$, then
$${\textstyle\sup_{m,n \geq 0}} |e^{\alpha m + \beta n}  (\BandI[k,f]v)_{mn}|
\leq (1-e^{-k \beta})^{-1} \; \textstyle{\sup_{m,n\geq 0}} |e^{\alpha m + \beta n}v_{mn}| < \infty$$
In other words, if $\alpha,\beta > 0$ work for $v$, then the same $\alpha,\beta$ also work for $\BandI[k,f]v$.
}
\thin
{\bf Notation for $\Band$, $\BandI$.} Three examples illustrate the notation. Here $i =\sqrt{-1}$.
\begin{alignat*}{4}
& \text{$\BandI[2,-1]$} && =\BandI_{2,-1} && = \BandI[2,f] \text{ with $f_{mn} = -1$}\\
& \text{$\Band[0,\delta_{0n}]$} && = \Band_{0,\delta_{0n}} && = \Band[0,f] \text{ with $f_{mn} = \delta_{0n}$ (Kronecker delta)}\\
& \text{$\Band[1,-im/2+\mu n]$} && = \Band_{1,-im/2+\mu n} && = \Band[1,f] \text{ with $f_{mn} = -im/2+\mu n$}
\end{alignat*}
\thin
 {\bf Basic linear operators.} Introduce the $\Space \to \Space$ linear operators
\begin{align*}
\XiParity & = \Band_{0,(-1)^n} \\
\TauDerivative  & = \Band_{0,im/2} \\
\TimesXi & = \tfrac{1}{2}(\BandEXEP+\Band_{1,1} + \Band_{1,\delta_{0n}}) \\
\OnePlusXiTimesXiDerivative & = \big(2 \BandI_{1,-1} - 1\big) \Band_{0,n}\\
\DivideByXiRegularized  & = 2\Band_{1,1}\BandI_{2,1}
\end{align*}
They satisfy
(here $P_{\text{Section \ref{2d4d}}}$ is \eqref{defpdefpdefp})
\begin{subequations} \label{3765444f4jgjgh}
\begin{align}
\label{eq:sidnew1} \TF(\XiParity v) & = P_{\text{Section \ref{2d4d}}}\TF(v)\\
\label{eq:sidnew2} \TF(\TauDerivative v) & = \tfrac{\p}{\p \tau} \TF(v)\\
\label{eq:sidnew3} \TF(\TimesXi v)& = \xi \TF(v)\\
\label{eq:sid3} \TF(\OnePlusXiTimesXiDerivative v) & = (1+\xi)\tfrac{\p}{\p \xi}\TF(v)\\
\notag \TF(\DivideByXiRegularized v) & = \xi^{-1} \big(\TF(v)\\
\label{eq:sid4} & \hskip 9mm - \TF(v)|_{\xi \to 0}\big)
\end{align}
\end{subequations}
The operator $\DivideByXiRegularized$ is only a left-inverse of $\TimesXi$:
\begin{subequations}
\begin{align}
\label{rhifhfhkrhr} \DivideByXiRegularized  \TimesXi & = 1\\
\TimesXi  \DivideByXiRegularized  & = 1 - (\Band_{0,\delta_{0n}}-\Band_{2,\delta_{0n}})\BandI_{2,1}
\end{align}
\end{subequations}
\thin
{\footnotesize
The identities \eqref{eq:sidnew1}, \eqref{eq:sidnew2} are checked directly, and for 
\eqref{eq:sidnew3} one only has to use $\xi = \cos \theta = \tfrac{1}{2}(e^{i\theta} + e^{-i\theta})$. To check \eqref{eq:sid3} and \eqref{eq:sid4}, observe that
\begin{subequations}
\begin{align}
\label{eq;sidaux1}  \sin \theta \tfrac{\p}{\p \theta} \TF(v) & = \TF(\tfrac{1}{2}(\BandEXEP-\Band_{1,1} - \Band_{1,\delta_{0n}})\Band_{0,n}v)\\
\label{eq;sidaux3}  \TF(v)|_{\xi \to 0} & = \TF((\Band_{0,\delta_{0n}}-\Band_{2,\delta_{0n}})\BandI_{2,1}v)
\end{align}
\end{subequations}
with $\xi = \cos \theta$ in \eqref{eq;sidaux1}.
Here $\sin \theta \tfrac{\p}{\p \theta}$ has to be moved inside $\TF$ as a whole, because the definition of $\Space $ only allows even functions of $\theta$.
Multiply \eqref{eq:sid3} by $(1- \xi)$ and use $(1-\xi^2)\tfrac{\p}{\p \xi} = -\sin \theta \tfrac{\p}{\p \theta}$
and \eqref{eq;sidaux1} and \eqref{eq:sidnew3}.
Multiply \eqref{eq:sid4} by $\xi$ and use \eqref{eq:sidnew3} and \eqref{eq;sidaux3}. This reduces \eqref{eq:sid3} and \eqref{eq:sid4} to the operator identities
\begin{align*}
(\BandEXEP + \Band_{1,1} + \Band_{1,\delta_{0n}}-2)(2\BandI_{1,-1} - 1)\Band_{0,n}
& = (\BandEXEP - \Band_{1,1} - \Band_{1,\delta_{0n}})\Band_{0,n}\\
(\BandEXEP + \Band_{1,1} + \Band_{1,\delta_{0n}})\Band_{1,1}\BandI_{2,1} & = 1 - (\Band_{0,\delta_{0n}} - \Band_{2,\delta_{0n}})\BandI_{2,1}
\end{align*}
The next step is to remove all $\BandI$'s. In the second line, just multiply from the right by the invertible  $1+\Band_{2,1}$ to remove $\BandI_{2,1}$. The first line requires an extra step: add the vanishing term $2 \Band_{0,\delta_{n0}}\Band_{0,n}$ to the left hand side, obtain a sufficient identity by removing the three occurrences of $\Band_{0,n}$, only then multiply from the right by the invertible $1 + \Band_{1,-1} = 1-\Band_{1,1}$ to remove $\BandI_{1,-1}$. Check the resulting quadratic $\Band$ identities using
$\BandEXEP\Band_{1,1} = 1 - \Band_{0,\delta_{n0}}$
and
$\Band_{k,f}\Band_{\ell,g}
= \Band_{k+\ell,h}$ with $h_{mn} = f_{mn} g_{m,n+k}$.
}
\thin
{\bf Convolution.} Define the $\R$-bilinear convolution $\ast: \Space \times \Space \to \Space $ by
$$\forall m,n\in \Z:\qquad (v \ast w)_{mn} = \textstyle\sum_{\substack{m_1,n_1,m_2,n_2 \in \Z: \\ m_1+m_2 = m,\, n_1+n_2=n}} v_{m_1n_1} w_{m_2n_2}$$
It satisfies $\TF(v\ast w) = \TF(v)\TF(w)$.
\thin
{\bf The $\OOA$ and $\OOB$ operators.} 
Let $\mu > 0$. Introduce the $\Space  \to \Space$ linear operators
\begin{align*}
\DD{\OOA} & = \OnePlusXiTimesXiDerivative + 1\\
\OOA_{\mu} & = \TauDerivative  + \mu\DD{\OOA} \displaybreak[0]\\
\DD{\OOB} & =
\tfrac{1}{2}(\DD{\OOA}+P\DD{\OOA}P)\\
\OOB_{\mu} & = \TauDerivative  + \mu\DD{\OOB}
 = \tfrac{1}{2}(\OOA_{\mu}+P\OOA_{\mu}P)
\end{align*}
with 
$\TF(\OOA_{\mu} v) = (D^+_{\text{Section \ref{2d4d}}}+\mu)\TF(v)$. Note that
\begin{equation}\label{jdjlkjf}
\tfrac{1}{2}(1 + \sigma \XiParity) \OOA_{\mu}  \tfrac{1}{2}(1 + \sigma \XiParity) = \OOB_{\mu} \tfrac{1}{2}(1 +\sigma \XiParity)
\qquad \sigma = \pm
\end{equation}
Equivalent definitions of the above operators, and expressions for the inverses of $\OOA$ and $\OOB$, with $\LINDEC_{\mu} = \Band_{0,im/2+\mu(n+1)}$
and
$\LINDEC_{\mu}^{-1} = \Band[0,\frac{1}{im/2+\mu(n+1)}]$, are:
\begin{subequations}\label{dkhddjjjjjff}
\begin{equation}
\begin{aligned}
\DD{\OOA} & = \BandI_{1,-1} \big(\Band_{0, n+1} + \Band_{1,n}\big) \\
\OOA_{\mu} & = \BandI_{1,-1} \big(\Band_{0,im/2+\mu(n+1)} + \Band_{1,-im/2+\mu n}\big) \\
&= \BandI_{1,-1} \LINDEC_{\mu} \big(1 + \Band[1,\,\tfrac{-im/2+\mu n}{+im/2+\mu(n+1)}]\big) \\
\OOA_{\mu}^{-1} & = \BandI[1,\,\tfrac{-im/2+\mu n}{+im/2+\mu(n+1)}]\, \LINDEC_{\mu}^{-1} \,\big(1+\Band_{1,-1}\big)
\end{aligned}
\end{equation}
and
\begin{equation}\label{z484ihrhr}
\begin{aligned}
\DD{\OOB} & = \BandI_{2,-1} \big(\Band_{0, n+1} + \Band_{2,n+1}\big)\\
\OOB_{\mu} & = \BandI_{2,-1} \big(\Band_{0,+im/2+\mu(n+1)} + \Band_{2,-im/2+\mu(n+1)}\big)\\
& = \BandI_{2,-1} \LINDEC_{\mu} \big(1 + \Band[2,\,\tfrac{-im/2+\mu(n+1)}{+im/2+\mu(n+1)}]\big) \\
\OOB_{\mu}^{-1} & = \BandI[2,\,\tfrac{-im/2+\mu(n+1)}{+im/2+\mu(n+1)}]\,\LINDEC_{\mu}^{-1}  \,\big(1+\Band_{2,-1}\big)
\end{aligned}
\end{equation}
\end{subequations}
\thin
{\footnotesize
To check $\DD{\OOB}$ in \eqref{z484ihrhr}, note that
$P \Band_{k,f} P = \Band_{k,(-1)^kf}$
and 
$P \BandI_{k,f} P = \BandI_{k,(-1)^kf}$, and
\begin{align*}
2\DD{\OOB}  = \DD{\OOA} + P\DD{\OOA}P
& = \BandI_{1,-1} (\Band_{0, n+1} + \Band_{1,n}) 
+ (P \BandI_{1,-1} P)(P (\Band_{0, n+1} + \Band_{1,n}) P) \\
& = \BandI_{1,-1} (\Band_{0, n+1} + \Band_{1,n})
+ \BandI_{1,1} (\Band_{0, n+1} - \Band_{1,n})\\
& = \BandI_{2,-1}(1+ \Band_{1,1}) (\Band_{0, n+1} + \Band_{1,n})
+ \BandI_{2,-1}(1+\Band_{1,-1}) (\Band_{0, n+1} - \Band_{1,n})
\end{align*}
Last step: $\BandI_{2,-1}
= \BandI_{1,\sigma}\BandI_{1,-\sigma}$ for $\sigma = \pm$, because $1+\Band_{2,-1} = (1+\Band_{1,-\sigma})(1+\Band_{1,\sigma})$.
}
\thin
{\bf Linear operators for multifields.} Set
\begin{alignat*}{4}
\SELECTION & = \DivideByXiRegularized \circ \begin{pmatrix}
\tfrac{1}{2}(1+\XiParity) & 0 & 0 & 0 & 0\\
0 & 1 & 0 & 0 & 0\\
0 & 0 & 0 & 1 & 0\\
0 & 0 & 0 & 0 & 1
\end{pmatrix}\\
& = \begin{pmatrix}
\tfrac{1}{2}(1-\XiParity) & 0 & 0 & 0 & 0\\
0 & 1 & 0 & 0 & 0\\
0 & 0 & 0 & 1 & 0\\
0 & 0 & 0 & 0 & 1
\end{pmatrix} \circ \DivideByXiRegularized 
&\quad& : \SPACE \to \SSPACE\displaybreak[0]\\
\rule{0pt}{25pt} \SELECTION^{\sharp} & = \begin{pmatrix}
\tfrac{1}{2}(1-\XiParity) & 0 & 0 & 0 & 0\\
0 & 0 & -P & 0 & 0 
\end{pmatrix}&\quad& : \SPACE \to \SSPACE^{\sharp}
\end{alignat*}
The operators $\SELECTION$ and $\SELECTION^{\sharp}$ will be used to split the equations into two parts. The part selected by  $\SELECTION$ has `as many equations as unknowns', the part selected by $\SELECTION^{\sharp}$ are the constraint equations.
For $\mu > 0$ introduce
\begin{alignat*}{4}
\HH_{\mu} & = \begin{pmatrix}
\OOA_{\mu} & 0 & 0 & 0\\
0 & \OOA_{\mu} & 0 & 0\\
0 & - \OOA_{\mu} \XiParity & 0 & 0\\
0 & 0 & \OOA_{\mu} & 0\\
0 & 0 & 0 & \OOA_{\mu}
\end{pmatrix} &&: \SSPACE \to \SPACE\displaybreak[0]\\
\DD{\OO} & = \diag(\DD{\OOB},\DD{\OOA},\DD{\OOA},\DD{\OOA})&&: \SSPACE \to \SSPACE\\
\OO_{\mu} & = \diag(\OOB_{\mu}, \OOA_{\mu}, \OOA_{\mu},  \OOA_{\mu} )\\
& = \TauDerivative + \mu \DD{\OO} &&: \SSPACE \to \SSPACE
\end{alignat*}
and note that $\OO_{\mu}^{-1} = \diag(\OOB_{\mu}^{-1}, \OOA_{\mu}^{-1}, \OOA_{\mu}^{-1},  \OOA_{\mu}^{-1} )$.
Note that $\DD{\OO}, \OO_{\mu}$ map $\SSPACE\to \SSPACE$ by
$[\XiParity,\DD{\OOB}]=[\XiParity,\OOB_{\mu}]=0$. 
 The definition of $\SSPACE$
 and \eqref{rhifhfhkrhr}, \eqref{jdjlkjf} 
 imply $\OO_{\mu} = \SELECTION \TimesXi \HH_{\mu}$.
\thin
\newcommand{\unknownone}{\mathbf{v}}%
\newcommand{\unknowntwo}{\mathbf{w}}%
{\bf The operators $\Gamma_1$ and $\Gamma_2$.} Define the linear operator
$\Gamma_1: \SSPACE\to \SPACE$ by
\begin{align*}
\Gamma_1 \mathbf{v} & = \begin{pmatrix}
2  v_1\\
v_1  +  v_2  -  P  v_2\\
2  P  v_3-2  P  v_2\\
- v_1\\
 v_4  -  P  v_4
\end{pmatrix}
\end{align*}
with $\mathbf{v}=(v_1,\ldots,v_4)$.
As required, $\Gamma_1 \mathbf{v} \in \SPACE$. Define the symmetric, bilinear operator
$\Gamma_2: \SSPACE\times \SSPACE\to \SPACE$ by (equation may continue on the next page)
\begin{align*}
&\Gamma_2(\mathbf{v},\mathbf{w}) =\\
& \frac{1}{4}\begin{pmatrix} 3 Pv_2-2 Pv_3+v_1-v_2\\Pv_2+v_1-v_2\\0\\-Pv_2-v_1+v_2\\0\end{pmatrix} \ast w_1 \displaybreak[0]\\
&+\frac{1}{2}\begin{pmatrix}Pv_2+2 Pv_3+v_1+v_2\\2 Pv_2+2 v_2\\-2 Pv_2+2 Pv_3\\-Pv_2-v_2\\Pv_4+v_4\end{pmatrix}\hskip-2pt \ast\hskip-1pt \frac{w_2+Pw_2}{2}+
\frac{1}{2}\begin{pmatrix}-2 v_1\\ \hskip-2pt Pv_2-v_1-v_2\hskip-2pt\\ \hskip-2pt 2 Pv_2-2 Pv_3\hskip-2pt\\v_1\\Pv_4-v_4\end{pmatrix}\hskip-2pt\ast \hskip-1pt\frac{w_2-Pw_2}{2}\displaybreak[0]\\
&+\frac{1}{2}\begin{pmatrix}Pv_2-v_1+v_2\\0\\2 Pv_2\\0\\0\end{pmatrix}\ast \frac{w_3+Pw_3}{2}+
\frac{1}{2}\begin{pmatrix}v_1-Pv_2-v_2\\0\\-2 Pv_2\\0\\0\end{pmatrix}\ast \frac{w_3-Pw_3}{2}\displaybreak[0]\\
&+\frac{1}{2}\begin{pmatrix}-2 Pv_4\\0\\-2 Pv_4\\Pv_4+v_4\\Pv_2+v_2\end{pmatrix}\ast \frac{w_4+Pw_4}{2}+
\frac{1}{2}\begin{pmatrix}2 Pv_4\\0\\2 Pv_4\\Pv_4-v_4\\Pv_2-v_2\end{pmatrix}\ast \frac{w_4-Pw_4}{2}
\end{align*}
where $\ast$ is convolution.
As required, $\Gamma_2(\mathbf{v},\mathbf{w})
= \Gamma_2(\mathbf{w},\mathbf{v})$ and $\Gamma_2(\mathbf{v},\mathbf{w}) \in \SPACE$.\\
For later reference (equation may continue on the next page):
\begin{multline}\label{skdfhkfdjhdkhjdgfdjh33hk}
 \SELECTION \TimesXi \Gamma_2(\mathbf{v},\mathbf{w}) = 
 \frac{1}{4}\begin{pmatrix}
Pv_2+v_2-Pv_3-v_3\\
v_1-v_2+Pv_2\\
-v_1+v_2-Pv_2\\
0
\end{pmatrix} \ast w_1\displaybreak[0]\\
 + \frac{1}{2}\begin{pmatrix}
v_1+Pv_3-v_3\\
2v_2+2Pv_2\\
-v_2-Pv_2\\
Pv_4+v_4
\end{pmatrix} \ast \frac{w_2+Pw_2}{2}
+ \frac{1}{2}\begin{pmatrix}
0\\
-v_1+Pv_2-v_2\\
v_1\\
Pv_4-v_4
\end{pmatrix} \ast \frac{w_2-Pw_2}{2} \displaybreak[0]\\
 +\frac{1}{2}\begin{pmatrix}
-v_1\\
0\\
0\\
0
\end{pmatrix} \ast \frac{w_3+Pw_3}{2}
+\frac{1}{2} \begin{pmatrix}
-v_2-Pv_2\\
0\\
0\\
0
\end{pmatrix} \ast \frac{w_3-Pw_3}{2} \displaybreak[0]\\
 + \frac{1}{2}\begin{pmatrix}
v_4 -Pv_4\\
0 \\
Pv_4+v_4\\
Pv_2+v_2
\end{pmatrix} \ast \frac{w_4+Pw_4}{2}
+
\frac{1}{2}\begin{pmatrix}
v_4 +Pv_4\\
0 \\
Pv_4-v_4\\
Pv_2-v_2
\end{pmatrix} \ast \frac{w_4-Pw_4}{2}
\end{multline}
\thin
{\bf The nonlinear operator $\Omega^+:(0,\infty)\times \Gauged{\SSPACE} \to \SPACE$.} Define
\begin{equation}\label{fkdkkdkkkeieiruus}
\Omega^+(\lambda,\mathbf{v}) = \TimesXi \mathbf{H}_{\lambda} \mathbf{v}
+ \lambda \Gamma_1 \mathbf{v} + \TimesXi \Gamma_2(\mathbf{v},\mathbf{v})
\end{equation}
with components denoted $\Omega^+ = (\Omega_1^+,\Omega_2^+,\Omega_{2\sharp}^+,\Omega_3^+,\Omega_4^+)$. By construction, 
\begin{equation}\label{f23ljf3lfj}
\TF(\Omega^+(\lambda,\mathbf{v})) = \Omega^+_{\text{Section \ref{2d4d}}}(\lambda, \TF(\mathbf{v}))
\end{equation}
where $\TF$ is applied component by component. Note that
$\Omega^+_{\text{Section \ref{2d4d}}}$ redirects to \eqref{eezuuirzr},
but that \eqref{eezuuirzr} must be interpreted as in Section \ref{2d4d}.
\thin
{\bf Periodic 2D problem for Fourier-Chebyshev series (2DprobSeries).} \label{dd384g8r784rh49} \emph{Find a solution $(\mu,\omega) \in (0,\infty) \times \Gauged{\SSPACE}$ to $\Omega^+(\mu,\omega)  = 0$.}
\step
By construction, in particular \eqref{f23ljf3lfj}:
\step
{\bf Claim.} \emph{If $(\mu,\omega)$ solves {\bf 2DprobSeries},
then the pair $(\mu,\TF(\omega))$, as a map on
$\Zylinder_{\text{Section \ref{2d4d}}}$ for some $\xi_{\ast}>1$, solves {\bf 2Dpre} and {\bf 2Dprob} in Section \ref{2d4d}, except for (b) and (d), which must be checked separately.
}
\thin
{\bf Remark.} It is convenient to split $\Omega^+=0$ into two parts. We have:
$$\SELECTION\Omega^+=0\;\;\text{and}\;\;
\SELECTION^{\sharp}\Omega^+=0
\qquad \Longleftrightarrow
\qquad 
\Omega^+=0$$
The direction $\Longrightarrow$ would be immediate from the definitions of $\SELECTION$ and $\SELECTION^{\sharp}$, \emph{if $\DivideByXiRegularized$ had a left inverse}. Nevertheless, the following observations yield $\Longrightarrow$:
\begin{itemize}
\item The first and third terms on the right hand side of \eqref{fkdkkdkkkeieiruus} are in the image of $\TimesXi$, and $\TimesXi \DivideByXiRegularized \TimesXi = \TimesXi$.
\item The components $1,2,4,5$ of $\Gamma_1\mathbf{v}$ are all in the image of $1-P$, and $\TimesXi \DivideByXiRegularized (1-P)=(1-P)$. Component $3$ of $\Gamma_1\mathbf{v}$ is annihilated by $\SELECTION$.
\end{itemize}
\thin
{\bf The $\sharp$ system.} 
Introduce the $\SSPACE^{\sharp} \to \SSPACE^{\sharp}$ linear operators
\begin{align*}
\DD{\OO}^{\sharp} & = \mathrm{diag}(\DD{\OOB},\DD{\OOA}) \\
\OO_{\mu}^{\sharp} & = \mathrm{diag}(\OOB_{\mu},\OOA_{\mu}) = \TauDerivative + \mu \DD{\OO}^{\sharp}\\
\Gamma_1^{\sharp}\mathbf{w} & = \begin{pmatrix} 0 \\ w_1
\end{pmatrix}
\end{align*}
and the $\Gamma_2^{\sharp}: \SSPACE \times \SSPACE^{\sharp} \to \SSPACE^{\sharp}$  bilinear operator
\begin{align*}
\Gamma_2^{\sharp}(\mathbf{v},\mathbf{w}) & = 
\frac{1}{2}\begin{pmatrix}
 v_2 
+Pv_2
-Pv_3
-v_3\\
v_1-v_2+Pv_2
\end{pmatrix} \ast w_1\\
&\qquad + \frac{1}{2}\begin{pmatrix}
-v_1+Pv_3-v_3\\
-v_1 +v_2 +3 Pv_2
\end{pmatrix}
\ast \frac{w_{2\sharp} + Pw_{2\sharp}}{2}\\
&\qquad
+ \frac{1}{2}\begin{pmatrix}
v_2+Pv_2+Pv_3+v_3\\
-v_1 +v_2 +3 Pv_2
\end{pmatrix}
\ast \frac{w_{2\sharp} - Pw_{2\sharp}}{2}
\end{align*}
where, $\mathbf{v}=(v_1,v_2,v_3,v_4)$ and $\mathbf{w}=(w_1,w_{2\sharp})$.
\step
{\bf Claim.} \emph{If $(\mu,\omega)\in (0,\infty) \times \Gauged{\SSPACE}$ is a solution to $\SELECTION\Omega^+(\mu,\omega) = 0$, then
$\omega^{\sharp} = \SELECTION^{\sharp} \Omega^+(\mu,\omega)$
satisfies the linear homogeneous identity}
\begin{equation}\label{6383hi3h39z}
\OO_{\mu}^{\sharp} \omega^{\sharp} + \mu \DivideByXiRegularized \Gamma_1^{\sharp}\omega^{\sharp} + \Gamma_2^{\sharp}(\omega,\omega^{\sharp}) = 0
\end{equation}
See \eqref{kdjskkkkksjsjhd}.
Note that
$\TF(\omega^{\sharp}) = (\Omega_1,\Omega_{2\sharp})_{\text{Section \ref{dkhfkhdfkhfieuhfkdf}}}$
by \eqref{f23ljf3lfj} and $\SELECTION \Omega^+=0$.
\subsection{Approximate solution yields true solution}\label{sec:ANALYSISanalysis}
This section contains the analysis.
We rearrange $\mathbf{S}\Omega^+=0$, solve it with the contraction mapping principle, and then use the $\sharp$ system to show that $\Omega^+=0$.
\emph{In other words, we solve {\bf 2DprobSeries}.}
The calculations are modulo definitions (e.g.~choice of explicit parameter values) and assumptions (e.g.~computer assisted results) that we postpone to
 Section \ref{fhufdhkdfhfdkjhdhdf}. For convenience, things that we postpone to Section \ref{fhufdhkdfhfdkjhdhdf} are tagged by $\tagforcomp$.
\thin
{\bf Cutoff operator.} Fix integers $m(\ks) \geq 2$ and $n(\ks)\geq 1$ $\tagforcomplabel{fixcutoffs}$. Define  $\ks: \Space \to \Space$ and
$\KS: \SSPACE \to \SSPACE$ by $\KS = \mathrm{diag}(\ks,\ks,\ks,\ks)$ and
$$(\ks v)_{mn} = \begin{cases} v_{mn} & \text{if $|m|<m(\ks)$ and $|n| < n(\ks)$}\\
0 & \text{otherwise}
\end{cases}$$
\thin
{\bf Reference.}
Fix a reference $(\Ref{\mu},\Ref{\omega}) \in (0,\infty)\times \Gauged{\SSPACE}$ $\tagforcomplabel{fixref}$.
Set
\begin{subequations}\label{ref5ref5ref5ref5}
\begin{equation}\label{covcovcovcov}
(\mu,\omega) = (\Ref{\mu}, \Ref{\omega}) + (\Corr{\mu},\Corr{\omega})
\end{equation}
This is a \emph{change of variables}
between the unknown
$(\mu,\omega)$ and the new unknown $(\Corr{\mu},\Corr{\omega}) \in (-\Ref{\mu},\infty) \times \Gauged{\SSPACE}$.
Both will be used simultaneously, as convenient, with the understanding that they are always given in terms of one another by \eqref{covcovcovcov}.  The condition
$\Corr{\mu}  > - \Ref{\mu}$
must be kept in mind. It is convenient to split the reference field itself into two parts, i.e.~to fix 
$(\RefA{\mu},\RefA{\omega}) \in (0,\infty)\times \Gauged{\SSPACE}$
$\tagforcomplabel{fixrefa}$
and to fix
$(\RefB{\mu},\RefB{\omega}) \in (-\RefA{\mu},\infty)\times \Gauged{\SSPACE}$
$\tagforcomplabel{fixrefb}$, and to set
\begin{equation}\label{covcovcovcov2}
(\Ref{\mu}, \Ref{\omega})
= (\RefA{\mu}, \RefA{\omega}) + (\RefB{\mu}, \RefB{\omega})
\end{equation}
\end{subequations}
Informally, refA is the main part, refB is for fine tuning. The calculations are organized in such a way that the most time consuming computer calculations only involve refA. If the reader wanted to rigorously construct more accurate approximations to the Choptuik solution than those stated in this paper, he or she could use the \emph{same} refA, but a \emph{better} refB. The assumption $\RefA{\mu} > 0$ implies that $\OO_{\RefA{\mu}}$
and $\OO_{\RefA{\mu}}^{-1}$ are defined. For convenience, suppose
\begin{equation}\label{srefassps}
(1-\KS) \RefA{\omega} = 0\;\;\tagforcomplabel{supposesupportrefa}
\end{equation}
In particular,
$\RefA{\omega}$ has only finitely many nonzero components.
\thin
{\bf Rearranging $\SELECTION\Omega^+(\mu,\omega)=0$: informal version.} Set $C = \SELECTION\Omega^+(\mu,\omega) - \SELECTION\Omega^+(\Ref{\mu},\Ref{\omega})$,
and use the replacements \eqref{covcovcovcov}
and
\eqref{covcovcovcov2} to exhaustion. Use the distributive law to multiply everything out. Clearly, $C$ is quadratic in $(\Corr{\mu},\Corr{\omega})$ without constant term. Split $C= C_1+C_2$ where $C_1$ are all those terms in $C$ that satisfy both of:
\begin{itemize}
\item They are homogeneous of degree 1 in $(\Corr{\mu},\Corr{\omega})$.
\item They do not involve $(\RefB{\mu}, \RefB{\omega})$.
\end{itemize}
All the remaining terms go into $C_2$. At this point, write
$$C_1 = \SELECTION\Omega^+(\mu,\omega) - \SELECTION\Omega^+(\Ref{\mu},\Ref{\omega}) - C_2$$
By construction, $C_1$ can be decomposed as $C_1 = \OO_{\RefA{\mu}} \Corr{\omega} + \Corr{\mu} a + A \Corr{\omega}$, where $a$ is a multifield, and $A$ is a linear operator. For high frequencies the term $\OO_{\RefA{\mu}}$ dominates. Therefore split $A$ into the low-to-low frequency part $\KS A \KS$, and the rest $A - \KS A \KS$, and move the latter to the right:
\begin{multline*}
\OO_{\RefA{\mu}} \Corr{\omega} + \Corr{\mu} a + \KS A\KS \Corr{\omega} = \\
\SELECTION\Omega^+(\mu,\omega) - \SELECTION\Omega^+(\Ref{\mu},\Ref{\omega}) - C_2 - (A-\KS A \KS) \Corr{\omega}
\end{multline*}
We obtain a rearrangement of $\SELECTION\Omega^+ = 0$ that will be useful to  apply the contraction mapping principle, namely:
\begin{equation*}
\SELECTION\Omega^+(\mu,\omega) = 0
\; \Longleftrightarrow \; 
\left\{\begin{aligned}&
 \Corr{\omega} + \OO_{\RefA{\mu}}^{-1}\big( \Corr{\mu} a + \KS A\KS \Corr{\omega}\big) \\
 &= \OO_{\RefA{\mu}}^{-1}\Big(- \SELECTION\Omega^+(\Ref{\mu},\Ref{\omega}) - C_2 - (A-\KS A \KS) \Corr{\omega}\Big)
 \end{aligned}\right.
\end{equation*}
\thin
{\bf Rearranging $\SELECTION\Omega^+(\mu,\omega)=0$: rigorous version.}
Let
$C,C_1,C_2,a \in \SSPACE$ and the linear operator $A: \SSPACE \to \SSPACE$ be given by:
\begin{equation}\label{fdrrihifuhr78}
\begin{aligned}
C & = \OO_{\Ref{\mu}} \Corr{\omega}
+ \Corr{\mu} (\DD{\OO} + \SELECTION\Gamma_1) \Ref{\omega}
+ \Ref{\mu} \SELECTION \Gamma_1 \Corr{\omega}
+ 2 \SELECTION \TimesXi\Gamma_2(\Ref{\omega},\Corr{\omega}) \\
& \qquad 
+ \Corr{\mu} ( \DD{\OO}  + \SELECTION\Gamma_1)\Corr{\omega} + \SELECTION \TimesXi\Gamma_2(\Corr{\omega},\Corr{\omega})\\
C_1 & = \OO_{\RefA{\mu}} \Corr{\omega}
+ \Corr{\mu} (\DD{\OO} + \SELECTION\Gamma_1) \RefA{\omega}
+ \RefA{\mu} \SELECTION \Gamma_1 \Corr{\omega}
+ 2 \SELECTION \TimesXi\Gamma_2(\RefA{\omega},\Corr{\omega}) \\
C_2 & = \RefB{\mu} \DD{\OO} \Corr{\omega}
+ \Corr{\mu} (\DD{\OO} + \SELECTION\Gamma_1) \RefB{\omega}
+ \RefB{\mu} \SELECTION \Gamma_1 \Corr{\omega}
+ 2 \SELECTION \TimesXi\Gamma_2(\RefB{\omega},\Corr{\omega}) \\
& \qquad 
+ \Corr{\mu} ( \DD{\OO}  + \SELECTION\Gamma_1)\Corr{\omega} + \SELECTION \TimesXi\Gamma_2(\Corr{\omega},\Corr{\omega})\\
a & = (\DD{\OO} + \SELECTION\Gamma_1) \RefA{\omega}\\
A & =  \RefA{\mu} \SELECTION \Gamma_1
+ 2 \SELECTION \TimesXi\Gamma_2(\RefA{\omega},\,\cdot\,)
\end{aligned}
\end{equation}
It is now convenient to introduce the linear map (there are several equivalent ways to write this definition, because
$a = \KS a$ and $ \OO_{\RefA{\mu}}^{-1}\KS = \KS \OO_{\RefA{\mu}}^{-1} \KS$):
\begin{alignat*}{4}
\mathbbm{U}: \quad \R & \oplus \Gauged{\SSPACE} && \,\to\, \SSPACE\\
\lambda  & \oplus \mathbf{v} && \,\mapsto \, 
\mathbf{v} + \OO_{\RefA{\mu}}^{-1} \big(\lambda a + \KS A \KS \mathbf{v}\big)
\end{alignat*}
and the nonlinear map 
$\bigstar: \R \oplus \Gauged{\SSPACE} \to \SSPACE$ by
\begin{equation}\label{bigstarkdhkdhk494}
\begin{aligned}
& \bigstar(\lambda,\mathbf{v}) = \\
& - \OO_{\RefA{\mu}}^{-1}\SELECTION\Omega^+(\Ref{\mu},\Ref{\omega})\\
& - \OO_{\RefA{\mu}}^{-1} \big\{ \RefB{\mu} \DD{\OO} \mathbf{v}
+ \lambda (\DD{\OO} + \SELECTION\Gamma_1) \RefB{\omega}
+ \RefB{\mu} \SELECTION \Gamma_1 \mathbf{v}
+ 2 \SELECTION \TimesXi\Gamma_2(\RefB{\omega},\mathbf{v})\big\} \\
& - \OO_{\RefA{\mu}}^{-1} \big\{ \lambda ( \DD{\OO}  + \SELECTION\Gamma_1)\mathbf{v} + \SELECTION \TimesXi\Gamma_2(\mathbf{v},\mathbf{v})\big\}\\
& - \OO_{\RefA{\mu}}^{-1}(A-\KS A \KS) \mathbf{v}
\end{aligned}
\end{equation}
By construction,
\begin{equation}\label{equivdkhjdhdkkkdk}
\SELECTION\Omega^+(\mu,\omega) = 0
\qquad  \Longleftrightarrow \qquad
\mathbbm{U}(\Corr{\mu}\oplus \Corr{\omega}) = \bigstar(\Corr{\mu},\Corr{\omega})
\end{equation}
\thin
{\bf Remark.}
Domain and range
of $\mathbbm{U}$ decompose as
\begin{equation} \label{dfdhfkhfdfkskkskkks}
\big(\R \oplus (\Gauged{\SSPACE} \cap \image \KS )\big)
\oplus \image (1-\KS) \to \big(\image \KS \big) \oplus \big(\image(1-\KS) \big)
\end{equation}
Note that $\image (1-\KS) \subset \Gauged{\SSPACE}$ by $m(\ks) \geq 2$ and  $n(\ks) \geq 1$. By construction, $\mathbbm{U}$ is block diagonal with respect to this decomposition.
Since the high-to-high frequency block $\image (1-\KS) \to \image (1-\KS)$ is the identity, the operator $\mathbbm{U}$ is invertible if and only if its low-to-low frequency block 
$\R \oplus (\Gauged{\SSPACE} \cap \image \KS ) \to \image \KS$ is invertible, \emph{a finite square matrix}.
\thin
{\bf Weighted $\ell^1$ norms.} Fix constants $\kappa_1,\kappa_2 > 1$ $\tagforcomplabel{fixk1k2}$ and $\kappa_{\ast} > 0$ $\tagforcomplabel{fixkstar}$. Set 
\begin{align*}
\|z\|_{\C} & =  |\RE z| + |\IM z|\\
\|v\|_{\Space} & = \textstyle\sum_{m,n\in \Z} (\kappa_1)^{|m|}(\kappa_2)^{|n|} \|v_{mn}\|_{\C}\\
& =\textstyle\sum_{m,n\geq 0} (2-\delta_{m0})(2-\delta_{n0}) (\kappa_1)^{m}(\kappa_2)^{n} \|v_{mn}\|_{\C}\\
\|\mathbf{v}\|_{\SSPACE} & = \|v_1\|_{\Space} + \|v_2\|_{\Space} + \|v_3\|_{\Space} + \|v_4\|_{\Space}\\
\|\lambda \oplus \mathbf{v}\|_{\R \oplus \SSPACE} & = \kappa_{\ast} |\lambda| + \| \mathbf{v}\|_{\SSPACE}
\end{align*}
with $\mathbf{v} = (v_1,v_2,v_3,v_4) \in \SSPACE$. 
Note that $\|\cdot\|_{\C}$ is a norm on $\C$, viewed as a real vector space, and
$|z| \leq \|z\|_{\C} \leq \sqrt{2}\, |z|$ and $\|z_1z_2\|_{\C} \leq \|z_1\|_{\C} \|z_2\|_{\C}$.
\step
Note that $\|\cdot\|_{\Space}$ is not a norm on $\Space$, but only on 
$\{v \in \Space \;|\; \| v \|_{\Space} < \infty\} \subset \Space$. We do not introduce a special name for this subspace. 
\thin
{\bf Operator norm.} 
For every linear operator $O: X \to Y$ set
$$\| O \|_{Y \leftarrow X} = \sup\big\{\,\|Ox\|_Y\;\big|\; x \in X\,\text{with}\,\|x\|_X=1\big\}$$
Then 
$\| O'O\|_{Z \leftarrow X} \leq \| O'\|_{Z \leftarrow Y}\,\| O\|_{Y \leftarrow X}$. 
\thin
{\bf Convolution estimate.} The estimate $\| v \ast w \|_{\Space} \leq \|v\|_{\Space}\, \|w\|_{\Space}$ follows from
\begin{align*}
\| v \ast w \|_{\Space} & = \textstyle \sum_{m,n\in \Z} (\kappa_1)^{|m|} (\kappa_2)^{|n|}
\big\| \sum_{m_1+m_2 = m,n_1+n_2=n} v_{m_1n_1} w_{m_2n_2}\big\|_{\C}\\
& \leq \textstyle \sum_{m,n\in \Z} (\kappa_1)^{|m|} (\kappa_2)^{|n|}
 \sum_{m_1+m_2 = m,n_1+n_2=n} \|v_{m_1n_1}\|_{\C}\,\|w_{m_2n_2}\|_{\C}\\
& \leq \textstyle \sum_{m_1,n_1,m_2,n_2 \in \Z}
(\kappa_1)^{|m_1+m_2|} (\kappa_2)^{|n_1+n_2|} \|v_{m_1n_1}\|_{\C}\,\|w_{m_2n_2}\|_{\C}\\
& \leq \textstyle \sum_{m_1,n_1,m_2,n_2 \in \Z}
(\kappa_1)^{|m_1|+|m_2|} (\kappa_2)^{|n_1|+|n_2|} \|v_{m_1n_1}\|_{\C}\,\|w_{m_2n_2}\|_{\C}
\end{align*}
Here, $\|z_1z_2\|_{\C} \leq \|z_1\|_{\C} \| z_2\|_{\C}$ and the assumption $\kappa_1,\kappa_2 > 1$ are used.\\
This convolution estimate is sharp, in the sense that $\| (v \ast) \|_{\Space \leftarrow \Space}
= \| v \|_{\Space}$.
\thin
{\bf Estimate for $\Gamma_2$.}
\begin{equation}\label{dkhfkjhffjjjsjs}
\begin{aligned}
& \|\SELECTION\TimesXi \Gamma_2(\mathbf{v},\,\cdot\,)\|_{\SSPACE \leftarrow \SSPACE}\\
&  \leq \max \left\{ \left\|\frac{1}{4}\begin{pmatrix}
Pv_2+v_2-Pv_3-v_3\\
v_1-v_2+Pv_2\\
-v_1+v_2-Pv_2\\
0
\end{pmatrix}\right\|_{\SSPACE},
\left\|\frac{1}{2}\begin{pmatrix}
v_1+Pv_3-v_3\\
2v_2+2Pv_2\\
-v_2-Pv_2\\
Pv_4+v_4
\end{pmatrix} \right\|_{\SSPACE}, \right. \\
&\qquad \qquad \hskip 5pt\left\|\frac{1}{2}\begin{pmatrix}
0\\
-v_1+Pv_2-v_2\\
v_1\\
Pv_4-v_4
\end{pmatrix} \right\|_{\SSPACE},
\left\|\frac{1}{2}\begin{pmatrix}
-v_1\\
0\\
0\\
0
\end{pmatrix} \right\|_{\SSPACE},\\
&\qquad \qquad\hskip 4pt\left. \left\|\frac{1}{2}\begin{pmatrix}
-v_2-Pv_2\\
0\\
0\\
0
\end{pmatrix} \right\|_{\SSPACE},
\left\|\frac{1}{2}\begin{pmatrix}
v_4 -Pv_4\\
0 \\
Pv_4+v_4\\
Pv_2+v_2
\end{pmatrix} \right\|_{\SSPACE},
\left\|\frac{1}{2}\begin{pmatrix}
v_4 +Pv_4\\
0 \\
Pv_4-v_4\\
Pv_2-v_2
\end{pmatrix} \right\|_{\SSPACE}\right\}
\end{aligned}
\end{equation}
To prove this estimate, use \eqref{skdfhkfdjhdkhjdgfdjh33hk} and
\begin{multline*}
\|\mathbf{v}\|_{\SSPACE} = 
\|v_1\|_{\Space}
+ \tfrac{1}{2}\|v_2+Pv_2\|_{\Space} + \tfrac{1}{2}\|v_2-Pv_2\|_{\Space}\\
+ \tfrac{1}{2}\|v_3+Pv_3\|_{\Space} + \tfrac{1}{2}\|v_3-Pv_3\|_{\Space}
+ \tfrac{1}{2}\|v_4+Pv_4\|_{\Space} + \tfrac{1}{2}\|v_4-Pv_4\|_{\Space}
\end{multline*}
As a corollary,
\begin{equation}\label{dkhfdkdhk3iuzi3i}
\|\SELECTION\TimesXi \Gamma_2(\mathbf{v},\mathbf{w})\|_{\SSPACE} \leq 3 \|\mathbf{v}\|_{\SSPACE}\|\mathbf{w}\|_{\SSPACE}
\end{equation}
\thin
{\bf Estimates for the auxiliary operators $\Band$ and $\BandI$.}
\begin{subequations}\label{fdkhdfkjhk464huih}
\begin{align}
\| \Band[k,f]\|_{\Space \leftarrow \Space}
& \leq (\kappa_2)^{-k} \textstyle\sup_{m,n\geq 0} \|f_{mn}\|_{\C} 
\leq C_f\,(\kappa_2)^{-k} \textstyle\sup_{m,n \geq 0} |f_{mn}|\\
\| \BandI[k,f]\|_{\Space \leftarrow \Space}
& 
\leq C_f\, (1- (\kappa_2)^{-k})^{-1}
\end{align}
\end{subequations}
with $C_f = 1$ if $f_{mn} \in \R$ for all $m,n$, otherwise $C_f = \sqrt{2}$.\\
For $\BandI$ we assume, \emph{as always}, that $k > 0$ and $|f_{mn}| \leq 1$ for all $m,n$.
\thin {\footnotesize Proof: Let $c_{mn} = (2-\delta_{m0})(2-\delta_{n0})$ for all $m,n\geq 0$. Then
\begin{align*}
\| \Band[k,f] v\|_{\Space}
& = \textstyle\sum_{m,n\geq 0} c_{mn} (\kappa_1)^{m}(\kappa_2)^{n} \| f_{mn} v_{m,n+k}\|_{\C}\\
& \leq \big((\kappa_2)^{-k}\,\textstyle\sup_{m,n} \|f_{mn}\|_{\C}\big) \textstyle\sum_{m,n\geq 0} c_{m,n+k} (\kappa_1)^{m}(\kappa_2)^{n+k} \| v_{m,n+k}\|_{\C}\\
& \leq \big((\kappa_2)^{-k}\,\textstyle\sup_{m,n} \|f_{mn}\|_{\C}\big) \|v\|_{\Space}
\end{align*}
Here $c_{mn}\leq c_{m,n+k}$ for all $m,n,k\geq 0$ has been used. Now
\begin{align*}
\| \BandI[k,f]\|_{\Space \leftarrow \Space} & = \| \textstyle\sum_{L=0}^{\infty} \Band[kL,f_L]\|_{\Space \leftarrow \Space}\\
& \leq \textstyle\sum_{L=0}^{\infty} \| \Band[kL,f_L]\|_{\Space \leftarrow \Space}\\
& \leq \textstyle\sum_{L=0}^{\infty} C_{f_L} (\kappa_2)^{-kL} \textstyle \sup_{m,n}|(f_L)_{mn}|\\
& \leq C_f (1-(\kappa_2)^{-k})^{-1}
\end{align*}
with $(f_L)_{mn} = (-1)^L \prod_{\ell=0}^{L-1}f_{m,n+k\ell}$.
Here $C_{f_L} \leq C_f$ and $|(f_L)_{mn}|\leq 1$ have been used.
}
\thin
{\bf Estimate for $\SELECTION \Gamma_1$.} By \eqref{fdkhdfkjhk464huih},
\begin{equation}\label{fdhkjhjdh4z894z89rhiu}
\| \SELECTION \Gamma_1 \|_{\SSPACE \leftarrow \SSPACE}
\leq 4 \|\Band_{1,1} \BandI_{2,1}\|_{\Space \leftarrow \Space}
\leq \mathcal{K}_1 \stackrel{\text{def}}{=} 4 (\kappa_2)^{-1} (1- (\kappa_2)^{-2})^{-1}
\end{equation}
To check this, note that the first component of
$\SELECTION \Gamma_1\mathbf{v}$ vanishes, for all $\mathbf{v}\in \SSPACE$.
\thin
{\bf Auxiliary estimates for $\mathbbm{U}^{-1}\bigstar$.} By \eqref{fdkhdfkjhk464huih},
\begin{subequations}\label{fdkhfdkhfjfjjfjjf}
\begin{align} 
\notag\| \OO_{\RefA{\mu}}^{-1}\|_{\SSPACE \leftarrow \SSPACE}
& \leq \max \{\|\OOB_{\RefA{\mu}}^{-1}\|_{\Space \leftarrow \Space},\|\OOA_{\RefA{\mu}}^{-1}\|_{\Space \leftarrow \Space}\}\\
\label{fdkhf873448748g} & \leq \mathcal{K}_2 \stackrel{\text{def}}{=} 2 \RefA{\mu}^{-1} (1-(\kappa_2)^{-1})^{-1} (1+(\kappa_2)^{-1})\\
\label{fdkhfdkhfjfjjfjjfxxx} \|\OO_{\RefA{\mu}}^{-1} \DD{\OO}\|_{\SSPACE \leftarrow \SSPACE} & \leq \mathcal{K}_2\phantom{ \stackrel{\text{def}}{=}}
\end{align}
\end{subequations}
For \eqref{fdkhfdkhfjfjjfjjf}, note that
\begin{equation} \label{dhfkjhfkdhfkjdff34}
(1-(\kappa_2)^{-2})^{-1} (1+(\kappa_2)^{-2})
\leq (1-(\kappa_2)^{-1})^{-1} (1+(\kappa_2)^{-1})
\end{equation}
For \eqref{fdkhfdkhfjfjjfjjfxxx}, note that
$\OOA_{\mu}^{-1} \DD{\OOA}
= \mu^{-1} \BandI[1,f](\Band[0,\frac{\mu(n+1)}{im/2+\mu(n+1)}]+\Band[1,\frac{\mu n}{im/2+\mu(n+1)}])$ for some $f$ with $|f_{mn}|\leq 1$. Similar for
$\OOB_{\mu}^{-1} \DD{\OOB}$.
\thin
Suppose $\mathbbm{U}$ is invertible $\tagforcomplabel{supposeUinvertible}$.
Fix an invertible linear operator ${\mathfrak A}: \SSPACE \to \SSPACE$ with
${\mathfrak A} = \KS{\mathfrak A} \KS + (1-\KS)$ $\tagforcomplabel{fixgothicA}$. (The operator ${\mathfrak A}$ is only introduced for an important operator norm estimate of the form
$\|O_1O_2\| \leq \|O_1 {\mathfrak A}^{-1}\| \| {\mathfrak A} O_2\|$, to improve over the naive $\leq \|O_1\| \|O_2\|$. See the left hand sides of \eqref{dkjfhk4} and \eqref{dkjfhk5} below.)
Fix constants $0 < \mathcal{L}_1,\ldots,\mathcal{L}_6 < \infty$ $\tagforcomplabel{fixl1l6}$. Suppose
\begin{subequations} \label{dih3ehiiz8736873hi3h}
\begin{align}
\label{dkjfhk1}
\| \SELECTION\TimesXi \Gamma_2(\RefA{\omega},\,\cdot\,)\|_{\SSPACE \leftarrow \SSPACE} & \leq \mathcal{L}_1\;\; \tagforcomplabel{supposededicatedgamma2}\\
\label{dkjfhk2}\| \RefB{\omega}\|_{\SSPACE} & \leq \mathcal{L}_2\;\; \tagforcomplabel{supposenormB}\\
\label{dkjfhk3}\|\SELECTION \Omega^+(\Ref{\mu},\Ref{\omega})\|_{\SSPACE} & \leq \mathcal{L}_3\;\; \tagforcomplabel{supposeerrorAB} \\
\label{dkjfhk4} \| {\mathfrak A} \OO_{\RefA{\mu}}^{-1} (A - \KS A \KS ) \|_{\SSPACE \leftarrow \SSPACE} & \leq \mathcal{L}_4\;\; \tagforcomplabel{supposeextest}\\
\label{dkjfhk5} \| \mathbbm{U}^{-1} {\mathfrak A}^{-1}\|_{\R \oplus \SSPACE \leftarrow \SSPACE} & \leq \mathcal{L}_5\;\; \tagforcomplabel{supposeinvnorm1} \\
\label{dkjfhk6} \| \mathbbm{U}^{-1} \|_{\R \oplus \SSPACE \leftarrow \SSPACE} & \leq \mathcal{L}_6
\;\; \tagforcomplabel{supposeinvnorm2}
\end{align}
\end{subequations}
Then (this estimate is used only in Section \ref{dflfklfjkfhkkkkkkkkkkkkdhkhdk})
$$\|A\|_{\SSPACE \leftarrow \SSPACE} \leq \mathcal{K}_3
 \stackrel{\text{def}}{=} \RefA{\mu} \mathcal{K}_1 + 2 \mathcal{L}_1$$
 Fix constants $0<\mathcal{S}_1,\ldots,\mathcal{S}_5 < \infty$ $\tagforcomplabel{fixs1s5}$.
Suppose
\begin{subequations}\label{sdldkdkjfjdkdddd}
\begin{alignat}{5}
&\mathcal{L}_6 \mathcal{K}_2 \mathcal{L}_3 & \leq \mathcal{S}_1\;\; \tagforcomplabel{supposes1} \\
& \mathcal{L}_6 \mathcal{K}_2(1+\mathcal{K}_1) \mathcal{L}_2/\kappa_{\ast} & \leq \mathcal{S}_2\;\;  \tagforcomplabel{supposes2}\\
 &\mathcal{L}_6 \mathcal{K}_2(\RefB{\mu} + \RefB{\mu} \mathcal{K}_1 + 6 \mathcal{L}_2) + \mathcal{L}_5\mathcal{L}_4\quad & \leq \mathcal{S}_3\;\;  \tagforcomplabel{supposes3}\\
 &\mathcal{L}_6 \mathcal{K}_2(1+\mathcal{K}_1)/\kappa_{\ast}& \leq \mathcal{S}_4\;\;  \tagforcomplabel{supposes4}\\
 &3 \mathcal{L}_6 \mathcal{K}_2 & \leq \mathcal{S}_5\;\;  \tagforcomplabel{supposes5}
\end{alignat}
\end{subequations}
Set $\mathcal{S}_{23} = \max \{\mathcal{S}_2,\mathcal{S}_3\}$
and $\mathcal{S}_{45} = \max\{\mathcal{S}_4,2\mathcal{S}_5\}$.
\thin
{\bf Estimates for $\mathbbm{U}^{-1}\bigstar$.} We have
\begin{subequations}\label{fkdhkjdhfkfhfk3k3hkj3}
\begin{equation}
\begin{aligned}
\| \mathbbm{U}^{-1} \bigstar(\lambda,\mathbf{v})\|_{\R \oplus \SSPACE}
& \leq
\mathcal{S}_1 + \mathcal{S}_2 \kappa_{\ast} |\lambda|
+ \mathcal{S}_3 \|\mathbf{v}\|_{\SSPACE}
+ \mathcal{S}_4 \kappa_{\ast}|\lambda| \|\mathbf{v}\|_{\SSPACE}
+ \mathcal{S}_5 \big(\|\mathbf{v}\|_{\SSPACE}\big)^2\\
& \leq \mathcal{S}_1 + \mathcal{S}_{23} \|\lambda \oplus \mathbf{v}\|_{\R \oplus \SSPACE}
+ \tfrac{1}{2}\mathcal{S}_{45} \big(\|\lambda \oplus \mathbf{v}\|_{\R \oplus \SSPACE}\big)^2
\end{aligned}
\end{equation}
by
\eqref{bigstarkdhkdhk494}, \eqref{dkhfdkdhk3iuzi3i}, \eqref{fdhkjhjdh4z894z89rhiu}, \eqref{fdkhfdkhfjfjjfjjf}, \eqref{dih3ehiiz8736873hi3h},
\eqref{sdldkdkjfjdkdddd}. Furthermore,
\begin{equation}
\begin{aligned}
&
\| \mathbbm{U}^{-1} \bigstar(\lambda_1,\mathbf{v}_1)
- 
\mathbbm{U}^{-1} \bigstar(\lambda_2,\mathbf{v}_2)
\|_{\R \oplus \SSPACE} \\
& \leq \Big(\mathcal{S}_{23} + \tfrac{1}{2}\mathcal{S}_{45}
\big(
\| \lambda_1 \oplus \mathbf{v}_1\|_{\R \oplus \SSPACE}
+
\| \lambda_2 \oplus \mathbf{v}_2\|_{\R \oplus \SSPACE}
\big)
\Big)\| \lambda_1 \oplus \mathbf{v}_1-\lambda_2 \oplus \mathbf{v}_2\|_{\R \oplus \SSPACE}
\end{aligned}
\end{equation}
\end{subequations}
\thin
{\bf Solution to $\SELECTION\Omega^+(\mu,\omega) = 0$.} Fix a constant $\mathcal{R}>0$ $\tagforcomplabel{fixr}$. Suppose
\begin{subequations}\label{sdldkdkjfjdkdddd22}
\begin{alignat}{4}
\mathcal{S}_1 + \mathcal{S}_{23} \mathcal{R} + \tfrac{1}{2}\mathcal{S}_{45} \mathcal{R}^2 & \leq \mathcal{R} && \tagforcomplabel{supposer1}\\
\mathcal{S}_{23} + \mathcal{S}_{45} \mathcal{R} & < 1 && \tagforcomplabel{supposer2}\\
\label{sdldkdkjfjdkdddd22c} \mathcal{R} & < \kappa_{\ast} \Ref{\mu}\;\;&& \tagforcomplabel{supposer3}
\end{alignat}
\end{subequations}
Then $\lambda \oplus \mathbf{v} \mapsto \mathbbm{U}^{-1} \bigstar(\lambda,\mathbf{v})$ maps $B\to B$ with
$$B  =\Big\{\lambda \oplus \mathbf{v} \,\in\, \R \oplus \Gauged{\SSPACE} \;\Big|\; \|\lambda \oplus \mathbf{v}\|_{\R \oplus \SSPACE} \leq \mathcal{R}\Big\}$$
and is a contraction on the nonempty, complete metric space $(B,\|\,\cdot\,\|_{\R\oplus \SSPACE})$ by \eqref{fkdhkjdhfkfhfk3k3hkj3}. Therefore, it has a unique fixed point $\Corr{\mu}\oplus \Corr{\omega}$.
Equation \eqref{sdldkdkjfjdkdddd22c} implies that $\Corr{\mu} > -\Ref{\mu}$, as required.
The corresponding $(\mu,\omega)$ satisfies $\SELECTION\Omega^+(\mu,\omega) = 0$ by \eqref{equivdkhjdhdkkkdk}. Note that
\begin{equation}\label{finnormfinnorm}
\| \omega \|_{\SSPACE} < \infty
\end{equation}
This follows from \eqref{srefassps} and \eqref{dkjfhk2}, and from
\begin{equation}\label{corrinb}
\|\Corr{\mu}\oplus \Corr{\omega}\|_{\R \oplus \SSPACE} \leq \mathcal{R}
\end{equation}
\thin
{\bf Rearranging the $\sharp$ system.} Introduce $A^{\sharp},\mathbbm{U}^{\sharp},\bigstar^{\sharp}: \SSPACE^{\sharp}\to \SSPACE^{\sharp}$ linear maps
\begin{equation}\label{3i3hj4kh4}
\begin{aligned}
A^{\sharp} & = \RefA{\mu} \DivideByXiRegularized \Gamma_1^{\sharp} + \Gamma_2^{\sharp}(\RefA{\omega},\,\cdot\,)\\
\mathbbm{U}^{\sharp} & = 1 + (\OO_{\RefA{\mu}}^{\sharp})^{-1} \KS  A^{\sharp} \KS \\
\bigstar^{\sharp} & = - (\OO_{\RefA{\mu}}^{\sharp})^{-1}\big\{
(\RefB{\mu}+\Corr{\mu}) \big(
\DD{\OO}^{\sharp} 
+ \DivideByXiRegularized \Gamma_1^{\sharp} \big) + \Gamma_2^{\sharp}(\RefB{\omega}+\Corr{\omega},\,\cdot\,)
\big\}\\
& \qquad - (\OO_{\RefA{\mu}}^{\sharp})^{-1}\big(A^{\sharp} - \KS A^{\sharp}\KS \big)
\end{aligned}
\end{equation}
Here, $\KS = \diag(\ks,\ks): \SSPACE^{\sharp}\to \SSPACE^{\sharp}$. This is an abuse of notation, but it will always be clear if $\KS: \SSPACE \to \SSPACE$ or $\KS: \SSPACE^{\sharp}\to \SSPACE^{\sharp}$. Since $(\mu,\omega)$ is now a solution to $\SELECTION\Omega^+(\mu,\omega) = 0$,
it follows from \eqref{6383hi3h39z} and the construction \eqref{3i3hj4kh4} that
\begin{equation}\label{ddhfkdhkdkkdkkkd1}
\mathbbm{U}^{\sharp}\omega^{\sharp} = \bigstar^{\sharp} \omega^{\sharp}
\end{equation}
Note that $\mathbbm{U}^{\sharp} = \KS \mathbbm{U}^{\sharp} \KS  + (1-\KS)$. Therefore, $\mathbbm{U}^{\sharp}$ is invertible if and only if the finite square matrix $\KS \mathbbm{U}^{\sharp} \KS: \image \KS \to \image \KS$ is invertible.
\thin
{\bf Norm on (a subspace of) $\SSPACE^{\sharp}$.} Set
$
\|\mathbf{w}\|_{\SSPACE^{\sharp}} = \|w_1\|_{\Space} + \|w_{2\sharp}\|_{\Space}
$ for all $\mathbf{w}=(w_1,w_{2\sharp}) \in \SSPACE^{\sharp}$. This is a norm on $\{\mathbf{w}\in \SSPACE^{\sharp}\;|\; \|\mathbf{w}\|_{\SSPACE^{\sharp}} < \infty\} \subset \SSPACE^{\sharp}$.
\thin
{\bf Estimate for $\Gamma_2^{\sharp}$.}
\begin{equation}\label{fdkhfjhfkdhslsdlss}
\begin{aligned}
& \|\Gamma_2^{\sharp}(\mathbf{v},\,\cdot\,)\|_{\SSPACE^{\sharp}\leftarrow \SSPACE^{\sharp}}
 \leq  \max\left\{
  \left\| \frac{1}{2}\begin{pmatrix}
   v_2  +Pv_2 -Pv_3 -v_3\\
  v_1-v_2+Pv_2
  \end{pmatrix} \right\|_{\SSPACE^{\sharp}}, \right.\\
& \hskip10mm
  \left.
  \left\| \frac{1}{2}\begin{pmatrix}
  -v_1+Pv_3-v_3\\
  -v_1 +v_2 +3 Pv_2
  \end{pmatrix} \right\|_{\SSPACE^{\sharp}},
  \left\| \frac{1}{2}\begin{pmatrix}
  v_2+Pv_2+Pv_3+v_3\\
  -v_1 +v_2 +3 Pv_2
\end{pmatrix} \right\|_{\SSPACE^{\sharp}}
\right\}
\end{aligned}
\end{equation}
\thin
{\bf Auxiliary estimates for $(\mathbbm{U}^{\sharp})^{-1} \bigstar^{\sharp}$.} 
We have
\begin{align*}
\| \Gamma_2^{\sharp}(\mathbf{v},\,\cdot\,)\|_{\SSPACE^{\sharp} \leftarrow \SSPACE^{\sharp}} &  \leq 3 \| \mathbf{v}\|_{\SSPACE}\\
 \| \DivideByXiRegularized \Gamma_1^{\sharp}\|_{\SSPACE^{\sharp}\leftarrow \SSPACE^{\sharp}} & \leq \mathcal{K}_1^{\sharp} \stackrel{\text{def}}{=} 2 (\kappa_2)^{-1}(1-(\kappa_2)^{-2})^{-1}\\
\| (\OO_{\RefA{\mu}}^{\sharp})^{-1}\|_{\SSPACE^{\sharp}\leftarrow \SSPACE^{\sharp}} & \leq \mathcal{K}_2^{\sharp} \stackrel{\text{def}}{=}  
2 \RefA{\mu}^{-1}(1-(\kappa_2)^{-1})^{-1}(1+(\kappa_2)^{-1})
\\
\| (\OO_{\RefA{\mu}}^{\sharp})^{-1}\DD{\OO}^{\sharp} \|_{\SSPACE^{\sharp}\leftarrow \SSPACE^{\sharp}} & \leq \mathcal{K}_2^{\sharp} \phantom{\stackrel{\text{def}}{=}}
\end{align*}
Suppose $\mathbbm{U}^{\sharp}$ is invertible $\tagforcomplabel{sharpuinvertible}$.
Fix an invertible linear operator ${\mathfrak A}^{\sharp}: \SSPACE^{\sharp} \to \SSPACE^{\sharp}$ with
${\mathfrak A}^{\sharp} = \KS {\mathfrak A}^{\sharp} \KS  + (1-\KS)$ $\tagforcomplabel{fixsharpa}$.
 Fix
$0<\mathcal{L}_1^{\sharp},\mathcal{L}_2^{\sharp},\mathcal{L}_3^{\sharp},\mathcal{L}_4^{\sharp},\mathcal{L}_5^{\sharp} < \infty$ $\tagforcomplabel{fixl1sharpl5sharp}$. Suppose
\begin{subequations}
\begin{align}
 \label{dhdkfhkdhdkkkk1}\|\Gamma_2^{\sharp}(\RefA{\omega},\,\cdot\,)\|_{\SSPACE^{\sharp}\leftarrow \SSPACE^{\sharp}}
 & \leq \mathcal{L}_1^{\sharp}\;\;\tagforcomplabel{supposel1sharp}\\
\label{dhdkfhkdhdkkkk2}\|\RefB{\omega} + \Corr{\omega}\|_{\SSPACE} & \leq \mathcal{L}_2^{\sharp}\;\;\tagforcomplabel{supposel2sharp}\\
\label{dhdkfhkdhdkkkk3}  \| {\mathfrak A}^{\sharp}(\OO_{\RefA{\mu}}^{\sharp})^{-1}\big(A^{\sharp} - \KS A^{\sharp}\KS \big)\|_{\SSPACE^{\sharp}\leftarrow \SSPACE^{\sharp}}
 & \leq \mathcal{L}_3^{\sharp}\;\;\tagforcomplabel{supposel3sharp}\\
\label{dhdkfhkdhdkkkk4} \|(\mathbbm{U}^{\sharp})^{-1}({\mathfrak A}^{\sharp})^{-1}\|_{\SSPACE^{\sharp}\leftarrow \SSPACE^{\sharp}}
 & \leq \mathcal{L}_4^{\sharp}\;\;\tagforcomplabel{supposel4sharp}\\
\label{dhdkfhkdhdkkkk5}\|(\mathbbm{U}^{\sharp})^{-1}\|_{\SSPACE^{\sharp}\leftarrow \SSPACE^{\sharp}}
 & \leq \mathcal{L}_5^{\sharp}\;\;\tagforcomplabel{supposel5sharp}
\end{align}
\end{subequations}
Then
$$ \|A^{\sharp}\|_{\SSPACE^{\sharp}\leftarrow \SSPACE^{\sharp}} \leq
\mathcal{K}_3^{\sharp} \stackrel{\text{def}}{=}
\RefA{\mu} \mathcal{K}_1^{\sharp} + \mathcal{L}_1^{\sharp}$$
\thin
{\bf Estimate for $(\mathbbm{U}^{\sharp})^{-1} \bigstar^{\sharp}$.}
\begin{equation}\label{weieuiuehhjkd}
\|(\mathbbm{U}^{\sharp})^{-1} \bigstar^{\sharp}\|_{\SSPACE^{\sharp}\leftarrow \SSPACE^{\sharp}} \leq 
|\RefB{\mu}+\Corr{\mu}| \mathcal{L}_5^{\sharp}\mathcal{K}_2^{\sharp}(1+\mathcal{K}_1^{\sharp})
+ 3 \mathcal{L}_5^{\sharp}\mathcal{K}_2^{\sharp}\mathcal{L}_2^{\sharp}
+ \mathcal{L}_4^{\sharp}\mathcal{L}_3^{\sharp}
\end{equation}
\thin
Suppose the right hand side of \eqref{weieuiuehhjkd} is $<1$ $\tagforcomplabel{supposelessthan1}$.
Then $\|(\mathbbm{U}^{\sharp})^{-1} \bigstar^{\sharp}\|_{\SSPACE^{\sharp}\leftarrow \SSPACE^{\sharp}} < 1$. We show that this implies $\omega^{\sharp}=0$, i.e.~that $(\mu,\omega)$ solves {\bf 2DprobSeries}.
\step
{\footnotesize 
The claim seems to follow from \eqref{ddhfkdhkdkkdkkkd1}, which yields
$(1- \|O\|_{\SSPACE^{\sharp}\leftarrow \SSPACE^{\sharp}})\|\omega^{\sharp}\|_{\SSPACE^{\sharp}} \leq 0$ with $O=(\mathbbm{U}^{\sharp})^{-1} \bigstar^{\sharp}$, and which seems to imply that $\|\omega^{\sharp}\|_{\SSPACE^{\sharp}}$ must vanish. This reasoning is invalid, because $\|\omega^{\sharp}\|_{\SSPACE^{\sharp}}$ is \emph{not known to be finite}. We have to work a little more. 
\step
For every $\nu$ with $\max \{1/\kappa_1,1/\kappa_2 \}<\nu\leq 1$, set:
$$\|v\|_{(\Space,\nu)}= \textstyle\sum_{m,n \geq 0} (2-\delta_{m0})(2-\delta_{n0}) (\nu \kappa_1)^{m} (\nu \kappa_2)^{n} \|v_{mn}\|_{\C}$$
Define $\|\,\cdot\,\|_{(\SSPACE,\nu)}$
and $\|\,\cdot\,\|_{(\SSPACE^{\sharp},\nu)}$ accordingly.
Note that $\|\,\cdot\,\|_{(\Space,1)} = \|\,\cdot\,\|_{\Space}$. 
If $\nu < 1$, then \eqref{finnormfinnorm} implies $\|\omega^{\sharp}\|_{(\SSPACE^{\sharp},\nu)}<\infty$. Therefore, if we can show that $\|O\|_{(\SSPACE^{\sharp},\nu)\leftarrow (\SSPACE^{\sharp},\nu)} < 1$ for \emph{some} $\nu < 1$, then $\omega^{\sharp}$ vanishes. Since $\|O\|_{\SSPACE^{\sharp}\leftarrow \SSPACE^{\sharp}}<1$ by assumption, it suffices to prove
\begin{equation}\label{dlhdhkfjhfncnncnc}
\textstyle\limsup_{\nu \uparrow 1}\|O\|_{(\SSPACE^{\sharp},\nu)\leftarrow (\SSPACE^{\sharp},\nu)}
\leq \|O\|_{\SSPACE^{\sharp}\leftarrow \SSPACE^{\sharp}}
\end{equation}
To check this, let $e=(e_1,e_{2\sharp}) \in \SSPACE^{\sharp}$ be a unit vector, i.e.~there are $d(e) \in \{1,2\sharp\}$ and $k(e) \in \{0,1\}$ and $m(e),n(e) \geq 0$ such that $(e_{d(e)})_{m(e),n(e)} = 1$ if $k(e)=0$ and 
$(e_{d(e)})_{m(e),n(e)} = i$ if $k(e)=1$, and all other components of $e$ are zero. Unit vectors are useful, because
$$\|O\|_{(\SSPACE^{\sharp},\nu)\leftarrow (\SSPACE^{\sharp},\nu)}
= \sup_{\text{$e$: unit vector}} \frac{\|Oe\|_{(\SSPACE^{\sharp},\nu)}}{\|e\|_{(\SSPACE^{\sharp},\nu)}} \qquad \text{(if the left hand side is $<\infty$)}$$
This is a special property of the $\ell^1$ operator norm.  Fix a unit vector $e$. For $N \geq 0$, decompose $Oe = (Oe)^{N+}+ (Oe)^{N-}$ with
$(Oe)^{N+},(Oe)^{N-} \in \SSPACE^{\sharp}$ and
$$
d\in \{1,2\sharp\},\;\; m,n\geq 0: \qquad ((Oe)^{N+}_d)_{mn} = \begin{cases}
((Oe)_d)_{mn} & \text{if $m+n > m(e)+n(e)- N$}\\
0 & \text{otherwise}
\end{cases}
$$
Suppose $\nu_0,\nu$ satisfy $\max \{1/\kappa_1,1/\kappa_2 \}<\nu_0 < \nu < 1$. Then
\begin{align*}
\|Oe\|_{(\SSPACE^{\sharp},\nu)} & \leq \|(Oe)^{N+}\|_{(\SSPACE^{\sharp},\nu)} + \|(Oe)^{N-}\|_{(\SSPACE^{\sharp},\nu)}\\
& \leq \nu^{m(e)+n(e)-N}\|(Oe)^{N+}\|_{\SSPACE^{\sharp}} + (\nu/\nu_0)^{m(e)+n(e)-N}\|(Oe)^{N-}\|_{(\SSPACE^{\sharp},\nu_0)}\\
& \leq \nu^{m(e)+n(e)-N}\|Oe\|_{\SSPACE^{\sharp}} + (\nu/\nu_0)^{m(e)+n(e)-N}\|Oe\|_{(\SSPACE^{\sharp},\nu_0)}
\end{align*}
Since $\|e\|_{(\SSPACE^{\sharp},\nu)} = \nu^{m(e)+n(e)} \|e\|_{\SSPACE^{\sharp}} = (\nu/\nu_0)^{m(e)+n(e)} \|e\|_{(\SSPACE^{\sharp},\nu_0)}$, it follows that
\begin{equation*}
\frac{\|Oe\|_{(\SSPACE^{\sharp},\nu)}}{\|e\|_{(\SSPACE^{\sharp},\nu)}} \leq \nu^{-N}\frac{\|Oe\|_{\SSPACE^{\sharp}}}{\|e\|_{\SSPACE^{\sharp}}} + (\nu/\nu_0)^{-N}\frac{\|Oe\|_{(\SSPACE^{\sharp},\nu_0)}}{\|e\|_{(\SSPACE^{\sharp},\nu_0)}}
\end{equation*}
Since this holds for all $e$,
and since $\|O\|_{(\SSPACE^{\sharp},\nu) \leftarrow (\SSPACE^{\sharp},\nu)}<\infty$ (see the remark below),
\begin{equation*}
\|O\|_{(\SSPACE^{\sharp},\nu) \leftarrow (\SSPACE^{\sharp},\nu)} \leq \nu^{-N}\|O\|_{\SSPACE^{\sharp} \leftarrow \SSPACE^{\sharp}} + (\nu/\nu_0)^{-N}\|O\|_{(\SSPACE^{\sharp},\nu_0) \leftarrow (\SSPACE^{\sharp},\nu_0)}
\end{equation*}
Fix $\nu_0$, set $N(\nu) = (1-\nu)^{-1/2}$ and let $\nu \uparrow 1$.
Then $\nu^{-N(\nu)} \to 1$ and  $(\nu/\nu_0)^{-N(\nu)} \to 0$, and 
inequality \eqref{dlhdhkfjhfncnncnc} follows,
because $\|O\|_{(\SSPACE^{\sharp},\nu_0) \leftarrow (\SSPACE^{\sharp},\nu_0)}<\infty$ (see the remark below).
\step
\emph{Remark:} We show that $\|O\|_{(\SSPACE^{\sharp},\nu) \leftarrow (\SSPACE^{\sharp},\nu)} < \infty$ if $\max \{1/\kappa_1,1/\kappa_2 \} < \nu \leq 1$. Separately,
$$
\|(\mathbbm{U}^{\sharp})^{-1}\|_{(\SSPACE^{\sharp},\nu) \leftarrow (\SSPACE^{\sharp},\nu)} < \infty
\qquad
\|\bigstar^{\sharp}\|_{(\SSPACE^{\sharp},\nu) \leftarrow (\SSPACE^{\sharp},\nu)} < \infty
$$
The first by $(\mathbbm{U}^{\sharp})^{-1}
= \KS (\mathbbm{U}^{\sharp})^{-1} \KS + (1-\KS)$. The second by direct inspection of \eqref{3i3hj4kh4}, using $\|\RefA{\omega}\|_{(\SSPACE,\nu)} \leq \|\RefA{\omega}\|_{\SSPACE}<\infty$
and $\|\RefB{\omega}+\Corr{\omega}\|_{(\SSPACE,\nu)} \leq \|\RefB{\omega}+\Corr{\omega}\|_{\SSPACE}<\infty$, see \eqref{finnormfinnorm}, and
$\|(\OO_{\RefA{\mu}}^{\sharp})^{-1}\DD{\OO}^{\sharp}\|_{(\SSPACE^{\sharp},\nu) \leftarrow (\SSPACE^{\sharp},\nu)} < \infty$, among others.}
\subsubsection{Exterior estimates: \eqref{dkjfhk4} and \eqref{dhdkfhkdhdkkkk3}} \label{dflfklfjkfhkkkkkkkkkkkkdhkhdk}
\newcommand{\kt}{\mathfrak{t}}
\newcommand{\KT}{\boldsymbol{\kt}}
\newcommand{\kq}{\mathfrak{q}}
\newcommand{\KQ}{\boldsymbol{\kq}}
Fix cutoff operators $\kt$, $\kq$ $\tagforcomplabel{fixcutoffstandq}$ similar to $\ks$. Suppose
\begin{subequations}\label{cutoffrestr}
\begin{alignat}{5}
2 &\leq\;& m(\ks) & \leq\;& m(\kq) & \leq\;& m(\kt) - m(\ks) + 1 &\leq\;& m(\kt)\;\; \tagforcomplabel{supposeTSQ1}\\
1 &\leq\;& n(\ks) &\leq& n(\kq) & \leq& n(\kt) - n(\ks) + 1 &\leq&n(\kt) \;\;\tagforcomplabel{supposeTSQ2}
\end{alignat}
\end{subequations}
and $3 \leq n(\kq)$ $\tagforcomplabel{supposeTSQ3}$.
It follows that $\ks \kq = \kq\ks = \ks$ and $\kq \kt = \kt\kq = \kq$ and $\ks \kt = \kt\ks = \ks$.
For convolution, $\kq((\overline{\kt}\,\cdot\,)\ast (\ks \,\cdot\,))=0$ and $\KQ \SELECTION \Xi \Gamma_2(\overline{\KT} \,\cdot\,,\KS \,\cdot\,)=0$, 
where a dot is any argument. In this section only, we abbreviate
$$\overline{\ks} = 1-\ks,\quad  \overline{\KS} = 1 - \KS, \quad \text{etc}  \qquad \text{(the bar \emph{is not} for complex conjugation)}$$
\thin
For every linear operator $O: \SSPACE \to \SSPACE$, the $\ell^1$ operator norm satisfies
$$\|O\|_{\SSPACE\leftarrow \SSPACE} = \max \big\{\|O\KT\|_{\SSPACE\leftarrow \SSPACE},\|O\overline{\KT}\|_{\SSPACE\leftarrow \SSPACE}\big\}$$
Similar for $\SSPACE^{\sharp}$. Suppose the $\KT$ 
parts in \eqref{dkjfhk4}, \eqref{dhdkfhkdhdkkkk3} are under control:
\begin{subequations} \label{dh33guzg4jdf}
\begin{align}
\label{dh33guzg4jdf1} \|{\mathfrak A} \OO_{\RefA{\mu}}^{-1} (A - \KS A \KS )\KT\|_{\SSPACE\leftarrow \SSPACE} & \leq \mathcal{L}_4\;\; \tagforcomplabel{l4EXTRA1}\\
\label{dh33guzg4jdf2} \|{\mathfrak A}^{\sharp}(\OO_{\RefA{\mu}}^{\sharp})^{-1}\big(A^{\sharp} - \KS A^{\sharp}\KS \big)\KT\|_{\SSPACE^{\sharp}\leftarrow\SSPACE^{\sharp}} & \leq \mathcal{L}_3^{\sharp}\;\; \tagforcomplabel{l3sharpEXTRA1}
\intertext{Suppose that in addition,}
\label{dh33guzg4jdf3} \|{\mathfrak A}\|_{\SSPACE\leftarrow \SSPACE}(\mathcal{K}_2 \mathcal{M}_1
+ \mathcal{M}_2  \mathcal{K}_3)
+ \mathcal{M}_3 \mathcal{K}_3 & \leq \mathcal{L}_4\;\; \tagforcomplabel{l4EXTRA2}\\
\label{dh33guzg4jdf4} \|{\mathfrak A}^{\sharp}\|_{\SSPACE^{\sharp}\leftarrow\SSPACE^{\sharp}}\,\big(\mathcal{K}_2^{\sharp} {\mathcal M}_1
+ {\mathcal M}_2 \mathcal{K}_3^{\sharp} \big)
+ {\mathcal M}_3 \mathcal{K}_3^{\sharp} & \leq \mathcal{L}_3^{\sharp}\;\; \tagforcomplabel{l3sharpEXTRA2}
\end{align}
\end{subequations}
with
\begin{align*}
{\mathcal M}_1 & = 4 \RefA{\mu}(\kappa_2)^{-(n(\kt)-n(\kq)+1)}(1-(\kappa_2)^{-2})^{-1}\\
{\mathcal M}_2 & =  4 \RefA{\mu}^{-1} (\kappa_2)^{-(n(\kq)-n(\ks)+1)}(1-(\kappa_2)^{-1})^{-1}\\
{\mathcal M}_3 & =  2 \big(1-(\kappa_2)^{-1}\big)^{-1} \max \big\{ 2\, (m(\kq ))^{-1}, \RefA{\mu}^{-1}(n(\kq)-1)^{-1}\big\} \big(1+(\kappa_2)^{-1}\big)
\end{align*}
In the rest of this section, we show that \eqref{dh33guzg4jdf3}, \eqref{dh33guzg4jdf4} control the $\overline{\KT}$ parts.\\
In other words, we show that \eqref{dh33guzg4jdf} implies \eqref{dkjfhk4}, \eqref{dhdkfhkdhdkkkk3}.
\thin
For \eqref{dh33guzg4jdf3}, set $O={\mathfrak A} \OO_{\RefA{\mu}}^{-1} (A - \KS A \KS )$, a local notation.  Since $\KS \overline{\KT} = 0$, one has
\begin{align*}
O\overline{\KT} = {\mathfrak A} \OO_{\RefA{\mu}}^{-1} A \overline{\KT}
& =  \big(\KS {\mathfrak A} \KS + \overline{\KS}\big)
\OO_{\RefA{\mu}}^{-1} \big(\KQ + \overline{\KQ}\big)A \overline{\KT}\\
& =  {\mathfrak A}
\OO_{\RefA{\mu}}^{-1} \KQ A \overline{\KT} +
\KS {\mathfrak A} \KS
\OO_{\RefA{\mu}}^{-1}\overline{\KQ}A \overline{\KT}
+
\overline{\KS}
\OO_{\RefA{\mu}}^{-1}\overline{\KQ}A \overline{\KT}
\end{align*}
It follows that
$$
\|O\overline{\KT}\| \leq 
\|{\mathfrak A}\|\,\|\OO_{\RefA{\mu}}^{-1}\|\,\| \KQ A \overline{\KT}\|
+ \|{\mathfrak A}\|\,\|\KS \OO_{\RefA{\mu}}^{-1} \overline{\KQ}\|\, \|A\|
+ \|\OO_{\RefA{\mu}}^{-1} \overline{\KQ}\|\,\|A\|
$$
with $\|\cdot\| = \|\cdot\|_{\SSPACE \leftarrow \SSPACE}$, a local notation.
Recall \eqref{srefassps}. We have
\begin{align*}
\| \KQ A \overline{\KT}\| & = \RefA{\mu} \|\KQ \SELECTION\Gamma_1 \overline{\KT}\| \leq
\RefA{\mu}\,4\| \kq \Band_{1,1}\BandI_{2,1} \overline{\kt}\|_{\Space\leftarrow \Space}
 \leq {\mathcal M}_1\\
\|\KS \OO_{\RefA{\mu}}^{-1} \overline{\KQ}\| & \leq
\max\{
\|\ks \OOA_{\RefA{\mu}}^{-1} \overline{\kq}\|_{\Space\leftarrow \Space},
\|\ks \OOB_{\RefA{\mu}}^{-1} \overline{\kq}\|_{\Space\leftarrow \Space}
\} \leq
{\mathcal M}_2
\\
\|\OO_{\RefA{\mu}}^{-1} \overline{\KQ}\| & \leq
\max\{
\|\OOA_{\RefA{\mu}}^{-1} \overline{\kq}\|_{\Space\leftarrow \Space},
\|\OOB_{\RefA{\mu}}^{-1} \overline{\kq}\|_{\Space\leftarrow \Space}
\} \leq {\mathcal M}_3
\end{align*}
Therefore, \eqref{dh33guzg4jdf3} implies $\|O\overline{\KT}\| \leq \mathcal{L}_4$, as required.
\thin
{\footnotesize In {\bf (A)}, we estimate $\|\ks \OOA_{\RefA{\mu}}^{-1} \overline{\kq}\|_{\Space\leftarrow \Space}$.
Similar for $\|\ks \OOB_{\RefA{\mu}}^{-1} \overline{\kq}\|_{\Space\leftarrow \Space}$ and $\| \kq \Band_{1,1}\BandI_{2,1} \overline{\kt}\|_{\Space\leftarrow \Space}$. In {\bf (B)} we estimate 
$\|\OOB_{\RefA{\mu}}^{-1} \overline{\kq}\|_{\Space\leftarrow \Space}$.
Similar for 
$\|\OOA_{\RefA{\mu}}^{-1} \overline{\kq}\|_{\Space\leftarrow \Space}$.
\step
{\bf (A)} There are numbers $Q_{mnn'}\in \C$ such that $(\ks \OOA_{\RefA{\mu}}^{-1} \overline{\kq} v)_{mn} = \sum_{n'\geq 0} Q_{mnn'} v_{mn'}$ for all $v \in \Space$, and $\| Q_{mnn'}\|_{\C} \leq 4 \RefA{\mu}^{-1}$. Also
$$Q_{mnn'}\neq 0 \quad \Longrightarrow \quad
\Big((m<m(\ks)) \text{ and } (n<n(\ks))\Big) \text{ and } \Big((m \geq m(\kq) )\text{ or } (n'\geq n(\kq))\Big)$$
Since $m(\ks) \leq m(\kq)$, this simplifies to
$$Q_{mnn'}\neq 0 \quad \Longrightarrow \quad
(m<m(\ks)) \text{ and } (n<n(\ks)) \text{ and }  (n'\geq n(\kq))$$
Therefore, with $c_{mn} = (2-\delta_{m0})(2-\delta_{n0})$,
\begin{align*}
\|\ks \OOA_{\RefA{\mu}}^{-1} \overline{\kq}v\|_{\Space}
& = \textstyle\sum_{m,n\geq 0} c_{mn} (\kappa_1)^m(\kappa_2)^n \big\|\textstyle\sum_{n'\geq 0} Q_{mnn'} v_{mn'} \big\|_{\C}\\
& = \textstyle\sum_{0\leq m < m(\ks), 0\leq n < n(\ks)} c_{mn} (\kappa_1)^m(\kappa_2)^n \big\|\textstyle\sum_{n'\geq n(\kq)} Q_{mnn'} v_{mn'} \big\|_{\C}\\
& \leq 4\RefA{\mu}^{-1} \textstyle\sum_{0\leq m < m(\ks), 0\leq n < n(\ks)} c_{mn} (\kappa_1)^m(\kappa_2)^n \textstyle\sum_{n'\geq n(\kq)} \|v_{mn'} \|_{\C}\\
& \leq 4\RefA{\mu}^{-1} \textstyle\sum_{0\leq m < m(\ks), n'\geq n(\kq)} c_{mn'} (\kappa_1)^m(\kappa_2)^{n'} \|v_{mn'} \|_{\C} \textstyle \sum_{0\leq n < n(\ks)}(\kappa_2)^{(n-n')}\\
& \leq 4\RefA{\mu}^{-1} \big(\textstyle\sum_{m,n'\geq 0} c_{mn'} (\kappa_1)^m(\kappa_2)^{n'} \|v_{mn'} \|_{\C}\big)\big(\textstyle \sum_{0\leq n < n(\ks)}(\kappa_2)^{(n-n(\kq))}\big)\\
& \leq 4\RefA{\mu}^{-1} \|v\|_{\Space} (\kappa_2)^{n(\ks)-1-n(\kq)} (1-(\kappa_2)^{-1})^{-1}
\end{align*}
Here $c_{mn}\leq c_{mn'}$ has been used in one step.
\step
\newcommand{\kr}{\mathfrak r}%
{\bf (B)} Let $\kr$ be the cutoff with $m(\kr) = m(\kq)$ and $n(\kr) = n(\kq) - 2$, a local notation.
Then $\OOB_{\RefA{\mu}}^{-1} \overline{\kq}
= \BandI[2,f] \LINDEC_{\RefA{\mu}}^{-1} (1+\Band_{2,-1}) \overline{\kq}
= \BandI[2,f] \LINDEC_{\RefA{\mu}}^{-1} \overline{\kr} (1+\Band_{2,-1}) \overline{\kq}$
 for some $f$. Therefore,
 $$
 \|\OOB_{\RefA{\mu}}^{-1} \overline{\kq}\|_{\Space\leftarrow \Space}
 \leq 
\| \BandI[2,f]\|_{\Space\leftarrow \Space}
\| \LINDEC_{\RefA{\mu}}^{-1} \overline{\kr}\|_{\Space\leftarrow \Space}
(1 + \|\Band_{2,-1}\|_{\Space\leftarrow \Space})
 $$
 Use
$ \| \BandI[2,f]\|_{\Space\leftarrow \Space}
\leq \sqrt{2}(1-(\kappa_2)^{-2})^{-1}$
and $1+\|\Band_{2,-1}\|_{\Space\leftarrow \Space} \leq 1+(\kappa_2)^{-2}$ and
$$
\| \LINDEC_{\RefA{\mu}}^{-1} \overline{\kr}\|_{\Space\leftarrow \Space}
\leq \sqrt{2} \sup_{\text{see below}} \frac{1}{|im/2 + \RefA{\mu}(n+1)|}
\leq \sqrt{2}\,\max \left\{\frac{2}{m(\kq)},\, \frac{1}{\RefA{\mu}(n(\kq)-1)}\right\}
$$
where the $\sup$ is over all $m,n\geq 0$ with
$m \geq m(\kq)$ or $n \geq n(\kq)-2$. Now use \eqref{dhfkjhfkdhfkjdff34}.
 }
\thin
For \eqref{dh33guzg4jdf4}, set $O^{\sharp} =  {\mathfrak A}^{\sharp}(\OO_{\RefA{\mu}}^{\sharp})^{-1}\big(A^{\sharp} - \KS A^{\sharp}\KS \big)$. Then
\begin{align*}
\|O^{\sharp}\overline{\KT}\| & \leq 
\|{\mathfrak A}^{\sharp}\|\,\|(\OO_{\RefA{\mu}}^{\sharp})^{-1}\|\,\| \KQ A^{\sharp} \overline{\KT}\|
+ \|{\mathfrak A}^{\sharp}\|\,\|\KS (\OO_{\RefA{\mu}}^{\sharp})^{-1} \overline{\KQ}\|\, \|A^{\sharp}\|\\
& \hskip 70mm
+ \|(\OO_{\RefA{\mu}}^{\sharp})^{-1} \overline{\KQ}\|\,\|A^{\sharp}\|
\end{align*}
with $\|\cdot\| = \|\cdot\|_{\SSPACE^{\sharp} \leftarrow \SSPACE^{\sharp}}$. 
 Recall \eqref{srefassps}. One has
\begin{align*}
\| \KQ A^{\sharp} \overline{\KT}\| & = \RefA{\mu} \|\KQ \DivideByXiRegularized \Gamma_1^{\sharp} \overline{\KT}\| \leq
\RefA{\mu}\,2\| \kq \Band_{1,1}\BandI_{2,1} \overline{\kt}\|_{\Space\leftarrow \Space}
 \leq {\mathcal M}_1\\
\|\KS (\OO_{\RefA{\mu}}^{\sharp})^{-1} \overline{\KQ}\| & \leq
{\mathcal M}_2
\\
\|(\OO_{\RefA{\mu}}^{\sharp})^{-1} \overline{\KQ}\| & \leq
 {\mathcal M}_3
\end{align*}
Therefore, \eqref{dh33guzg4jdf4} implies $\|O^{\sharp}\overline{\KT}\| \leq \mathcal{L}_3^{\sharp}$, as required.
\subsection{Parameter values and Computer assisted results}\label{fhufdhkdfhfdkjhdhdf}
This section takes care of every individual $\tagforcomp$ tag in Section \ref{sec:ANALYSISanalysis}.
Set
\begin{align*}
\kappa_1 & = 65/64 &
\kappa_2 & = 5/4 &
\kappa_{\ast} & = 1\\
m(\ks) & = 250 & m(\kt) & = 1600 & m(\kq) & =1350 \\
n(\ks) & = 750 & n(\kt) & = 4800 & n(\kq) & =4050
\end{align*}
We have
$\tagforcomprefcheck{fixcutoffs}$
$\tagforcomprefcheck{fixk1k2}$
$\tagforcomprefcheck{fixkstar}$
$\tagforcomprefcheck{fixcutoffstandq}$
$\tagforcomprefcheck{supposeTSQ1}$
$\tagforcomprefcheck{supposeTSQ2}$
$\tagforcomprefcheck{supposeTSQ3}$.
The text files \texttt{RefA.dat}
and \texttt{RefAplusB.dat} contain  $(\RefA{\mu},\RefA{\omega})$
and $(\RefA{\mu}+\RefB{\mu},\RefA{\omega}+\RefB{\omega})$; the file format is self-explanatory from the discussion in Section \ref{sec:datastruc}
 $\tagforcomprefcheck{fixref}$
$\tagforcomprefcheck{fixrefa}$
$\tagforcomprefcheck{fixrefb}$ $\tagforcomprefcheck{supposesupportrefa}$. One finds, either by looking at the first few lines of these files, or with the command \texttt{choptuik LoadAplusB}:
\begin{align*}
\RefA{\mu} & = 722873400\cdot 2^{-32}\\
\RefA{\mu}+\RefB{\mu} & = 786320438575298[\text{\ldots180 final digits omitted}\ldots] \cdot 2^{-650}\tagforB\\
|\RefB{\mu}| & \leq 43\cdot 2^{-40} \tagforB
\end{align*}%
Most readers should ignore the $\tagforB$ tags. (They tag the things that must be revisited if one repeats the construction with a \emph{different} refB, but the \emph{same} refA, to get a better approximation to Choptuik's solution.) Set
\begin{align*}
\mathcal{L}_1 & = 10.5 &
\mathcal{L}_2 & = 2^{-25}\tagforB&
\mathcal{L}_3 & = 2^{-294}\tagforB \\
\mathcal{L}_4 & = 0.625&
\mathcal{L}_5 & = 1.004&
\mathcal{L}_6 & = 258\\
\mathcal{S}_1 & = 2^{-279}\tagforB&
\mathcal{S}_2 & = 0.01\tagforB&
\mathcal{S}_3 & = 0.64\tagforB\\
\mathcal{S}_4 & = 2^{19}&
\mathcal{S}_5 & = 2^{17}\\
\mathcal{L}_1^{\sharp} & = 8&
\mathcal{L}_2^{\sharp} & = 2^{-25}\tagforB \\
\mathcal{L}_3^{\sharp} & = 0.625 &
\mathcal{L}_4^{\sharp} & = 1.004&
\mathcal{L}_5^{\sharp} & = 258
\end{align*}
$\tagforcomprefcheck{fixl1l6}$
$\tagforcomprefcheck{fixs1s5}$
$\tagforcomprefcheck{fixl1sharpl5sharp}$.
Set ${\mathfrak A} = \mathrm{diag}({\mathfrak B},{\mathfrak B},{\mathfrak B},{\mathfrak B})$ and ${\mathfrak A}^{\sharp} = \mathrm{diag}({\mathfrak B},{\mathfrak B})$ where
$$
({\mathfrak B}v)_{mn}
= \begin{cases}
256v_{mn} & \text{if $m<50$ and $n<150$}\\
16 v_{mn} & \text{else, if $m<100$ and $n<300$}\\
4 v_{mn} & \text{else, if $m<150$ and $n<450$}\\
2 v_{mn} & \text{else, if $m<200$ and $n<600$}\\
1.5625 v_{mn} & \text{else, if $m<250$ and $n<750$}\\
v_{mn} & \text{else}
\end{cases}
$$
in particular,
${\mathfrak B} = \ks {\mathfrak B} \ks + (1-\ks)$ $\tagforcomprefcheck{fixgothicA}$
$\tagforcomprefcheck{fixsharpa}$.
It follows that
\begin{align*}
\mathcal{K}_1 & = \tfrac{80}{9} &
\mathcal{K}_2 & \leq 107 &
\mathcal{K}_3 & \leq 22.5\\
\mathcal{S}_{23} & = 0.64\tagforB &
\mathcal{S}_{45} & = 2^{19}\\
\mathcal{K}_1^{\sharp} & = \tfrac{40}{9} &
\mathcal{K}_2^{\sharp} & \leq 107 &
\mathcal{K}_3^{\sharp} & \leq 8.75\\
\mathcal{M}_1 & \leq 2^{-240} &
\mathcal{M}_2 & \leq 2^{-1055} &
\mathcal{M}_3 & \leq 0.027\\
\|{\mathfrak A}\|_{\SSPACE \leftarrow \SSPACE} & \leq 256 &
\|{\mathfrak A}^{\sharp}\|_{\SSPACE^{\sharp} \leftarrow \SSPACE^{\sharp}} & \leq 256
\end{align*}
Now $\tagforcomprefcheck{supposes1}$
$\tagforcomprefcheck{supposes2}$
$\tagforcomprefcheck{supposes3}$
$\tagforcomprefcheck{supposes4}$
$\tagforcomprefcheck{supposes5}$.
\thin
\eqref{dkjfhk1}: Use \eqref{dkhfkjhffjjjsjs} with $\mathbf{v}=\RefA{\omega}$. Use  \texttt{choptuik DedicatedGAMMA2} .\\
\eqref{dkjfhk2}: Use \texttt{choptuik LoadAplusB}. $\tagforB$\\
\eqref{dkjfhk3}: Use \texttt{choptuik FailureToBeSolAplusB}. $\tagforB$ One output line is
\begin{center}
\verb+(l1 norm of SBold(OmegaPlus(refAplusB))) <= 15312135*2^(-320)+
\end{center}
(This command requires $\sim 16$ GB RAM.) This takes care of $\tagforcomprefcheck{supposededicatedgamma2}$ $\tagforcomprefcheck{supposenormB}$ $\tagforcomprefcheck{supposeerrorAB}$.
\thin
{\footnotesize We note that the command \texttt{choptuik FailureToBeSolAplusB} also yields
\begin{center}
\verb+(l1 norm of OmegaPlus(refAplusB)) <= 24707998*2^(-320)+
\end{center}
which includes the constraint equations. We don't need this inequality.}
\thin
For \eqref{dkjfhk4}, 
note that \eqref{dh33guzg4jdf3} holds. To check
\eqref{dh33guzg4jdf1} use the command
\begin{equation}\label{dfhdkhfkh4izihrk}
\texttt{choptuik EXTERIOR\myunderbar ESTIMATE offm\myunderbar offn\myunderbar numm\myunderbar numn}
\end{equation}
where \texttt{offm}, \texttt{offn}, \texttt{numm}, \texttt{numn} have to be replaced by actual numbers that satisfy $\verb+offm+, \verb+offn+ \geq 0$ and
$\verb+numm+, \verb+numn+ > 0$ and
$\verb+offm+ + \verb+numm+ \leq m(\kt)$
and 
$\verb+offn+ + \verb+numn+ \leq n(\kt)$. For example, an actual command would be
\begin{center}
\texttt{choptuik EXTERIOR\myunderbar ESTIMATE 212\myunderbar748\myunderbar2\myunderbar4}
\end{center}
The command \eqref{dfhdkhfkh4izihrk} generates a file with a number of lines, one line for each unit vector in $\SSPACE$ whose Fourier-Chebyshev indices $(m,n)$ satisfy
$\verb+offm+ \leq m < \verb+offm+ + \verb+numm+$ and $\verb+offn+ \leq n < \verb+offn+ + \verb+numn+$. Each line begins with \verb+d_k_m_n+, for the unit vector $e=(e_1,e_2,e_3,e_4) \in \SSPACE$ with
$(e_{\text{\texttt{d}}+1})_{\text{\texttt{mn}}}=1$ if $\text{\texttt{k}}=0$, and $(e_{\text{\texttt{d}}+1})_{\text{\texttt{mn}}}=i$ if $\text{\texttt{k}}=1$, all other entries equal to zero. For example, the line
\begin{equation}\label{hfkh4krkhkfhf}
\verb+1_0_212_750 550*2^(-10) TRUE+
\end{equation}
for the unit vector $(e_2)_{212,750}=1$ asserts that
\begin{equation}\label{sdjdl839h9h933}
\|{\mathfrak A} \OO_{\RefA{\mu}}^{-1} (A - \KS A \KS )e\|_{\SSPACE}/\|e\|_{\SSPACE} \leq 550\cdot 2^{-10}
\end{equation}
The \texttt{TRUE} indicates that this number is $\leq 0.625$, as required by \eqref{dh33guzg4jdf1}.

\vskip 2mm
\noindent The file \verb+EXTERIOR_ESTIMATE_0_0_1600_4800.dat+ was generated on an HPC cluster, and contains the result of applying  \eqref{dfhdkhfkh4izihrk} to all 26868000 unit vectors in $\image \KT\subset \SSPACE$.
Many tools can be used to analyze the file. For example, every line of the file has the same structure as \eqref{hfkh4krkhkfhf}, because
the unix command
\begin{center}
\verb+grep -v -c '^[0123]_[01]_[0-9]*_[0-9]* [0-9]*\*2\^(-10) TRUE $'+\\
\hfill \verb+EXTERIOR_ESTIMATE_0_0_1600_4800.dat+
\end{center}
yields \verb+0+.  Without the \verb+-v+ option, it yields \verb+26868000+.
$\tagforcomprefcheck{supposeextest}$
$\tagforcomprefcheck{l4EXTRA1}$
$\tagforcomprefcheck{l4EXTRA2}$
\thin
For \eqref{dkjfhk5} and \eqref{dkjfhk6}, use the command
\begin{equation}\label{kh3k3494h}
\text{\texttt{choptuik INV offm\myunderbar offn\myunderbar numm\myunderbar numn}}
\end{equation}
with 
$\verb+offm+ + \verb+numm+ \leq m(\ks)$
and 
$\verb+offn+ + \verb+numn+ \leq n(\ks)$ (more details below). 
It shows that there exists a linear map $\mathbbm{V}: \SSPACE \to \R \oplus \Gauged{\SSPACE}$, block diagonal with respect to the decomposition (cf.~\eqref{dfdhfkhfdfkskkskkks})
$$ \big(\image \KS \big) \oplus \big(\image(1-\KS) \big)
\to \big(\R \oplus (\Gauged{\SSPACE} \cap \image \KS )\big)
\oplus \image (1-\KS)$$
such that the high-to-high frequency block $\image (1-\KS)\to \image (1-\KS)$ is the identity, and such that
\begin{equation}\label{dfkhdjdjdj33h4}
\| 1-\mathbbm{U}\mathbbm{V} \|_{\SSPACE \leftarrow \SSPACE} \leq 2^{-16}
\qquad \|\mathbbm{V}{\mathfrak A}^{-1}\|_{\R \oplus \SSPACE \leftarrow \SSPACE} \leq 1
\end{equation}
(The operator ${\mathfrak A}$ was constructed so that the second inequality would hold.)
It follows that $\mathbbm{U}$ is invertible, and given in terms of the Neumann series
\begin{equation}\label{hffjdsjkskskahhhh287383}
\mathbbm{U}^{-1} = \mathbbm{V} \frac{1}{1-(1-\mathbbm{U}\mathbbm{V})}
\quad \text{or}
\quad
\mathbbm{U}^{-1}= \mathbbm{V}{\mathfrak A}^{-1} \frac{1}{1-(1-{\mathfrak A}\mathbbm{U}\mathbbm{V}{\mathfrak A}^{-1})} {\mathfrak A}
\end{equation}
The second Neumann series converges by
\begin{multline*}
\|1-{\mathfrak A}\mathbbm{U}\mathbbm{V}{\mathfrak A}^{-1}\|
= \|{\mathfrak A}(1-\mathbbm{U}\mathbbm{V}){\mathfrak A}^{-1}\|\\
\leq 
\|{\mathfrak A}\|\,\|1-\mathbbm{U}\mathbbm{V}\|\,\|{\mathfrak A}^{-1}\|
\leq
256\cdot 2^{-16}\cdot 1 = 2^{-8} < 1
\end{multline*}
The estimates \eqref{dkjfhk5}, \eqref{dkjfhk6} follow from
$$\|\mathbbm{U}^{-1} {\mathfrak A}^{-1}\|_{\R \oplus \SSPACE \leftarrow \SSPACE}
\leq (1-2^{-8})^{-1} \qquad 
\|\mathbbm{U}^{-1}\|_{\R \oplus \SSPACE \leftarrow \SSPACE} \leq
(1-2^{-8})^{-1} 2^8$$
The file \verb+INV_0_0_250_750.dat+ was generated on an HPC cluster. It contains the result of applying \eqref{kh3k3494h} to all 654375 unit vectors $e \in \image \KS \subset \SSPACE$. For example, it contains a line for the unit vector with $(e_4)_{1,7}=i$,
\begin{center}
\verb+3_1_1_7 580*2^(-32) TRUE 1156600*2^(-13) TRUE+
\end{center}
 This line asserts the existence of a vector $f \in \R \oplus (\Gauged{\SSPACE} \cap \image \KS)$ with
\begin{equation}\label{dk6487674hfkhkf}
\|e- \mathbbm{U}f\|_{\SSPACE} / \|e\|_{\SSPACE} \leq 580\cdot 2^{-32} \qquad
\|f\|_{\R\oplus \SSPACE}/\|e\|_{\SSPACE} \leq 1156600\cdot 2^{-13}
 \end{equation}
The first \verb+TRUE+ indicates that $580\cdot 2^{-32} \leq 2^{-16}$, the second indicates that $1156600\cdot 2^{-13} \leq  \|{\mathfrak A}e\|_{\SSPACE}/\|e\|_{\SSPACE}$, where $\|{\mathfrak A}e\|_{\SSPACE}/\|e\|_{\SSPACE} = 256$ for this particular $e$.
Partially define $\mathbbm{V}$ by $\mathbbm{V}e = f$.
Analogous statements for all unit vectors $e$ define $\mathbbm{V}$ completely, and yield \eqref{dfkhdjdjdj33h4}.
$\tagforcomprefcheck{supposeUinvertible}$
$\tagforcomprefcheck{supposeinvnorm1}$
$\tagforcomprefcheck{supposeinvnorm2}$
\thin
Set
\begin{subequations}\label{kdkkkskkskksksks}
\begin{equation}
\mathcal{R}=2^{-277} \leq 10^{-83} \tagforB
\end{equation}
$\tagforcomprefcheck{fixr}$
$\tagforcomprefcheck{supposer1}$
$\tagforcomprefcheck{supposer2}$
$\tagforcomprefcheck{supposer3}$.
The resulting fixed point $(\mu,\omega)$ satisfies
\begin{equation}
|\mu - (\RefA{\mu} + \RefB{\mu})|
\leq 2^{-277}\; \tagforB
\qquad
\| \omega - (\RefA{\omega}+\RefB{\omega})\|_{\SSPACE}
 \leq 2^{-277}\; \tagforB
 \end{equation}
 \end{subequations}
\thin
We proceed analogously with the $\sharp$ system.\\
\eqref{dhdkfhkdhdkkkk1}: Use \eqref{fdkhfjhfkdhslsdlss} with $\mathbf{v}=\RefA{\omega}$. Use \texttt{choptuik DedicatedGAMMA2}.\\
\eqref{dhdkfhkdhdkkkk2}: Use the triangle inequality,
\texttt{choptuik LoadAplusB} and \eqref{kdkkkskkskksksks}. $\tagforB$\\
\eqref{dhdkfhkdhdkkkk3}: Note that
\eqref{dh33guzg4jdf4} holds, and to check
\eqref{dh33guzg4jdf2} use the command
 \begin{align*}
& \text{\texttt{choptuik SHARP\myunderbar EXTERIOR\myunderbar ESTIMATE offm\myunderbar offn\myunderbar numm\myunderbar numn}}\\
& \text{or see \texttt{SHARP\myunderbar EXTERIOR\myunderbar ESTIMATE\myunderbar 0\myunderbar 0\myunderbar 1600\myunderbar 4800.dat}}
 \end{align*}
$\tagforcomprefcheck{supposel1sharp}$
$\tagforcomprefcheck{supposel2sharp}$
$\tagforcomprefcheck{supposel3sharp}$
$\tagforcomprefcheck{l3sharpEXTRA1}$
$\tagforcomprefcheck{l3sharpEXTRA2}$
 \thin
For 
\eqref{dhdkfhkdhdkkkk4} and \eqref{dhdkfhkdhdkkkk5}, use
 \begin{align*}
& \text{\texttt{choptuik SHARP\myunderbar INV offm\myunderbar offn\myunderbar numm\myunderbar numn}} \\
& \text{or see \texttt{SHARP\myunderbar INV\myunderbar 0\myunderbar 0\myunderbar 250\myunderbar 750.dat}}
 \end{align*}
There exists a linear map $\mathbbm{V}^{\sharp}: \SSPACE^{\sharp} \to \SSPACE^{\sharp}$ with
$\mathbbm{V}^{\sharp} = \KS \mathbbm{V}^{\sharp} \KS + (1-\KS)$ and
$$\|1- \mathbbm{U}^{\sharp}\mathbbm{V}^{\sharp}\|_{\SSPACE^{\sharp} \leftarrow \SSPACE^{\sharp}} \leq 2^{-16}
\qquad \|\mathbbm{V}^{\sharp}({\mathfrak A}^{\sharp})^{-1}\|_{\SSPACE^{\sharp} \leftarrow \SSPACE^{\sharp}} \leq 1$$
$\tagforcomprefcheck{sharpuinvertible}$
$\tagforcomprefcheck{supposel4sharp}$
$\tagforcomprefcheck{supposel5sharp}$. Now
$\tagforcomprefcheck{supposelessthan1}$. 
We are done with all the  $\tagforcomp$ in Section \ref{sec:ANALYSISanalysis}, and we now have a solution to {\bf 2DprobSeries}.
\thin
To see that we also have a solution to {\bf 2Dprob}, i.e.~a solution to the Einstein equations with massless scalar field, we still have to check (b) and (d) in {\bf 2Dpre} with $\xi_{\ast} = \tfrac{1}{2}(\kappa_2+1/\kappa_2)=41/40$.
Let $\Zylinder$ be the cylinder \eqref{eueueziu33333}.
Then
$$\textstyle\sup_{(\tau,\xi) \in \Zylinder} | \TF(v)| \leq \| v\|_{\Space}$$
follows from $|z| \leq \|z\|_{\C}$, and
$|e^{\pm i\theta}| \leq \kappa_2$ for all $\theta \in \C$ with $\cos \theta \in (-\xi_{\ast},\xi_{\ast})$.
Since $\|\RefB{\mu}\oplus \RefB{\omega}+ \Corr{\mu}\oplus \Corr{\omega} \|_{\R\oplus \SSPACE}\leq 2^{-24}$ $\tagforB$, we have
$$|\mu - \RefA{\mu}| + \textstyle\sum_{i=1}^4
\sup_{(\tau,\xi) \in \Zylinder} | \TF(\omega_i - \omega_{\text{RefA},i})|
\leq 2^{-24}\;\;\tagforB$$
Now, (b) and (d) follow immediately by analyzing the data in \texttt{RefA.dat}.
  \thin
For \eqref{dnkdfkjhfkdhfkdhdfdk1} and \eqref{dnkdfkjhfkdhfkdhdfdk2}, note that $|\mu - (\RefA{\mu}+\RefB{\mu})| \leq 10^{-83}\;\tagforB$ and
  \begin{multline}\label{eq23eq23just}
  \left| K - \left(2\pi (\RefA{\mu}+\RefB{\mu}) - \int_0^{2\pi}\dd \tau\, (\omega_{\text{RefA},3}+\omega_{\text{RefB},3})(\tau,0)\right)\right|
  \\ \leq 2\pi \| \Corr{\mu}\oplus \Corr{\omega}\|_{\R \oplus \SSPACE}
   \leq 10^{-82}\;\tagforB
  \end{multline}
See \eqref{zetazeta1}. Now, the data in \texttt{RefAplusB.dat}, or more conveniently the command \texttt{choptuik LoadAplusB}, yield inequalities \eqref{dnkdfkjhfkdhfkdhdfdk1} and \eqref{dnkdfkjhfkdhfkdhdfdk2}.
\subsection{Data structures}\label{sec:datastruc}
To enable the interested reader to translate
between this paper and the accompanying C source code in the directory \texttt{sourcecode} (arXiv ancillary files), we introduce the basic data structures that we use, and explain what data they represent.

The type \verb+Sector+ contains four non-negative integers,
\begin{quote}
\begin{verbatim}
typedef struct
{
        unsigned long off_m;
        unsigned long off_n;
        unsigned long num_m;
        unsigned long num_n;    
} Sector;
\end{verbatim}
\end{quote}
and represents a rectangular region of Fourier-Chebyshev indices $(m,n)$:
\begin{alignat*}{4}
\text{\texttt{off\myunderbar m}} & \leq &\; m & < \text{\texttt{off\myunderbar m}} + \text{\texttt{num\myunderbar m}}\\
\text{\texttt{off\myunderbar n}} & \leq & n & < \text{\texttt{off\myunderbar n}} + \text{\texttt{num\myunderbar n}}
\end{alignat*}
The type \verb+GI+ contains two arbitrary precision integers (\verb+mpz_t+ from \verb+gmp.h+):
\begin{quote}
\begin{verbatim}
typedef struct
{
        mpz_t Re;
        mpz_t Im;
} GI;
\end{verbatim}
\end{quote}
and represents the Gaussian integer $\text{\texttt{Re}}+i\,\text{\texttt{Im}} \in \Z + i \Z$. The type
\begin{quote}
\begin{verbatim}
typedef struct
{
        long TwoExp;
        Sector sec;
        GI** data;
} Field;
\end{verbatim}
\end{quote}
represents the field $v = (v_{mn}) \in \Space$ given by
\begin{align*}
& v_{\text{\texttt{sec.off\myunderbar m}}+\text{\texttt{u}},\,\text{\texttt{sec.off\myunderbar n}}+\text{\texttt{v}}} = 2^{\text{\texttt{TwoExp}}}\, \text{\texttt{data[u][v]}} \\
& \text{for all \texttt{u} and \texttt{v} with:}\;\;
0 \leq \text{\texttt{u}} < \text{\texttt{sec.num\myunderbar m}}
\;\;\text{and}\;\;
0 \leq \text{\texttt{v}} < \text{\texttt{sec.num\myunderbar n}}
\end{align*}
and $v_{mn}=0$ for all $m,n \geq 0$ not in \texttt{sec}.
The offsets 
$\verb+off_m+$, $\verb+off_n+$ allow us to efficiently store and calculate with fields $v$ that are `strongly localized' in frequency space.
For example, for tasks (2a), (2b), (3) in Section \ref{allthingscomp}, we often use fields with $\verb+off_m+ \gg \verb+num_m+$
 or $\verb+off_n+ \gg \verb+num_n+$.
The type
\begin{quote}
\begin{verbatim}
typedef struct
{
        long TwoExp;
        mpz_t coeff;
} DyadicQ;
\end{verbatim}
\end{quote}
represents the dyadic rational number $\text{\texttt{coeff}}\cdot 2^{\text{\texttt{TwoExp}}}$. The type
\begin{quote}
\begin{verbatim}
typedef struct
{
        unsigned long num_comp;
        Field* comp;
} MultiField;
\end{verbatim}
\end{quote}
represents the multifield $(\text{\texttt{comp[0]}},\ldots,\text{\texttt{comp[num\myunderbar comp-1]}}) \in \Space^{\;\text{\texttt{num\myunderbar comp}}}$, and
\begin{quote}
\begin{verbatim}
typedef struct
{
        DyadicQ mu;
        MultiField mf;  
} mu_MultiField;
\end{verbatim}
\end{quote}
represents $(\text{\texttt{mu}},\text{\texttt{mf}}) \in \R \oplus \Space^{\;\text{\texttt{mf.num\myunderbar comp}}}$. One can now also read the text files \texttt{RefA.dat} and \texttt{RefAplusB.dat}, each contains a \texttt{mu\myunderbar MultiField}.
\let\oldbibliography\thebibliography
\renewcommand{\thebibliography}[1]{\oldbibliography{#1}\setlength{\itemsep}{-3pt}}

\end{document}

%% file: penrose.pstex_t
\begin{picture}(0,0)%
\includegraphics{penrose.pstex}%
\end{picture}%
\setlength{\unitlength}{3947sp}%
\begingroup\makeatletter\ifx\SetFigFont\undefined%
\gdef\SetFigFont#1#2#3#4#5{%
  \reset@font\fontsize{#1}{#2pt}%
  \fontfamily{#3}\fontseries{#4}\fontshape{#5}%
  \selectfont}%
\fi\endgroup%
\begin{picture}(5843,2944)(4036,-4559)
\put(6901,-2311){\makebox(0,0)[lb]{\smash{{\SetFigFont{12}{14.4}{\familydefault}{\mddefault}{\updefault}{\color[rgb]{0,0,0}\emph{critical light cone}, $u_{-}=0$}%
}}}}
\put(7726,-2986){\makebox(0,0)[lb]{\smash{{\SetFigFont{12}{14.4}{\familydefault}{\mddefault}{\updefault}{\color[rgb]{0,0,0}boundary of domain on}%
}}}}
\put(7726,-3436){\makebox(0,0)[lb]{\smash{{\SetFigFont{12}{14.4}{\familydefault}{\mddefault}{\updefault}{\color[rgb]{0,0,0}$u_{-}=-u_+/81>0$}%
}}}}
\put(7726,-3211){\makebox(0,0)[lb]{\smash{{\SetFigFont{12}{14.4}{\familydefault}{\mddefault}{\updefault}{\color[rgb]{0,0,0}which we prove existence}%
}}}}
\put(7051,-4486){\makebox(0,0)[lb]{\smash{{\SetFigFont{12}{14.4}{\familydefault}{\mddefault}{\updefault}{\color[rgb]{0,0,0}$u_+$ const: an outgoing light cone}%
}}}}
\put(7051,-4261){\makebox(0,0)[lb]{\smash{{\SetFigFont{12}{14.4}{\familydefault}{\mddefault}{\updefault}{\color[rgb]{0,0,0}$u_-$ const: an incoming light cone}%
}}}}
\put(6901,-2536){\makebox(0,0)[lb]{\smash{{\SetFigFont{12}{14.4}{\familydefault}{\mddefault}{\updefault}{\color[rgb]{0,0,0}spacetime is analytic across it}%
}}}}
\put(6151,-1786){\makebox(0,0)[lb]{\smash{{\SetFigFont{12}{14.4}{\familydefault}{\mddefault}{\updefault}{\color[rgb]{0,0,0}future singular point $(u_-,u_+)=(0,0)$}%
}}}}
\put(4051,-3886){\makebox(0,0)[lb]{\smash{{\SetFigFont{12}{14.4}{\familydefault}{\mddefault}{\updefault}{\color[rgb]{0,0,0}shaded region}%
}}}}
\put(4051,-2611){\makebox(0,0)[lb]{\smash{{\SetFigFont{12}{14.4}{\familydefault}{\mddefault}{\updefault}{\color[rgb]{0,0,0}$u_+=u_-$}%
}}}}
\put(4051,-2386){\makebox(0,0)[lb]{\smash{{\SetFigFont{12}{14.4}{\familydefault}{\mddefault}{\updefault}{\color[rgb]{0,0,0}\emph{central geodesic}}%
}}}}
\put(4051,-3661){\makebox(0,0)[lb]{\smash{{\SetFigFont{12}{14.4}{\familydefault}{\mddefault}{\updefault}{\color[rgb]{0,0,0}$u_{\pm}<0$ in the}%
}}}}
\end{picture}%

%% file: SIGMANEW.pstex_t
\begin{picture}(0,0)%
\includegraphics{SIGMANEW.pstex}%
\end{picture}%
\setlength{\unitlength}{3947sp}%
\begingroup\makeatletter\ifx\SetFigFont\undefined%
\gdef\SetFigFont#1#2#3#4#5{%
  \reset@font\fontsize{#1}{#2pt}%
  \fontfamily{#3}\fontseries{#4}\fontshape{#5}%
  \selectfont}%
\fi\endgroup%
\begin{picture}(5606,2165)(5657,-3983)
\put(6901,-2311){\makebox(0,0)[lb]{\smash{{\SetFigFont{12}{14.4}{\familydefault}{\mddefault}{\updefault}{\color[rgb]{0,0,0}$u_{-}=0$}%
}}}}
\put(6224,-3736){\makebox(0,0)[lb]{\smash{{\SetFigFont{12}{14.4}{\familydefault}{\mddefault}{\updefault}{\color[rgb]{0,0,0}$\Sigma$}%
}}}}
\put(10051,-2311){\makebox(0,0)[lb]{\smash{{\SetFigFont{12}{14.4}{\familydefault}{\mddefault}{\updefault}{\color[rgb]{0,0,0}$u_{-}=0$}%
}}}}
\put(9365,-3736){\makebox(0,0)[lb]{\smash{{\SetFigFont{12}{14.4}{\familydefault}{\mddefault}{\updefault}{\color[rgb]{0,0,0}$\Sigma'$}%
}}}}
\end{picture}%

%% file: complexdomain.pstex_t
\begin{picture}(0,0)%
\includegraphics{complexdomain.pstex}%
\end{picture}%
\setlength{\unitlength}{3947sp}%
\begingroup\makeatletter\ifx\SetFigFont\undefined%
\gdef\SetFigFont#1#2#3#4#5{%
  \reset@font\fontsize{#1}{#2pt}%
  \fontfamily{#3}\fontseries{#4}\fontshape{#5}%
  \selectfont}%
\fi\endgroup%
\begin{picture}(4708,1160)(-311,-2900)
\put(4276,-2836){\makebox(0,0)[lb]{\smash{{\SetFigFont{12}{14.4}{\familydefault}{\mddefault}{\updefault}{\color[rgb]{0,0,0}$\xi\in\C$}%
}}}}
\put(4126,-2536){\makebox(0,0)[lb]{\smash{{\SetFigFont{12}{14.4}{\familydefault}{\mddefault}{\updefault}{\color[rgb]{0,0,0}$1$}%
}}}}
\put(2701,-2536){\makebox(0,0)[lb]{\smash{{\SetFigFont{12}{14.4}{\familydefault}{\mddefault}{\updefault}{\color[rgb]{0,0,0}$-1$}%
}}}}
\put(491,-2761){\makebox(0,0)[lb]{\smash{{\SetFigFont{12}{14.4}{\familydefault}{\mddefault}{\updefault}{\color[rgb]{0,0,0}$\tau\in\C$}%
}}}}
\put(2094,-2366){\makebox(0,0)[lb]{\smash{{\SetFigFont{12}{14.4}{\familydefault}{\mddefault}{\updefault}{\color[rgb]{0,0,0}$\times$}%
}}}}
\end{picture}%

%% file: lowhigh.pstex_t
\begin{picture}(0,0)%
\includegraphics{lowhigh.pstex}%
\end{picture}%
\setlength{\unitlength}{4736sp}%
\begingroup\makeatletter\ifx\SetFigFont\undefined%
\gdef\SetFigFont#1#2#3#4#5{%
  \reset@font\fontsize{#1}{#2pt}%
  \fontfamily{#3}\fontseries{#4}\fontshape{#5}%
  \selectfont}%
\fi\endgroup%
\begin{picture}(2655,1186)(136,-4475)
\put(2776,-3436){\makebox(0,0)[lb]{\smash{{\SetFigFont{14}{16.8}{\familydefault}{\mddefault}{\updefault}{\color[rgb]{0,0,0}$n$}%
}}}}
\put(151,-4411){\makebox(0,0)[lb]{\smash{{\SetFigFont{14}{16.8}{\familydefault}{\mddefault}{\updefault}{\color[rgb]{0,0,0}$m$}%
}}}}
\end{picture}%

%% file: matrixX.pstex_t
\begin{picture}(0,0)%
\includegraphics{matrixX.pstex}%
\end{picture}%
\setlength{\unitlength}{3947sp}%
\begingroup\makeatletter\ifx\SetFigFont\undefined%
\gdef\SetFigFont#1#2#3#4#5{%
  \reset@font\fontsize{#1}{#2pt}%
  \fontfamily{#3}\fontseries{#4}\fontshape{#5}%
  \selectfont}%
\fi\endgroup%
\begin{picture}(2119,2128)(3894,-4374)
\put(5909,-2410){\makebox(0,0)[lb]{\smash{{\SetFigFont{12}{14.4}{\familydefault}{\mddefault}{\updefault}{\color[rgb]{0,0,0}$N$}%
}}}}
\put(3961,-4310){\makebox(0,0)[lb]{\smash{{\SetFigFont{12}{14.4}{\familydefault}{\mddefault}{\updefault}{\color[rgb]{0,0,0}$N$}%
}}}}
\put(4024,-2551){\makebox(0,0)[lb]{\smash{{\SetFigFont{12}{14.4}{\familydefault}{\mddefault}{\updefault}{\color[rgb]{0,0,0}$1$}%
}}}}
\put(5407,-3812){\makebox(0,0)[lb]{\smash{{\SetFigFont{12}{14.4}{\familydefault}{\mddefault}{\updefault}{\color[rgb]{0,0,0}\ding{203}}%
}}}}
\put(4319,-2718){\makebox(0,0)[lb]{\smash{{\SetFigFont{12}{14.4}{\familydefault}{\mddefault}{\updefault}{\color[rgb]{0,0,0}\ding{202}}%
}}}}
\put(5402,-2708){\makebox(0,0)[lb]{\smash{{\SetFigFont{12}{14.4}{\familydefault}{\mddefault}{\updefault}{\color[rgb]{0,0,0}\ding{204}}%
}}}}
\put(4318,-3806){\makebox(0,0)[lb]{\smash{{\SetFigFont{12}{14.4}{\familydefault}{\mddefault}{\updefault}{\color[rgb]{0,0,0}\ding{205}}%
}}}}
\put(3909,-2918){\makebox(0,0)[lb]{\smash{{\SetFigFont{12}{14.4}{\familydefault}{\mddefault}{\updefault}{\color[rgb]{0,0,0}$N'$}%
}}}}
\put(4517,-2393){\makebox(0,0)[lb]{\smash{{\SetFigFont{12}{14.4}{\familydefault}{\mddefault}{\updefault}{\color[rgb]{0,0,0}$N'$}%
}}}}
\put(4173,-2393){\makebox(0,0)[lb]{\smash{{\SetFigFont{12}{14.4}{\familydefault}{\mddefault}{\updefault}{\color[rgb]{0,0,0}$1$}%
}}}}
\end{picture}%

%% file: reduction.pstex_t
\begin{picture}(0,0)%
\includegraphics{reduction.pstex}%
\end{picture}%
\setlength{\unitlength}{3947sp}%
\begingroup\makeatletter\ifx\SetFigFont\undefined%
\gdef\SetFigFont#1#2#3#4#5{%
  \reset@font\fontsize{#1}{#2pt}%
  \fontfamily{#3}\fontseries{#4}\fontshape{#5}%
  \selectfont}%
\fi\endgroup%
\begin{picture}(6438,1289)(293,-1123)
\put(5221,-736){\makebox(0,0)[lb]{\smash{{\SetFigFont{12}{14.4}{\familydefault}{\mddefault}{\updefault}{\color[rgb]{0,0,0}2D formulation}%
}}}}
\put(1101,-441){\makebox(0,0)[lb]{\smash{{\SetFigFont{12}{14.4}{\familydefault}{\mddefault}{\updefault}{\color[rgb]{0,0,0}abstract 4D problem}%
}}}}
\put(521,-711){\makebox(0,0)[lb]{\smash{{\SetFigFont{12}{14.4}{\familydefault}{\mddefault}{\updefault}{\color[rgb]{0,0,0}$\Ric_g=2\,\dd \phi\otimes\dd\phi ,\;\;\;\; \Box_g \phi=0$}%
}}}}
\put(3706,-981){\makebox(0,0)[lb]{\smash{{\SetFigFont{12}{14.4}{\familydefault}{\mddefault}{\updefault}{\color[rgb]{0,0,0}Section \ref{2d4d}}%
}}}}
\put(3701, 19){\makebox(0,0)[lb]{\smash{{\SetFigFont{12}{14.4}{\familydefault}{\mddefault}{\updefault}{\color[rgb]{0,0,0}Section \ref{4d2d}}%
}}}}
\put(5506,-481){\makebox(0,0)[lb]{\smash{{\SetFigFont{12}{14.4}{\familydefault}{\mddefault}{\updefault}{\color[rgb]{0,0,0}concrete}%
}}}}
\end{picture}%

%% file: causal.pstex_t
\begin{picture}(0,0)%
\includegraphics{causal.pstex}%
\end{picture}%
\setlength{\unitlength}{3947sp}%
\begingroup\makeatletter\ifx\SetFigFont\undefined%
\gdef\SetFigFont#1#2#3#4#5{%
  \reset@font\fontsize{#1}{#2pt}%
  \fontfamily{#3}\fontseries{#4}\fontshape{#5}%
  \selectfont}%
\fi\endgroup%
\begin{picture}(2853,2401)(3538,-2975)
\put(6376,-2761){\makebox(0,0)[lb]{\smash{{\SetFigFont{12}{14.4}{\familydefault}{\mddefault}{\updefault}{\color[rgb]{0,0,0}$\xi$}%
}}}}
\put(3553,-2911){\makebox(0,0)[lb]{\smash{{\SetFigFont{12}{14.4}{\familydefault}{\mddefault}{\updefault}{\color[rgb]{0,0,0}$0$}%
}}}}
\put(4760,-2905){\makebox(0,0)[lb]{\smash{{\SetFigFont{12}{14.4}{\familydefault}{\mddefault}{\updefault}{\color[rgb]{0,0,0}$1$}%
}}}}
\put(3696,-736){\makebox(0,0)[lb]{\smash{{\SetFigFont{12}{14.4}{\familydefault}{\mddefault}{\updefault}{\color[rgb]{0,0,0}$\tau$}%
}}}}
\end{picture}%